\documentclass[preprint,notitlepage,superscriptaddress,frontmatterverbose,amsmath,amssymb,aps,prd,floatfix,nofootinbib]{revtex4-2}

\usepackage{graphicx}
\usepackage{listings}
\usepackage{hyperref}
\usepackage{bigints}
\usepackage{cleveref}
\usepackage{slashed}
\usepackage{orcidlink}
\usepackage{mathtools}
\usepackage{booktabs}

\allowdisplaybreaks

\usepackage{relsize}
\def\babar{\mbox{\slshape B\kern-0.1em{\smaller A}\kern-0.1em
    B\kern-0.1em{\smaller A\kern-0.2em R}}~}

\def \beq{\begin{equation}}
\def \eeq{\end{equation}}
\def \bea{\begin{eqnarray}}
\def \eea{\end{eqnarray}}
\def \bma{\begin{matrix}}
\def \ema{\end{matrix}}

\def \BDtaunu{{\overline B} \to D \tau^{-} {\bar\nu}_\tau}

\def \BDstartaunu{{\overline B} \to D^{*} \tau^{-} {\bar\nu}_\tau}

\def \BDstarlnu{{\overline B} \to D^{*} \ell^{-} {\bar\nu}_\ell}
\def \BDstarlN{{\bar B}\to D^*(\to D\pi)\ell^-{\bar N}_R}

\def \Amp{\mathcal{M}}
\def \BDslnu{{\bar B}\to D^*\ell^-{\bar\nu}_L}
\def \ew{\epsilon_{W}}
\def \eds{\epsilon_{D^*}}

\def \BDslN{\bar{B}\to D^*\ell^-{\bar N}_R}

\begin{document}
\title{A massive right-handed sterile neutrino in three-body ${\bar B}\to D^*\ell^-{\bar N}_R$ decays - a Monte Carlo simulation study}

\author{Bhubanjyoti Bhattacharya\,\orcidlink{0000-0003-2238-321X}}
\email{bbhattach@ltu.edu}
\affiliation{Department of Natural Sciences, Lawrence Technological University, Southfield, MI 48075, USA}

\author{Thomas E. Browder\,\orcidlink{0000-0001-7357-9007}}
\email{teb@phys.hawaii.edu}
\affiliation{Department of Physics and Astronomy, 2505 Correa Road, University of Hawaii, Honolulu,  HI, 96822, USA}

\author{Lucien M. Cremaldi\,\orcidlink{0000-0001-5550-7827}}
\email{cremaldi@phy.olemiss.edu}
\affiliation{Department of Physics and Astronomy, 108 Lewis Hall, University of Mississippi, Oxford, MS 38677-1848, USA}

\author{Horacio Crotte Ledesma\,\orcidlink{0000-0003-2670-5618}}
\email{horacio4@hawaii.edu}
\affiliation{Department of Physics and Astronomy, 2505 Correa Road, University of Hawaii, Honolulu,  HI, 96822, USA}

\author{Nilakshi Das\,\orcidlink{0000-0002-5824-3094}}
\email{nilakshidas0225@gmail.com}
\affiliation{Indian Institute of Technology Gandhinagar, Department of Physics, \\ Gujarat 382355, India}

\author{Alakabha Datta\,\orcidlink{0000-0001-8713-2783}}
\email{datta@phy.olemiss.edu}
\affiliation{Department of Physics and Astronomy, 108 Lewis Hall, University of Mississippi, Oxford, MS 38677-1848, USA}

\author{Tejhas Kapoor\,\orcidlink{0000-0001-5726-3037}}
\email{kapoor@lpccaen.in2p3.fr}
\affiliation{LPC Caen, Normandie Univ, ENSICAEN, UNICAEN, CNRS/IN2P3, 6 boulevard Maréchal Juin, Caen, 14050 France}

\author{Kumar Pandey\,\orcidlink{0009-0009-9944-3322}}
\email{kpandey@go.olemiss.edu}
\affiliation{Department of Physics and Astronomy, 108 Lewis Hall, University of Mississippi, Oxford, MS 38677-1848, USA}

\author{Alexei Sibidanov\,\orcidlink{0000-0001-8805-4895}}
\email{sibid@hawaii.edu}
\affiliation{Department of Physics and Astronomy, 2505 Correa Road, University of Hawaii, Honolulu,  HI, 96822, USA}

\begin{abstract}
We investigate the differential decay distributions of $\bar{B}\to D^*\ell^-{\bar N}_R $, where $N_R$ is a heavy right-handed neutrino (RHN) and $\ell = e, \mu$. We employ a newly developed Monte Carlo event generator based on the EvtGen framework that can simulate beyond-the-Standard-Model processes. We validate the implementation of the model by comparing the Monte-Carlo results with analytical results for several kinematic variables such as $q^2$ and missing mass, as well as signatures in the angular distributions such as $A_{FB}$, $S_{3}$, $S_{5}$, etc. This MC Generator will enable new experimental searches for RHNs. We discuss the sensitivity of various observables to the presence of an RHN and the feasibility of an experimental analysis at Belle II.
\end{abstract}

         \maketitle
\newpage

\section{Introduction}

The experimental evidence on neutrino masses and mixing indicates the need for beyond-the-Standard-Model (BSM) physics. The Standard Model (SM) only contains left-handed neutrino fields, $\nu_L$. A simple extension of the SM by adding a right-handed neutrino (RHN) can account for neutrino masses via the seesaw mechanism. While there are other mechanisms to introduce neutrino masses, these often involve a sterile heavy neutral lepton that can mix with the active neutrinos. In the seesaw picture, the active-neutrino mass matrix can be expressed as $m_\nu\approx- m_D M_{N_R}^{-1} m_D^T$, where $m_D$ is the Dirac mass matrix and $M_{N_R}$ is the sterile-neutrino mass matrix. Neutrino oscillations give $\Delta m_\nu^2 \sim m_\nu^2$, which can be accommodated for a wide range of $M_{N_R}$ by adjusting $m_D$. If $m_D$ is taken to be near the top quark mass, then $M_{N_R}$ is a high mass near the GUT scale, well above the electroweak scale of $\sim $ 100 GeV. However, GeV-scale or lighter RHNs have been used to understand baryogenesis and dark matter \cite{Abazajian:2012ys, Canetti:2012kh}. Hence, the mass scale of sterile RHNs can be above or below the electroweak scale. In this work, we will consider a sterile RHN with a mass in the MeV-GeV range, which may be accessible in $B$ decays or other experiments. We note that the particle $N_R$ can refer to any sterile heavy neutral lepton, and thus may also be a candidate particle in dark sector models. 

Sterile neutrinos may be produced in experiments if they interact with SM particles. The simplest mechanism for generating such an interaction is through the mixing of sterile neutrinos with the active SM neutrinos. However, many extensions of the SM contain sterile neutrinos that have new interactions with SM particles as well as additional new particles. A general model-independent framework for studying the interactions of sterile neutrinos with masses below the electroweak scale is the effective field theory called the \textit{Standard Model Neutrino Effective Field Theory} (SMNEFT)~\cite{delAguila:2008ir, Aparici:2009fh, Bhattacharya:2015vja, Liao:2016qyd, Bischer:2019ttk}. This effective theory satisfies the gauge symmetries of the SM, and only right-handed (RH) sterile neutrinos appear as additional particles in the interactions. It systematically organizes the interactions between RHNs and the SM in terms of higher-dimensional operators, starting with operators of mass dimension six. These operators allow us to calculate the production and decay of RHNs in transitions involving SM particles. In this work, we will discuss the production of an RHN in the decay ${\bar B}\to D^*\ell^-{\bar N}_R$, where $N_R$ is the sterile state and $\ell = e, \mu$. The produced $N_R$ may be invisible or long-lived, or, as considered in other work, decay promptly to visible SM states via mixing or other new interactions. In our study, we will assume $N_R$ to be invisible or sufficiently long-lived to decay outside the detector. Some experimental and theoretical studies of RHN production in semileptonic decays in which the RHN promptly decays into visible particles can be found in Ref.~\cite{Belle:2013ytx, Belle:2022tfo, Han:2022uho, Bolton:2026yrt}. Studies of RHN production in charged-current semileptonic $B$ decays with a light lepton have been discussed recently in Refs.~\cite{Datta:2022czw,Han:2022uho,Bernlochner:2024xiz, Becirevic:2024iyi,Hu:2024ozo,Boora:2024lso,Kolay:2026mgv} with a claim of a $\sim2.5\sigma$ excess for an additional neutrino state around 354 MeV \cite{Bernlochner:2024xiz}. RHNs in semileptonic $B$ decays with a $\tau$ lepton have been discussed recently in Refs.~\cite{Endo:2026dxr,Becirevic:2026tnk, Das:2026uyt}. 

An RHN has been invoked as a solution to the anomalies in the 
charged-current semileptonic $B$ decays with a $\tau$ final state. In these decays, measurements of the Lepton Flavor Universality (LFU) of the gauge interactions are made via the following ratios
\beq
R(D) ~\equiv~ \frac{{\cal B}(\BDtaunu)}{{\cal B}({\overline B}\to D\ell^-{\bar \nu}_\ell)} \qquad
R(D^*) ~\equiv~ \frac{{\cal B}(\BDstartaunu)}{{\cal B}(\BDstarlnu)}.
\label{RDexpt}
\eeq
Over the years, measurements of these ratios \cite{babaranomaly1, babaranomaly2, belleanomaly1, belleanomaly2, belleanomaly3, belleanomaly4, belle2anomaly1, lhcbanomaly1,lhcbanomaly2} have shown deviations from SM expectations. Including correlations, one finds that the current deviation from the SM expectation is at the $3.7\sigma$ level \cite{HeavyFlavorAveragingGroupHFLAV:2024ctg}.

The presence of an additional light RHN contribution always enhances the decay rate for $\BDstartaunu$ because even in the limit of vanishing active neutrino mass, the SM amplitude does not interfere with the RHN contribution. This feature has been used to naturally explain the enhancement in the measurements of $R(D^{(*)})$ \cite{He:2012zp, Cvetic:2017gkt, Asadi:2018wea, Greljo:2018ogz, Azatov:2018kzb, Babu:2018vrl, Mandal:2020htr} assuming lepton-generation non-universal NP effects.

Another motivation for sterile neutrinos is the first evidence for $B^+\to K^+ + \mathrm{inv}$ reported by the Belle II collaboration
~\cite{Belle-II:2023esi}:
\begin{align}
  \mathcal{B}(B^+ \to K^+ + \mathrm{inv}) = (2.3 \pm 0.5 \, (\text{stat})^{+0.5}_ {-0.4} \, (\text{syst})) \times 10^{-5}.
\end{align}
The measured value of this process exceeds the SM prediction, where it is
interpreted as ${\cal B}(B^+\to K^+\nu{\bar\nu})$, by approximately
$ 2.7\sigma$. This process, because of the invisible final states, is a very powerful
probe of weakly coupled particles and, in particular, sterile neutrinos that are invisible or very long-lived. 

In the case of $\BDslnu$ decays, there is no evidence of anomalies in branching ratios; here the measurements agree well with their SM predictions. However, shapes of various angular distributions have not yet been precisely measured. Therefore, signatures of an RHN may become visible in future precise measurements of such shape functions. Furthermore, the missing-mass distribution can be a valuable tool to find a signal of an RHN produced in $\BDslN$. While a light massive neutrino might be consistent with the published missing-mass distribution \cite{Belle:2023bwv}, experimental cuts must be optimized for a GeV-mass RHN. In general, for a more massive neutrino one expects the associated lepton to have less available phase space and hence also a softer lepton spectrum.

In this work, we first review the theoretical distribution for $\BDslnu$, including the RHN-associated decay $\BDslN$. We then simulate these angular distributions using a newly developed Monte Carlo (MC) event generator built upon the EvtGen framework. In an earlier work \cite{Bhattacharya:2022bdk}, we simulated NP signatures in the EvtGen framework with only left-handed neutrinos in the SM effective theory formalism. In this work, we extend our earlier work by including a light RHN. The development of this new MC generator will also aid future RHN searches. After matching the theoretical distributions with the simulated ones, we perform an
initial experimental feasibility study that includes the experimental efficiencies of an RHN signal in $\BDslnu$.

The paper is organized as follows. In Sec.~\ref{sec:effectivehamiltonian}, we describe the NP effective Hamiltonian. In Sec.~\ref{sec:angdistfullbdmunu}, we present the full angular distribution and discuss how to extract various angular observables. In Sec.~\ref{Section:RHN_EvtGEN}, we discuss the implementation of an RHN in EvtGen, while in Sec.~\ref{sec:Summarized plots}, we present a summary of the plots showing interesting correlations between observables. In Sec.~\ref{sec:exptanalysis}, we present our experimental analysis. Finally, in Sec.~\ref{sec:conclusion}, we summarize our work and present our conclusions.

\section{Effective Hamiltonian}
\label{sec:effectivehamiltonian}

At the $b$-quark mass scale, the relevant effective field theory for $b\to c\ell \nu_L$ (SM case) and $b\to c\ell N_R$ (NP case) contains the following dimension-six four-fermion operators:
\begin{eqnarray}
\label{eq:effective_hamiltonian}
\mathcal{H}_{\rm eff}
&=&
\frac{4G_F}{\sqrt{2}}\,V_{cb}
\left[
O_{V}^{LL}
+ C_{V}^{LR} O_{V}^{LR}
+ C_{V}^{RR} O_{V}^{RR}
+ C_{S}^{LR} O_{S}^{LR}
+ C_{S}^{RR} O_{S}^{RR}
+ C_{T}^{RR} O_{T}^{RR}
\right]\,,
\end{eqnarray}
where the operators are given as
\bea
O_{V}^{LL} ~=~ (\bar c\gamma^\mu P_L b)(\bar\ell\gamma_\mu\nu_L)\,,&&~~
O_{V}^{LR} ~=~ (\bar c\gamma^\mu P_L b)(\bar\ell\gamma_\mu N_R)\,,~~
O_{V}^{RR} ~=~ (\bar c\gamma^\mu P_R b)(\bar\ell\gamma_\mu N_R)\,,~ \nonumber \\
O_{S}^{LR} ~=~ (\bar c P_L b)(\bar\ell N_R)\,,&&~~
O_{S}^{RR} ~=~ (\bar c P_R b)(\bar\ell N_R)\,,~~
O_{T}^{RR} ~=~ (\bar c\sigma^{\mu\nu}P_R b)(\bar\ell\sigma_{\mu\nu}N_R)\,,~
\eea
and their corresponding Wilson coefficients are $C^{AB}_X$ with $X = S, V$, or $T$ for scalar, vector, or tensor and $A,B$ can be $L$ for left-handed or $R$ for right-handed interaction. Here $O^V_{LL}$ represents the SM contribution to $b\to c\ell\nu_L$. Note that the operator $O_V^{LR}$ does not have a gauge-invariant dimension-six four-fermion representation in the SMNEFT ~\cite{Robinson:2018gza, Datta:2020ocb}. In this article, we only consider the operator $O_{V}^{RR}$ as it appears in several UV-complete models \cite{Robinson:2018gza}. We take its WC, $C^{RR}_V$, to be purely real throughout this analysis. Our analysis can easily be adapted to include other operators or combinations of operators.

\section{Differential distribution for $\BDstarlN$}
\label{sec:angdistfullbdmunu}

\begin{figure}[!htbp]
        \includegraphics[width=0.33\textwidth]{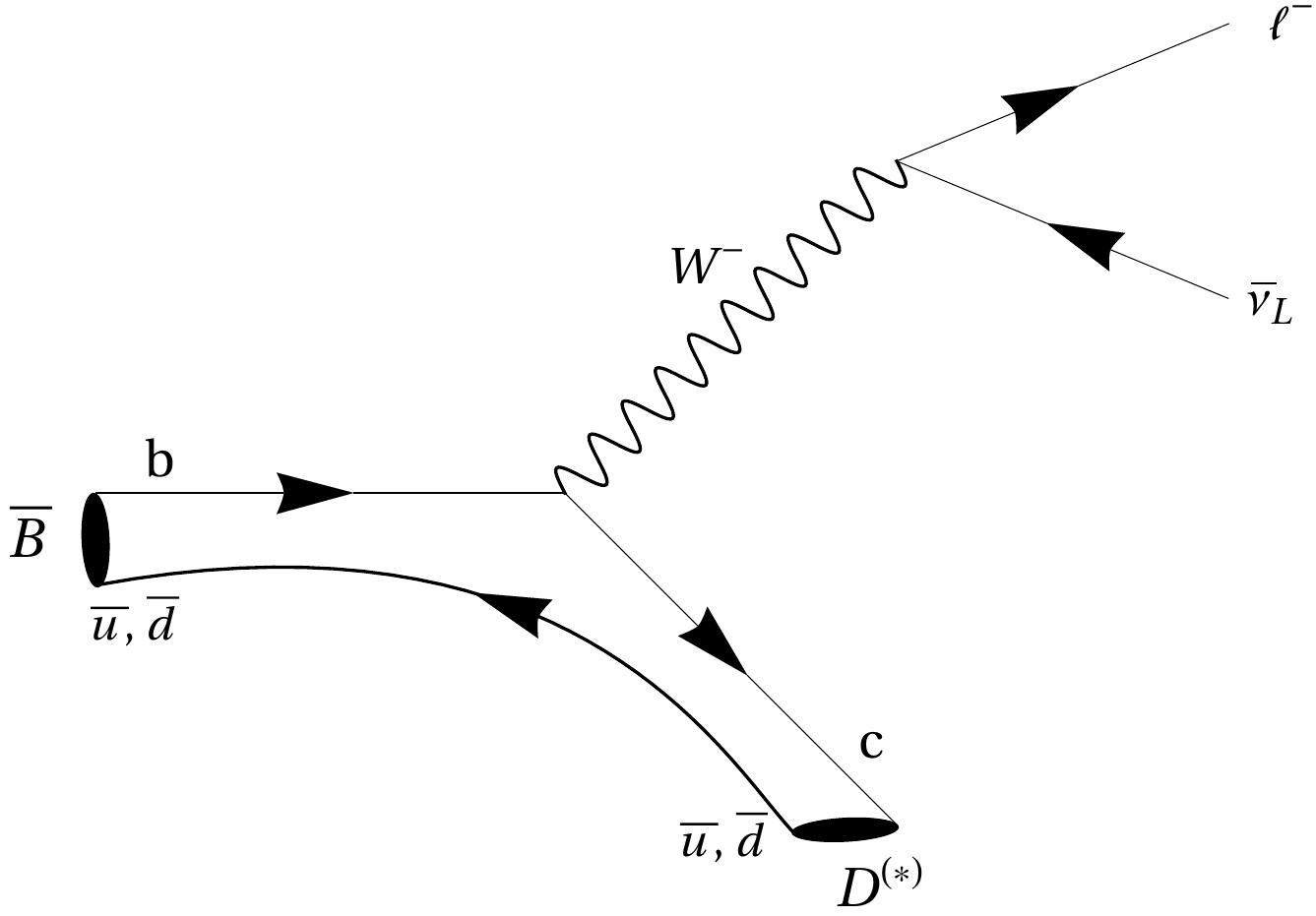} \hspace{15truemm}
        \includegraphics[width=0.33\textwidth]{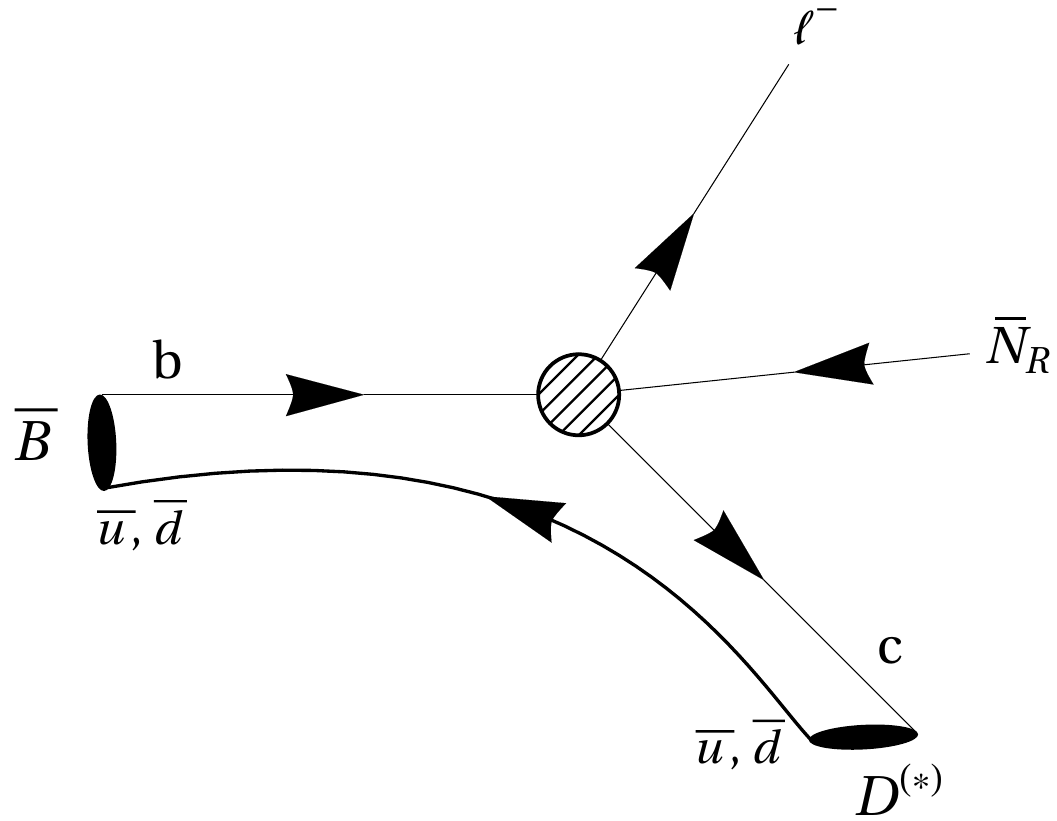} \\
        \includegraphics[width=0.33\textwidth]{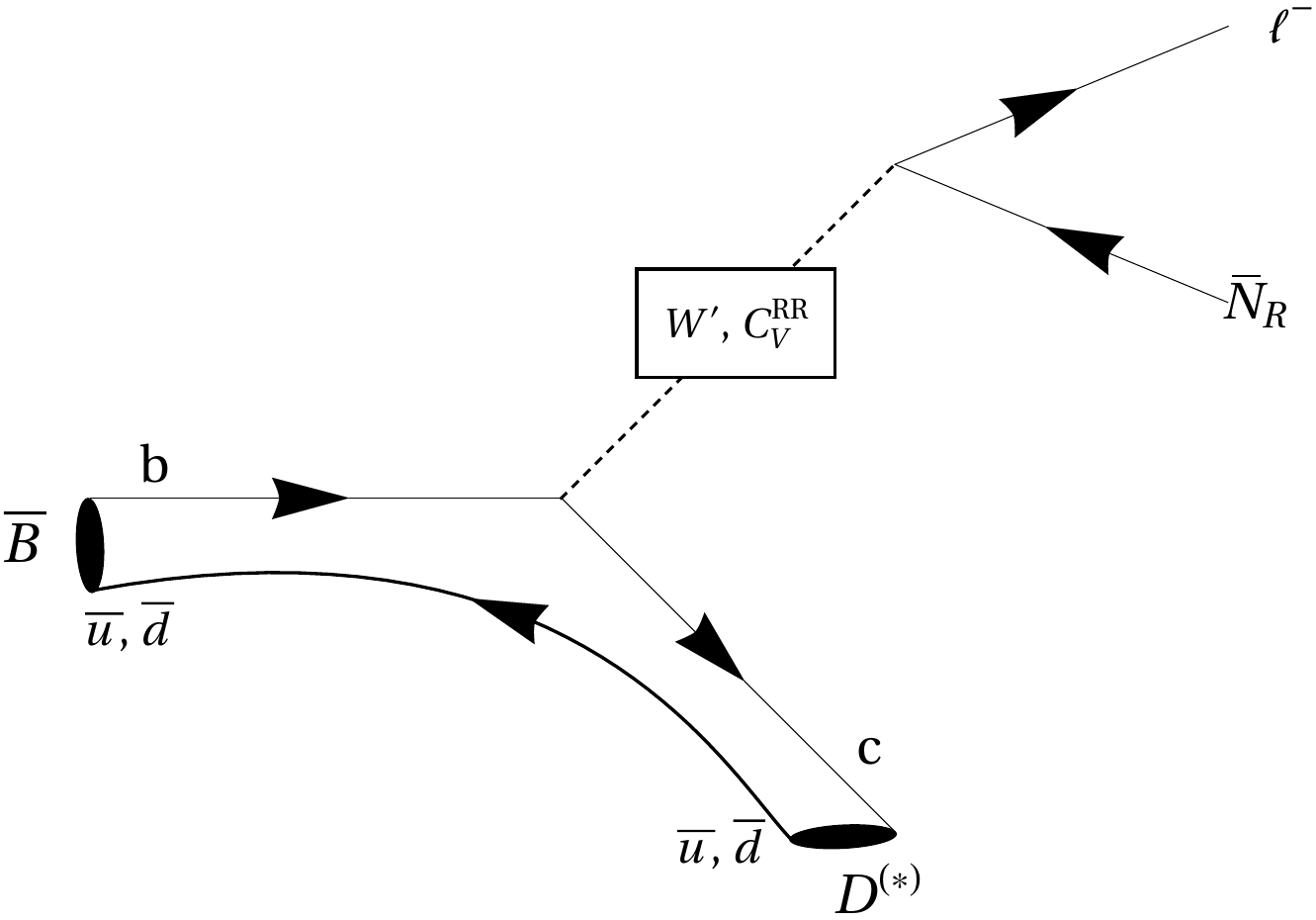} \hspace{15truemm}
        \includegraphics[width=0.33\textwidth]{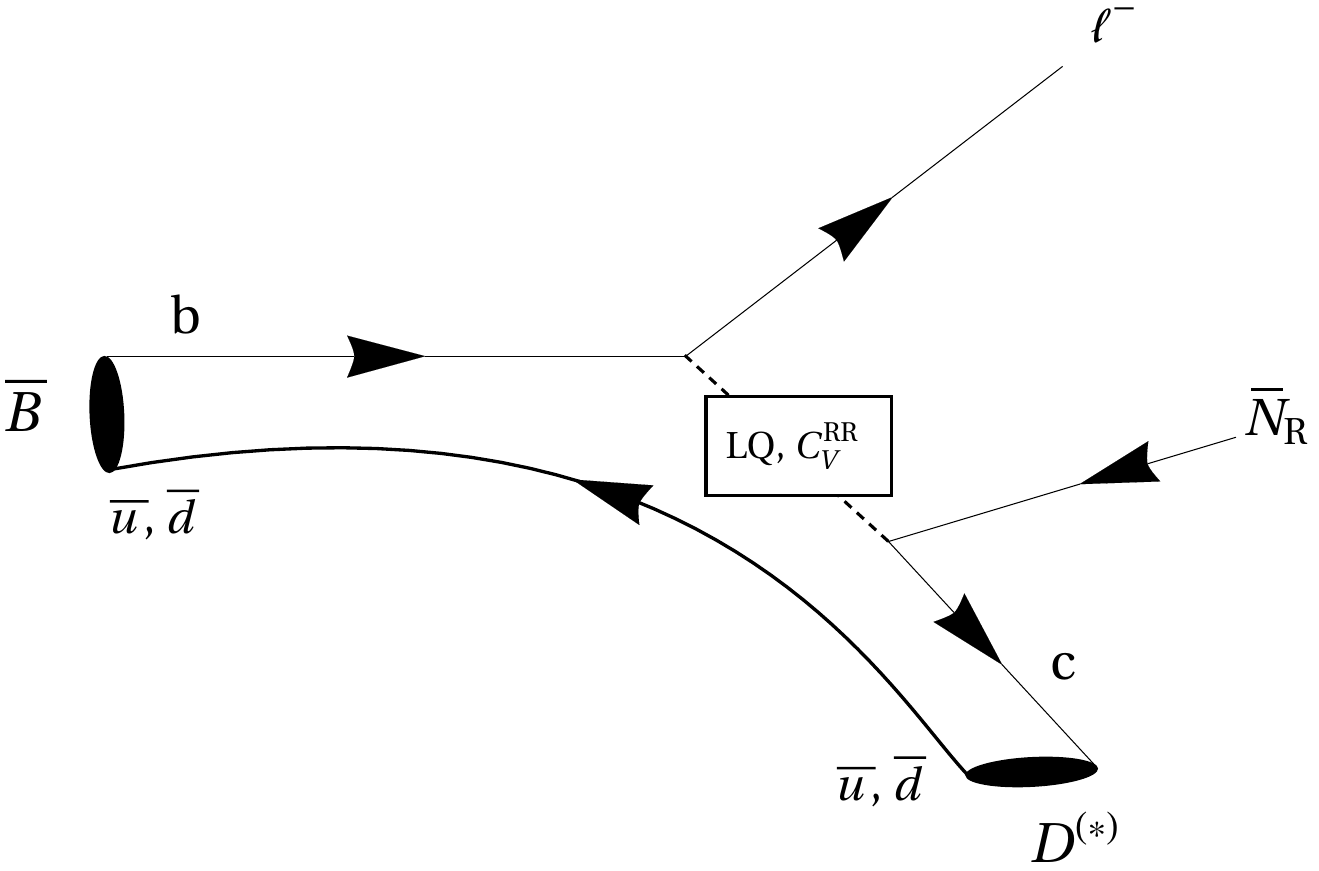}
\caption{Feynman diagrams for charged-current semileptonic $B$ decays considered in this article. The top-left panel shows the SM contribution to ${\bar B}\to D^{(*)}\ell^-{\bar\nu}_L$. The top-right panel shows an effective four-fermion contact interaction for ${\bar B}\to D^{(*)}\ell^-{\bar N}_R$. The bottom-left panel represents a right-handed vector interaction mediated by a $W'$, while the bottom-right panel shows a similar contribution mediated by a leptoquark. Here, $C^{RR}_V$ denotes the Wilson coefficient for the four-fermion contact interaction used in this paper.
\label{fig:four_diagrams}}
\end{figure}

We derive the differential distribution for the decay $\BDstarlN$ using the same helicity-amplitude formalism as in $\bar B\to D^*(\to D\pi)\ell\bar\nu_\ell$ \cite{Bhattacharya:2019olg}. The Feynman diagrams for the decay amplitudes for the SM and NP cases are shown in Fig.~\ref{fig:four_diagrams}. The total decay amplitude is the sum $\Amp=\Amp^{\rm SM}+\Amp^{\rm NP}$; since they represent different final states, the two amplitudes do not interfere. Complete expressions for the amplitude are given in Appendix \ref{sec:amplitudecalculationmu}, and the phase-space element is derived in Appendix \ref{sec:phase-space-muN}.

\begin{figure}[!htbp]
\centering\includegraphics[width=0.5\textwidth]{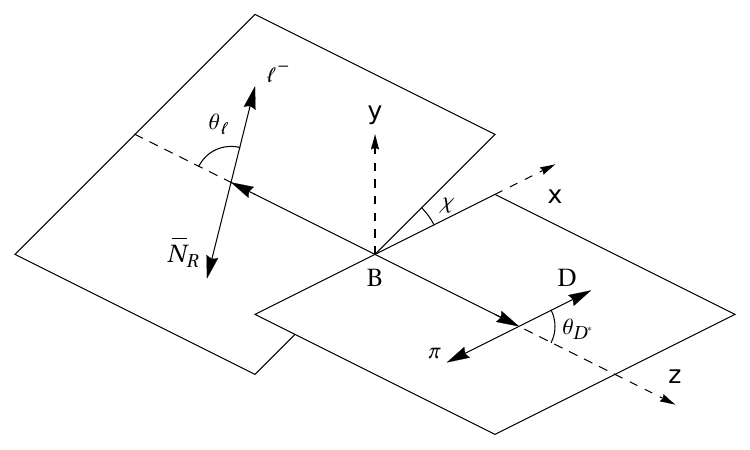}
\caption{Schematic diagram defining helicity angles in ${\bar B}\to D^{*}(\to D\pi)\ell^-{\bar N}_R$ . The coordinate axes are defined such that the $z$-axis aligns with the $D^*$ direction in the $B$ rest frame, and the $x$-$z$ plane coincides with the $D^*$ decay plane. $\theta_{D^*}$ ($\theta_\ell$) is the polar angle of the $D$($\ell^-$) direction relative to the positive (negative) $z$-axis in the rest frame of its parent $D^*$ (virtual $W^{(\prime)}$). $\chi$ is the azimuthal angle between the $D^*$ and $W^{(\prime)}$ in the $B$ rest frame.
\label{fig:angN-bar}}
\end{figure}

The differential rate for this four-body decay can be expressed in terms of four independent kinematic parameters. These are $q^2,\theta_{D^*},\theta_\ell,$ and $\chi$. $q^2$ is the invariant mass of the lepton-neutrino pair. $\theta_{D^*}$ is the polar angle of the $D$-meson direction relative to the direction opposite to the $B$ meson in the $D^*$ rest frame. Similarly, $\theta_\ell$ is the polar angle of the lepton direction relative to the direction opposite to the $B$ meson in the $q^2$ rest frame. $\chi$ is the azimuthal angle between the $D^*$ decay plane and the lepton-neutrino plane in the $B$ rest frame. Fig.~\ref{fig:angN-bar} provides a schematic diagram defining the three helicity angles $\theta_{D^*},\theta_\ell,$ and $\chi$.

Inserting the amplitude squared of Eq.~\eqref{eq:ampsquaredmuN} into Eq.~\eqref{eq:decayratewithphasespacemuN}, the four-fold differential distribution can be expressed in terms of these kinematic parameters as follows.
\bea
\frac{d\Gamma({\bar B}\to D^*(\to D\pi)\ell{\bar N}_R)}{dq^2\,d\cos\theta_{D^*}\,d\cos\theta_\ell\,d\chi} &=& \frac{ 3\,G_F^2|V_{cb}|^2|{\vec p}_{D^*}||{\vec p}_\ell|\,{\cal B}(D^*\to D\pi)}{2^{8}\pi^4\,m_B^2|{\vec p}_D|^2\sqrt{q^2}} \nonumber \\
&&\hspace{-30truemm}\times\Big\{J_{1s}\sin^2\theta_{D^*} + J_{1c}\cos^2\theta_{D^*} + \left(J_{2s}\sin^2\theta_{D^*} + J_{2c}\cos^2\theta_{D^*}\right)\cos2\theta_\ell \nonumber \\
&&\hspace{-25truemm} +~J_3\sin^2\theta_{D^*}\sin^2\theta_\ell\cos2\chi + J_4\sin 2\theta_{D^*}\sin 2\theta_\ell\cos\chi + J_5\sin 2\theta_{D^*}\sin\theta_\ell\cos\chi \nonumber \\
&&\hspace{-25truemm} +~\left(J_{6s}\sin^2\theta_{D^*} + J_{6c}\cos^2\theta_{D^*}\right)\cos\theta_\ell + J_7\sin 2\theta_{D^*}\sin\theta_\ell\sin\chi \nonumber \\
&&\hspace{-25truemm} +~J_8\sin 2\theta_{D^*}\sin 2\theta_\ell\sin\chi + J_9\sin^2\theta_{D^*}\sin^2\theta_\ell\sin 2\chi\Big\},
\label{eq:rateJ}
\eea
where ${\vec p}_X$ represents the three-momentum of particle $X$ with $X = D, D^*$, and $\ell$. Explicit expressions for these three momenta in terms of the kinematic parameters are given in Eq.~(\ref{eq:momentums}). The angular coefficients $J_i(q^2)$ for the SM and the different NP cases are collected in Appendix \ref{app:J functions}. We evaluate them following the conventions adopted in the Belle analysis \cite{Belle:2023xgj}, in which $\theta_\ell$ is defined as the angle between the direction of the charged lepton $\ell$ and the direction opposite to the $B$ meson in the $W$ rest frame; our results agree with those reported by Belle \cite{Belle:2023xgj}. Compared to Refs.~\cite{Colangelo:2018cnj, Bernlochner:2024xiz}, we find opposite signs for the coefficients $J_4$, $J_{6s}$, $J_{6c}$, and $J_8$, a difference that can be traced to the opposite convention adopted there for the lepton angle $\theta_\ell$.

We can construct several differential decay distributions by integrating Eq.~\eqref{eq:rateJ} over one or more kinematic variables. The first of these is obtained by integrating Eq.~\eqref{eq:rateJ} over the three helicity angles. This gives the differential decay distribution as a function of $q^2$,
\beq
\frac{d\Gamma}{dq^2} = \frac{G_F^2|V_{cb}|^2|\vec p_{D^*}||\vec p_\ell|{\cal B}(D^*\to D\pi)}{96\pi^3m_B^2|\vec p_D|^2\sqrt{q^2}}\left[3 J_{1c} - J_{2c} + 2\left(3 J_{1s}  - J_{2s}\right)\right]\,.
\eeq
Similarly, other differential decay distributions as functions of $q^2$ and other angular variables can be obtained by integrating Eq.~(\ref{eq:rateJ}) over the two remaining angles, 
\bea    
\frac{d^2\Gamma}{dq^2\,d\cos\theta_{D^*}}
&=&
\frac{3}{4}\frac{d\Gamma}{dq^2}
\left[
2F_L^{D^*}(q^2)\cos^2\theta_{D^*}
+
F_T^{D^*}(q^2)\sin^2\theta_{D^*}
\right]\,,
\label{eq:dq2dcosthetaDstar}
 \\
\frac{d\Gamma}{dq^2 d\cos\theta_\ell} &=& \frac{d\Gamma}{dq^2}\left[\frac{1}{2}+ A_{FB}\cos\theta_\ell + \frac{1-3\tilde{F_L^\ell}}{4}\frac{3\cos^2\theta_\ell - 1}{2}
\right]\,, \label{eq:dq2dcosthetal}
\\
 \frac{d^2\Gamma}{dq^2d\chi} &=& \frac{1}{2\pi}\frac{d\Gamma}{dq^2}\left[ 1 + S_3 \cos2\chi +S_9 \sin2\chi
  \right]\,, 
  \label{eq:dq2dchi}
\eea
where $F_L^{D^*}(q^2)$ and $F_T^{D^*}(q^2)$ are the longitudinal and transverse polarizations of the $D^{*}$ respectively, $A_{FB}$ is the charged-lepton forward-backward asymmetry, $\tilde{F_L^\ell}$ is the longitudinal polarization of the charged-leptons, and $S_9$ is a triple-product asymmetry \cite{Duraisamy:2013pia, Bhattacharya:2019olg}. Likewise, one can express angular observables appearing in Eqs.~\eqref{eq:dq2dcosthetaDstar}--\eqref{eq:dq2dchi}  in terms of the angular coefficients,
\bea
F_L^{D^*}(q^2) &=& \frac{3J_{1c}-J_{2c}}
{6J_{1s}+3J_{1c}-2J_{2s}-J_{2c}}, \nonumber \\
A_{FB}(q^2) &=& \frac{3}{2}\, \frac{2J_{6s}+J_{6c}}
{6J_{1s}+3J_{1c}-2J_{2s}-J_{2c}}, \nonumber\\
\widetilde{F}_L^\ell(q^2) &=& \frac{J_{1c}-3J_{2c}+2\left(J_{1s}-3J_{2s}\right)}
{6J_{1s}+3J_{1c}-2J_{2s}-J_{2c}}, \label{eq:angular-observables} \\
S_3(q^2) &=& \frac{8J_3}{\pi (6J_{1s}+3J_{1c}-2J_{2s}-J_{2c})}, \nonumber\\
S_5(q^2) &=& \frac{-3 J_5} {6J_{1s}+3J_{1c}-2J_{2s}-J_{2c}}\,. \nonumber
\eea

Among these angular observables, we focus on three: $A_{FB}$, $S_3$ and $S_5$. $A_{FB}$ and $S_3$ can be extracted from the angular distributions given in Eqs.\eqref{eq:dq2dcosthetal} and \eqref{eq:dq2dchi} using the following asymmetric integrals over $\cos\theta_\ell$ and $\chi$ respectively.
\bea
A_{FB}(q^2) &=& \left(\frac{d\Gamma}{dq^2}\right)^{-1}
\left[\int_0^1-\int_{-1}^0\right]
d\cos\theta_\ell\,
\frac{d^2\Gamma}{dq^2\,d\cos\theta_\ell},
\label{eq:AFBint} \\
S_3(q^2) &=& \left(\frac{d\Gamma}{dq^2}\right)^{-1}
\left[
\int_{0}^{\pi/4} - \int_{\pi/4}^{\pi/2} -\int_{\pi/2}^{3\pi/4} + \int_{3\pi/4}^{\pi}\right. \nonumber \\
&&\hspace{4truecm} \left. + \int_{\pi}^{5\pi/4} -\int_{5\pi/4}^{3\pi/2} - \int_{3\pi/2}^{7\pi/4}
+ \int_{7\pi/4}^{2\pi}\right] d\chi\,\frac{d^2\Gamma}{dq^2 d\chi}\,.
\label{eq:S3int}
\eea
The extraction of $S_5$ requires asymmetric integrals over $\chi$ and $\cos\theta_{D^*}$ of the angular distribution obtained by integrating  Eq.~\eqref{eq:rateJ} with respect to $\cos\theta_{\ell}$:
\bea
S_5(q^2) &=& \left( \frac{d\Gamma}{dq^2} \right)^{-1}
\left[ \int_0^{\pi/2} - \int_{\pi/2}^\pi-\int_{\pi/2}^{3\pi/2} + \int_{3\pi/2}^{2\pi} \right]
d\chi \nonumber \\
&&\hspace{65trueMm}\left[ \int_0^{1} - \int_{-1}^{0} \right] d\cos\theta_{D^*}\,
\frac{d^3\Gamma}{dq^2\, d\cos\theta_{D^*}\, d\chi}\,.~~~~~~~~\label{eq:S5}
\eea

\section{Right-handed Neutrino implementation in EvtGen}
\label{Section:RHN_EvtGEN}

EvtGen \cite{Lange:2001uf} is an MC framework for simulating decays of heavy particles containing $b$ or $c$ quarks. It is used by collaborations worldwide, including those at the Large Hadron Collider, BES III, and Belle II. EvtGen allows one to simulate decays of heavy-flavor particles such as $B$ mesons and model their decay products. The results of these simulations can then be used to obtain valuable information on the properties and behavior of the particles involved in those decays.

In our work, we develop a decay model called BTODSTARLNUNPRH \cite{B2DstRHN}. This model is built within the EvtGen framework and applies to RHN NP scenarios. The model contains two new-physics parameters: $C_V^{RR}$, the Wilson coefficient associated with the right-handed vector interaction, and $M_{N_R}$, the mass of the RHN. The RHN is modeled as a massive spinor with its mass, spin, and charge parameters set manually in an EvtGen file called evt.pdl (here pdl stands for ``Particle Definition List'') before executing the decay file. The file evt.pdl contains a database of the fundamental properties of all SM particles, such as their mass, spin, and charges. The evt.pdl file also allows for the introduction of new particles. The following is the evt.pdl snippet entry for the RHN model.
\begin{lstlisting}[
    basicstyle=\ttfamily\fontsize{7.5}{9}\selectfont,
    frame=single,
    columns=fullflexible,
    keepspaces=true,
    showstringspaces=false
]
                  name        id   mass/GeV   width/GeV   max_Dm/GeV   3*charge   2*spin   lifetime*c/mm   PythiaId
add p Particle anti-nu_M     -19      1.5         0           0           0          1          0            -19
add p Particle nu_M           19      1.5         0           0           0          1          0             19
end
\end{lstlisting}

The SM events used in this study are generated with the EvtGen decay file developed in Ref.~\cite{Bhattacharya:2022bdk}. To generate massive RHN events, the SM contribution can be removed for two reasons. First, the left- and right-handed neutrino amplitudes do not interfere. Second, the SM ${\bar B}\to D^*\ell^-{\bar\nu}_L$ background can be neglected in the region of phase space in which we are looking for a signal. However, the SM contribution is built into the EvtGen code and cannot be set to zero. Even so, in the EvtGen decay file developed in Ref.~\cite{Bhattacharya:2022bdk}, the operator $O_V^{LL}$, which is identical to the SM operator, was added with the left-handed vector Wilson coefficient, $C_V^{LL}$. To remove the default contribution from the SM, therefore, it is necessary to set $C_V^{LL} = -1$.  The following snippet of code provides an example user decay file illustrating how to use the NP MC generator.
\raggedbottom
\begin{verbatim}
## first argument is cartesian(0) or polar(1) 
## representation of NP coefficients which are three
## consecutive numbers {id, Re(C), Im(C)} or {coeff id, |C|, Arg(C)}
## id==0 C_VLL -- left-handed vector coefficient change from SM
## id==1 C_VRR -- right-handed right-handed vector coefficient
Decay B+
## B+ -> anti-D*0  e+  nu_M generated with the addition of NP 
## And nu_M is  massive RHN
1 anti-D*0    e+   nu_M   BTODSTARLNUNPRH 0 0 -1 0 1 1 0;
#1 anti-D*0    mu+   nu_M  BTODSTARLNUNPRH 0 0 -1 0 1 1 0;
Enddecay
End
\end{verbatim} 

We study two decay channels using our EvtGen implementation: $B^+\to{\bar D}^{*0}e^+ N_R$ and $B^+\to{\bar D}^{*0}\mu^+N_R$. For each decay channel, we consider three benchmark RHN masses: $M_{N_R} = 0.1, 0.5,$ and 1.5 GeV, and generate 10 million events for each process. Form factors are necessary to describe the hadronic transitions; for this, we use the Boyd-Grinstein-Lebed (BGL) parametrization of hadronic form factors, given in Appendix \ref{app:bglformfactors}. To generate events in EvtGen, we also need to provide a numerical value of $C_V^{RR}$. However, since existing experimental analyses are not optimized to observe an RHN signal, current published data cannot be used to put a constraint on $C_{V}^{RR}$. Therefore, we use $C_V^{RR} = 1$ in all our EvtGen simulations. Even so, we present our results as probability distributions or shape functions with appropriate normalizations. These normalized shape functions do not directly depend on the value of the WC when a single dimension-six operator is used. Thus, although we use $C_V^{RR} = 1$ to simulate the events, our plots remain independent of this input. In what follows, we present the results of our simulations. These include event distributions as functions of the four kinematic variables, dependence of interesting angular observables on $q^2$, and correlations between these angular observables. Whenever appropriate, we also compare and contrast the electron and muon channels to demonstrate the dependence of these distributions on the light-lepton mass. Note that the matrix elements of the neutral and charged $B$-decay modes are identical, but their experimental signatures are distinct since the neutral and charged $D^*$ decay to different final states.

\subsection{Results I: Event Distributions as Functions of Kinematic Variables}

\begin{figure}[!htpb]
\centering
\includegraphics[width=0.47\textwidth]{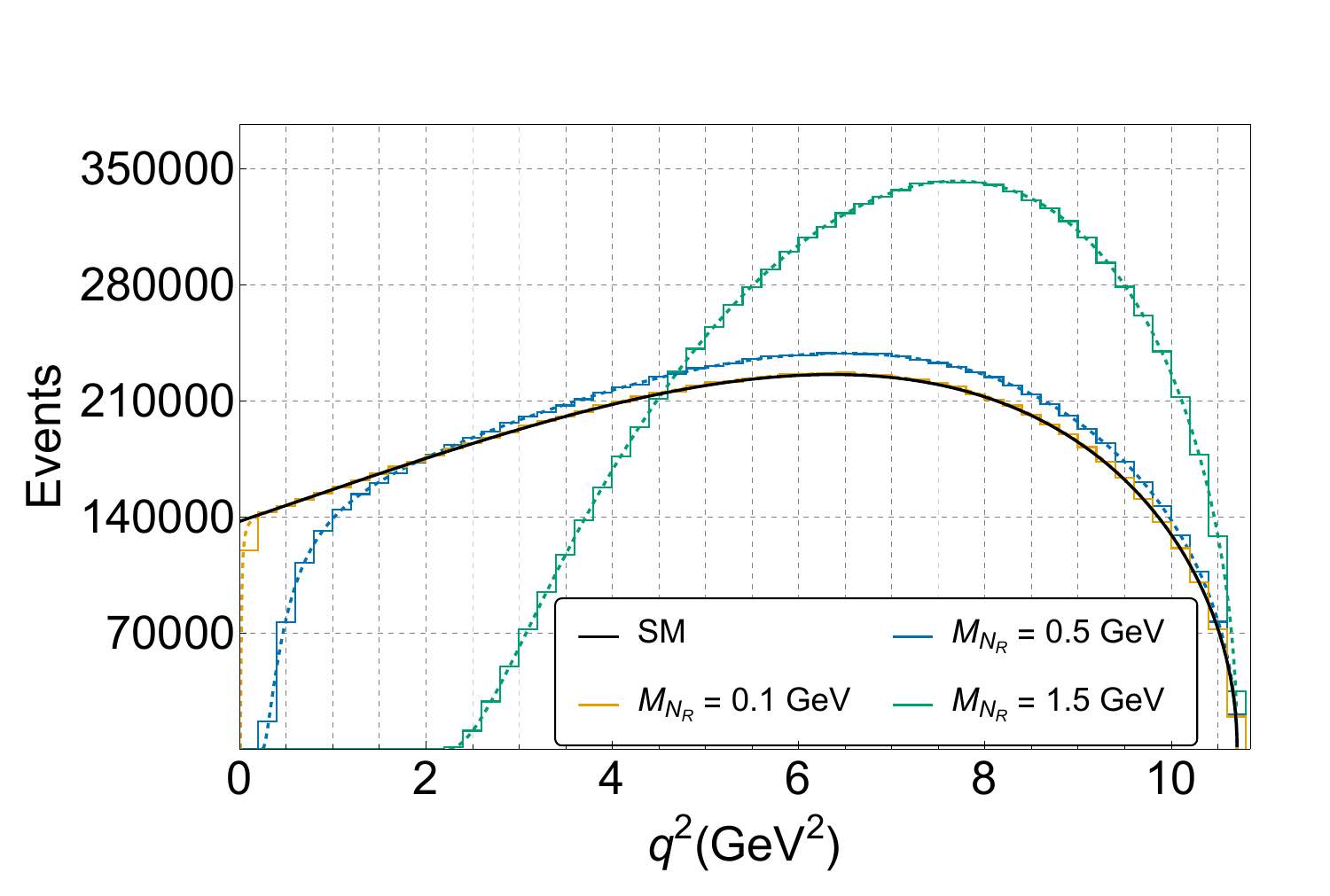}
\includegraphics[width=0.47\textwidth]{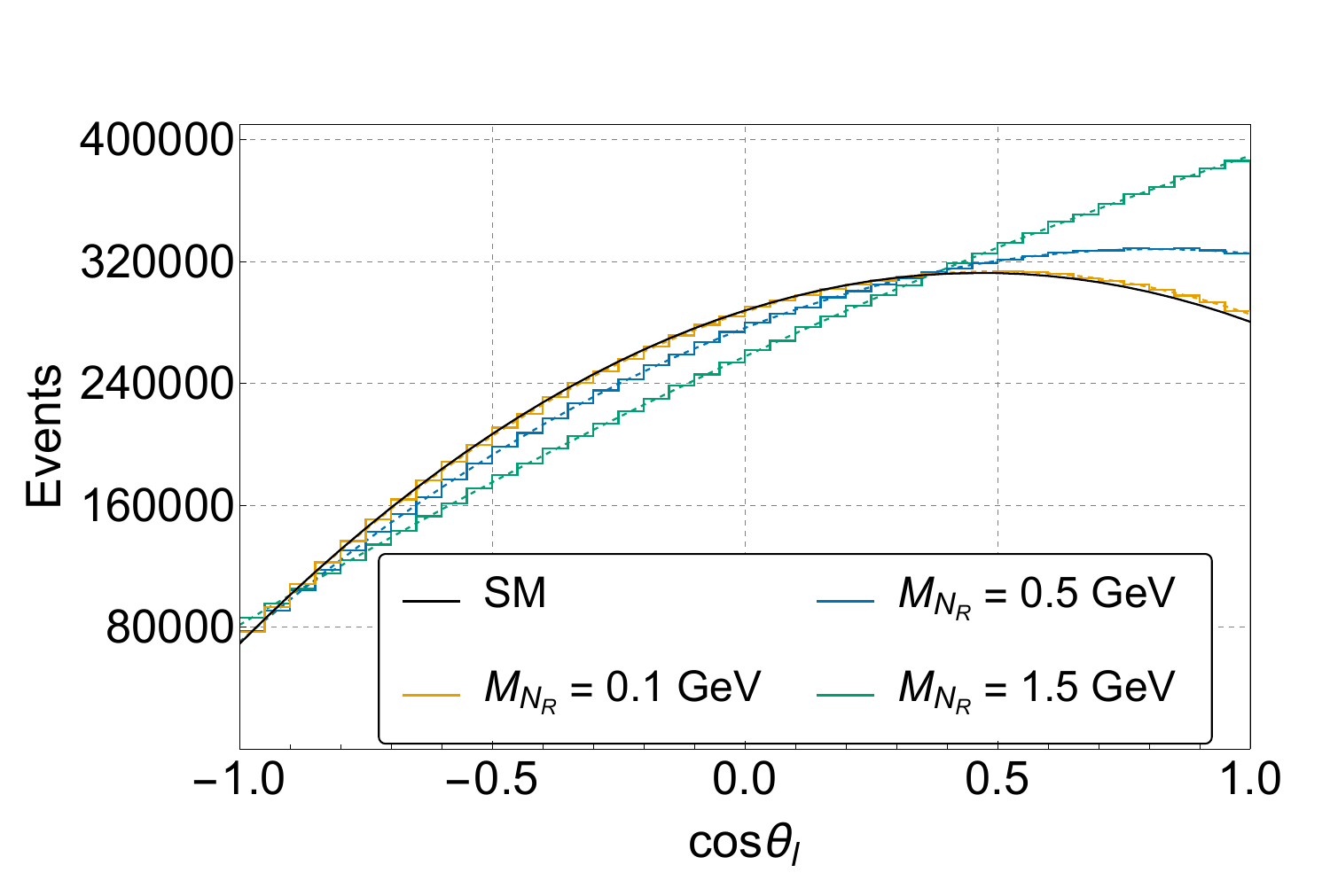}
\includegraphics[width=0.47\textwidth]{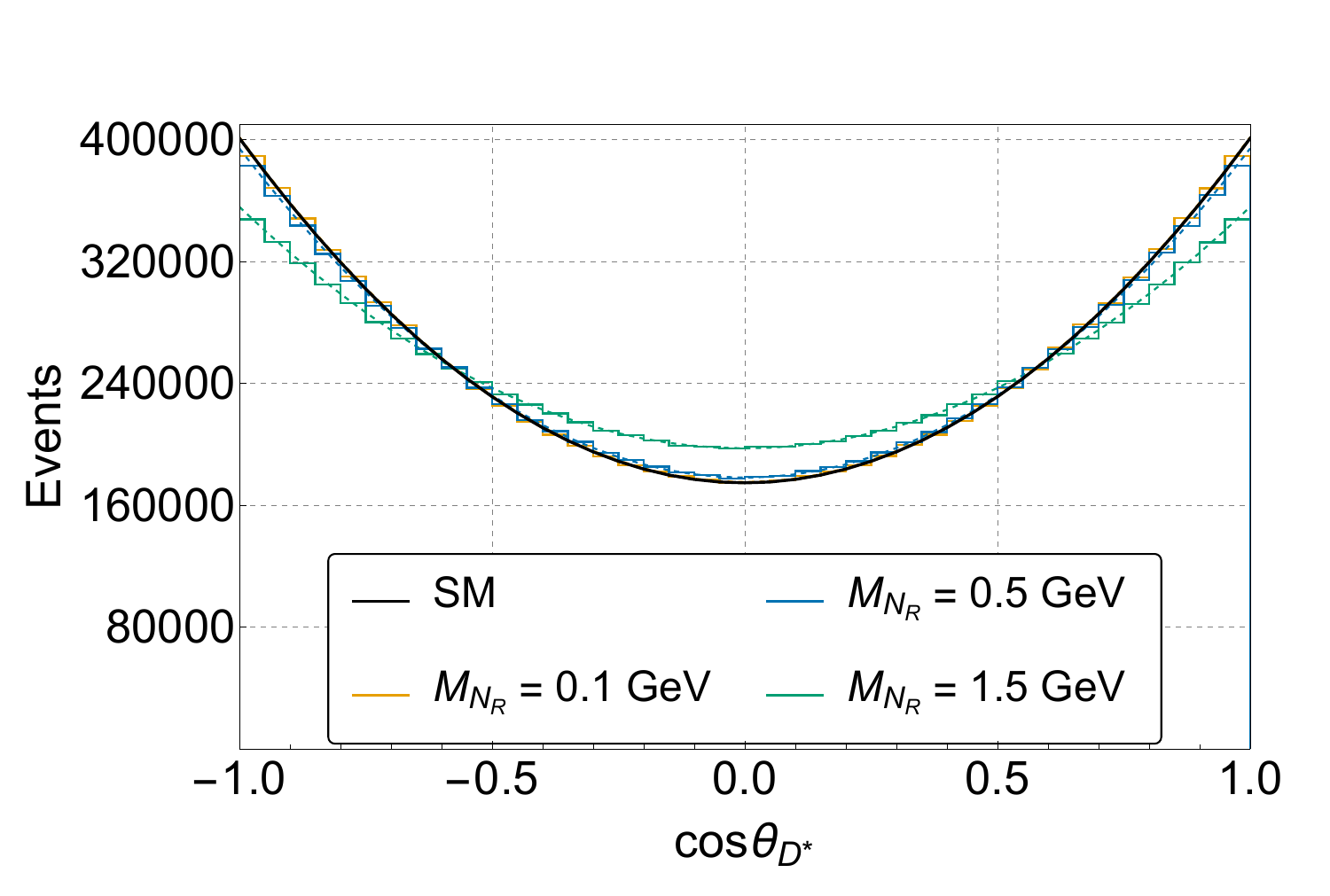}
\includegraphics[width=0.47\textwidth]{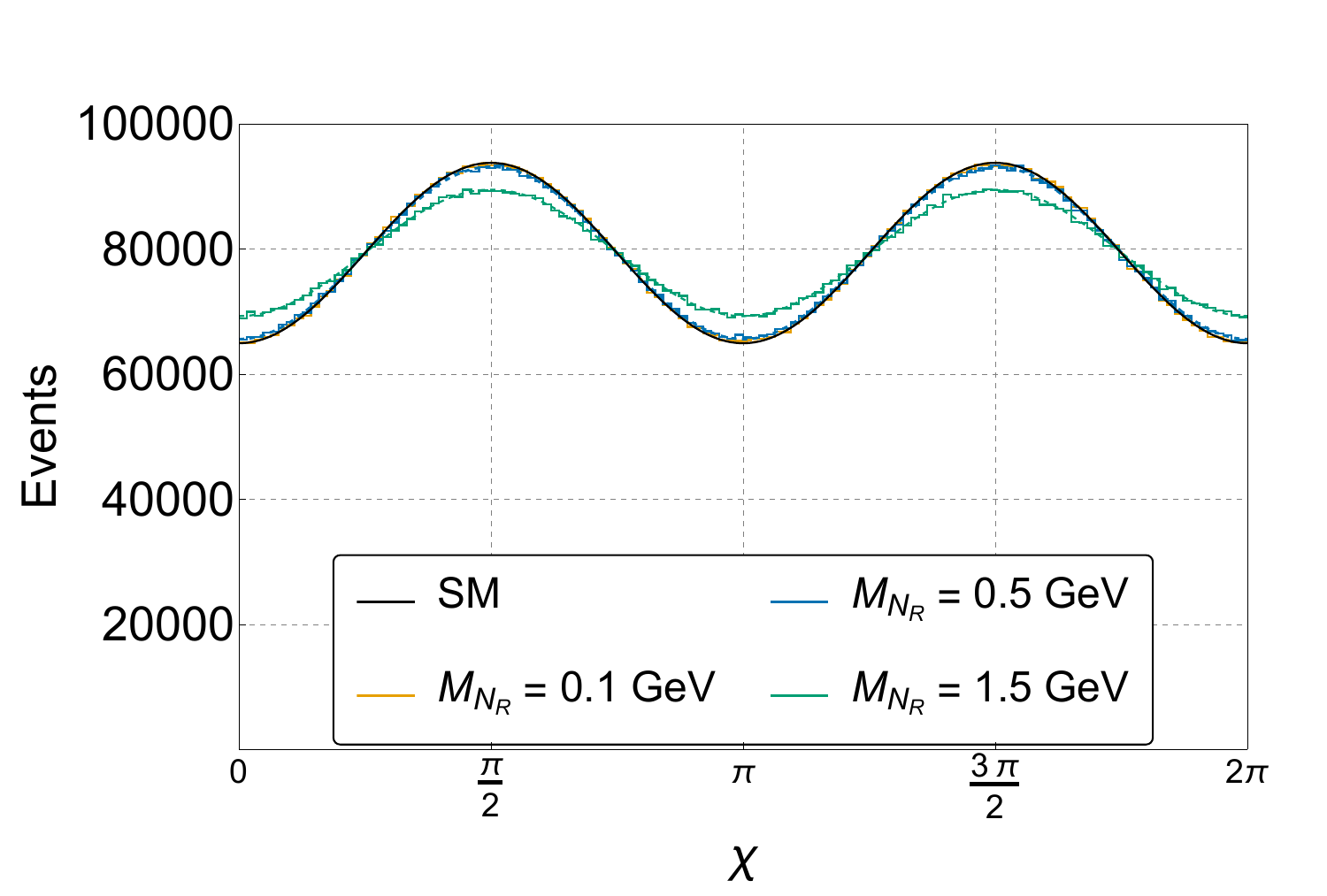}
\caption{\small Event distributions as functions of kinematic variables for $B^+\to{\bar D}^{*0}e^+\nu_L$ (SM) and $B^+\to{\bar D}^{*0}e^+N_R$ (NP). Clockwise from the top left panel, we present the distributions as functions of $q^2, \cos\theta_\ell, \chi,$ and $\cos\theta_{D^*}$. SM distributions (left-handed neutrino) are shown in black, while the remaining colored curves represent NP distributions with an RHN of mass $M_{N_R} = 0.1$ GeV (orange), 0.5 GeV (blue), and 1.5 GeV (teal). The histograms show simulated data from our EvtGen model, while the dashed curves represent theoretical distributions using analytical expressions.
\label{fig:elecdist}}
\end{figure}
We begin by plotting the event distributions obtained using our EvtGen model as functions of the four kinematic variables. Our results for the electron channel in the SM and NP with three different benchmark RHN masses, $M_{N_R} = 0.1, 0.5,$ and 1.5 GeV, are presented as histograms in Fig.~\ref{fig:elecdist}. Each distribution has been appropriately normalized so that the total number of events equals 10 million. The $q^2$ distributions were obtained by binning the data into equal-sized bins of 0.2 GeV$^2$ -- the total number of bins then depended on $M_{N_R}$. For the $\cos\theta_{D^*}$ and $\theta_\ell$ distributions, equal-sized bins were used, but the total number of bins was fixed to 40. The $\chi$ distribution was generated with 126 equal-sized bins. Fig.~\ref{fig:elecdist} also displays the theoretical curves for each distribution for comparison. An excellent match between the theoretical curves and those generated by EvtGen validates our EvtGen model.

\begin{figure}[!htpb]
\centering
\includegraphics[width=0.6\textwidth]{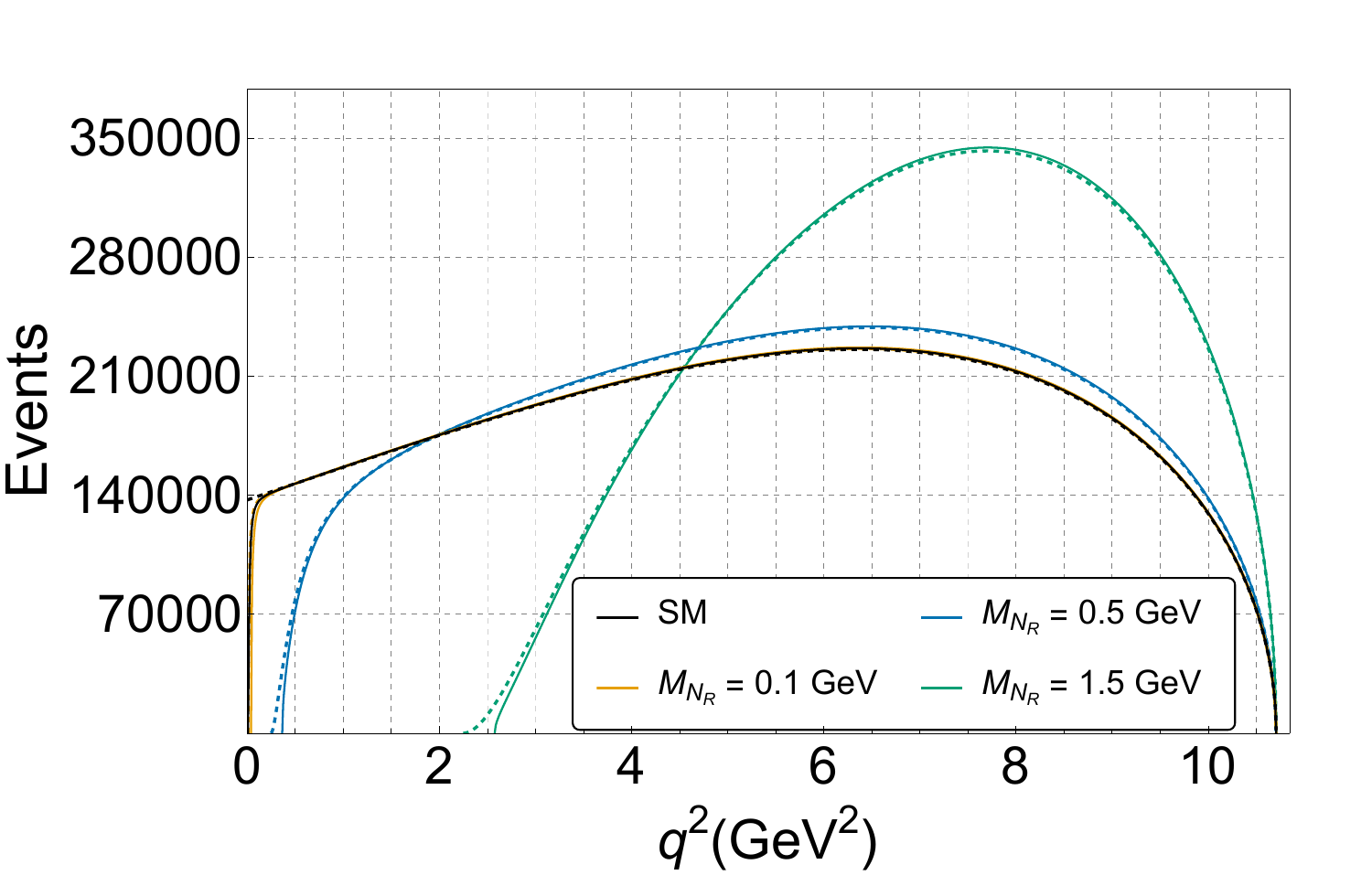}
\caption{\small Theoretical event distributions as functions of $q^2$ for $B^+\to{\bar D}^{*0}\ell^+\nu_L$ (SM) and $B^+\to{\bar D}^{*0}\ell^+N_R$ (NP) for $\ell = e$ (dashed curves), and $\mu$ (solid curves). The SM distribution (left-handed neutrino) is shown in black, while the remaining colored curves represent NP distributions with an RHN of mass $M_{N_R} = 0.1$ GeV (orange), 0.5 GeV (blue), and 1.5 GeV (teal). The only significant differences between the electron and muon curves are seen close to the low-$q^2$ threshold.
\label{fig:muelecdiff}}
\end{figure}

A similar set of plots was generated for the muon channel, which also demonstrates excellent agreement with theory. These plots can be found in Appendix \ref{app:EvtHistogram}. We find that the only noticeable differences between the electron and muon channel plots appear in the $q^2$ distributions, which are only present near the low-$q^2$ threshold. These small deviations are shown in the $q^2$ distributions of Fig.~\ref{fig:muelecdiff}.

The $q^2$ distributions shown in Fig.~\ref{fig:elecdist} help us probe the phase space available in each decay. The lower kinematic threshold in these plots, $q^2_{\rm min} = (m_\ell + M_{N_R})^2$, increases with the RHN mass. As a result, for a higher-mass RHN, the $q^2$ distribution peaks further to the left. For $m_{N_R} = 0.1$ GeV, we find that the $q^2$ distribution largely overlaps with the SM distribution for $\nu_L$. This similarity between the RHN with $m_{N_R} = 0.1$ GeV and the SM is also observed in the event distributions plotted as functions of $\cos\theta_{D^*}, \cos\theta_\ell,$ and $\chi$. For $M_{N_R} = 1.5$ GeV, however, all four kinematic distributions show significant deviations from the SM. The slope of the $\cos\theta_\ell$ distribution for $M_{N_R} = 1.5$ GeV appears to be roughly uniform and positive. In contrast, the SM distribution has a positive slope for $\cos\theta_\ell = -1$ (backward lepton) but a negative slope for $\cos\theta_\ell = +1$ (forward lepton).

The $\cos\theta_{D^*}$ distribution has a characteristic cosine-squared shape, while the $\chi$ distribution shows sinusoidal oscillations. For $M_{N_R} = 0.1$ and 0.5 GeV, these distributions show very little deviation from their SM counterparts. However, large deviations from the SM are seen for $M_{N_R} = 1.5$ GeV in both cases. As $M_{N_R}$ increases, the $\cos\theta_{D^*}$ curve flattens as the longitudinally-polarized component decreases. On the other hand, the phase of the sinusoidal $\chi$ distribution appears to be independent of $M_{N_R}$, but its oscillation amplitude decreases as $M_{N_R}$ increases.

\subsection{Results II: Angular Observables}
\label{sec:evtgenangobs}

\begin{figure}[!htbp]
\centering
\includegraphics[width=0.45\textwidth]{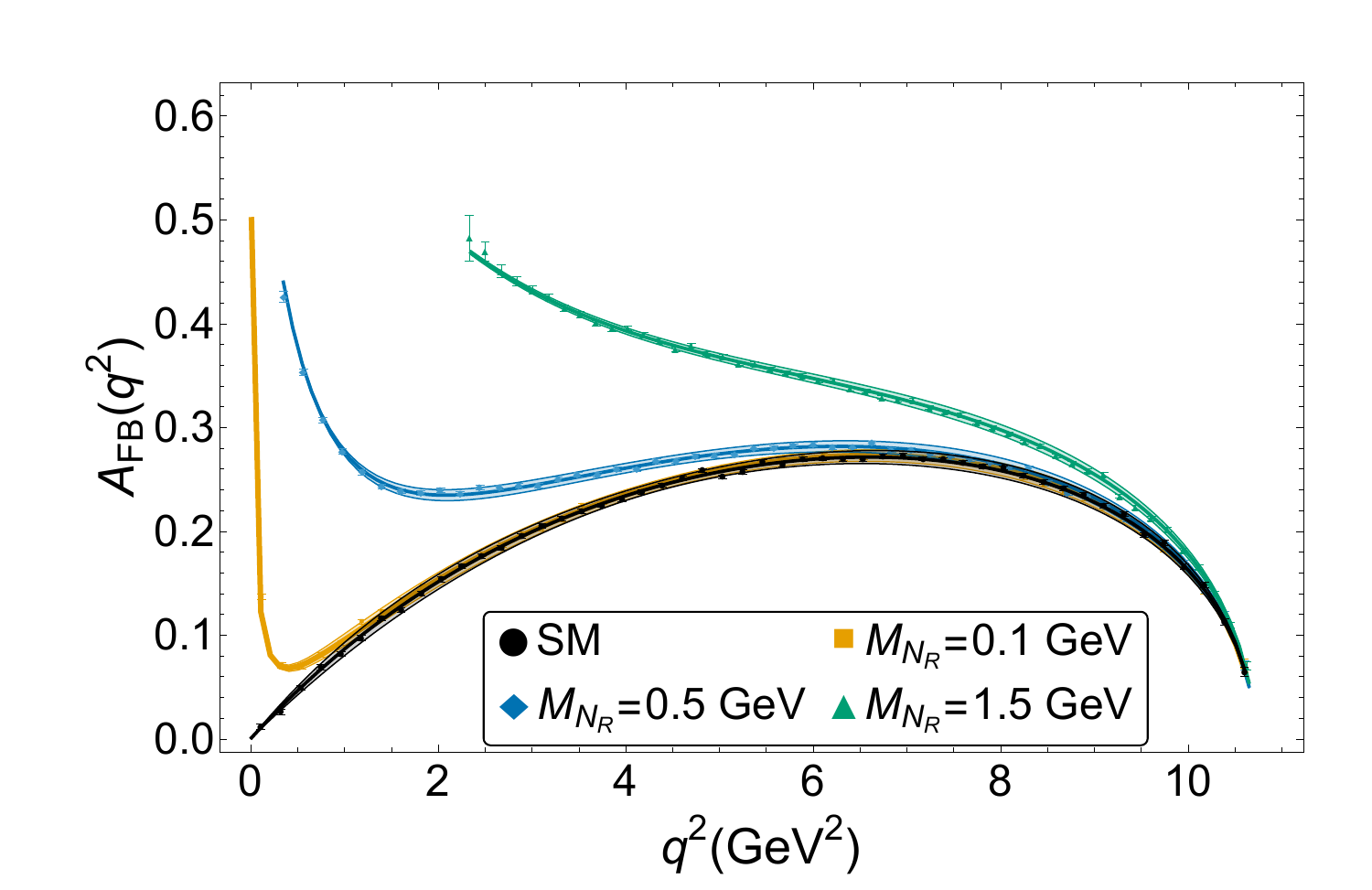}
\includegraphics[width=0.45\textwidth]{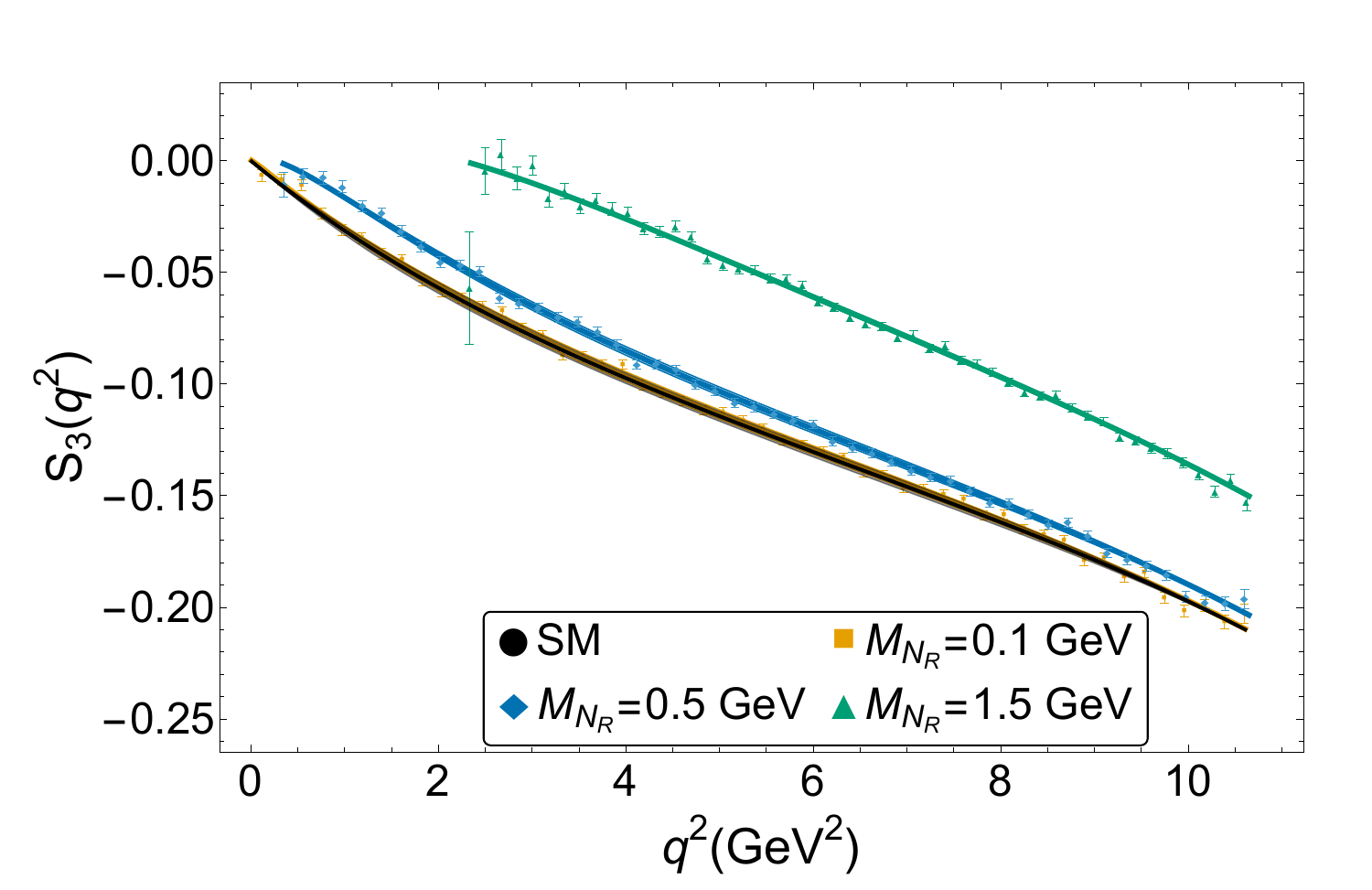}
\includegraphics[width=0.45\textwidth]{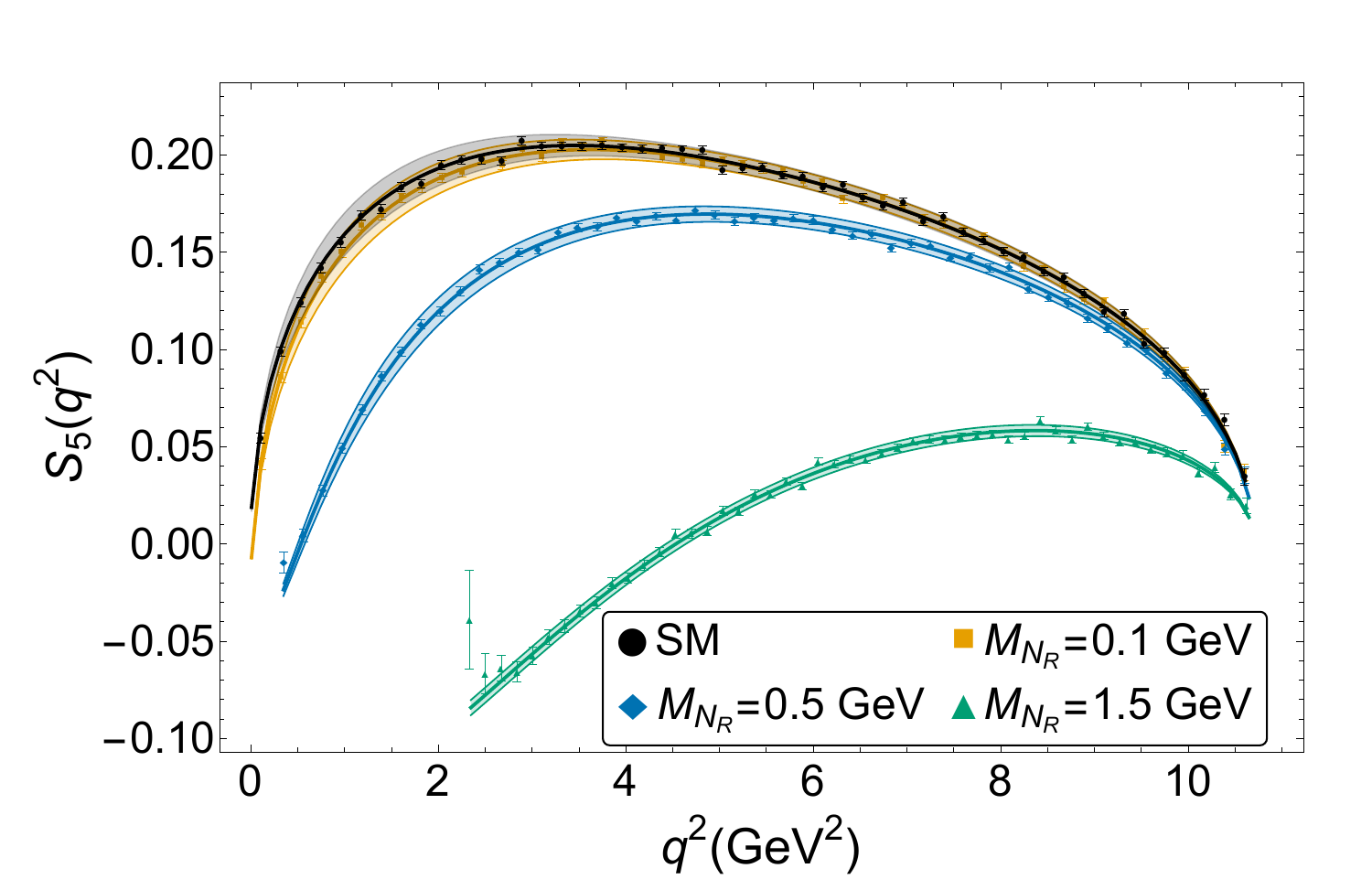}
\caption{Angular observables $A_{FB}$ (top left), $S_3$ (top right), and $S_5$ (bottom) as functions of $q^2$ for the decay $B^+\to{\bar D}^{*0}e^{+}N_R$. We use three benchmark RHN masses: $M_{N_R} = 0.1$ GeV (orange), 0.5 GeV (blue), and 1.5 GeV (teal). EvtGen data are plotted as points with error bars, while solid curves show the corresponding theoretical predictions; the SM expectation (left-handed neutrino) is shown in black.
\label{fig:AngObsEle}}
\end{figure}
Here we examine three angular observables as functions of $q^2$: the forward-backward asymmetry $A_{FB}$ and the angular coefficients $S_3$ and $S_5$. Each of these angular observables probes a different property of the underlying decay amplitude. The forward-backward asymmetry is sensitive to the parity structure of the leptonic current. The angular asymmetry $S_3$ probes the interference between the two transverse polarization amplitudes. The angular asymmetry $S_5$ tracks the interference between the transverse and longitudinal polarization amplitudes. To extract these asymmetries from the EvtGen MC samples, we evaluate the integrals in Eqs.~\eqref{eq:AFBint}, \eqref{eq:S3int}, and \eqref{eq:S5} using binned distributions of the appropriate angular variable. The physical $q^2$ range, $(m_\ell + M_{N_R})^2 \leq q^2 \leq (m_B - m_{D^*})^2$ is divided into 50 equal-sized bins. The $q^2$-bin size, therefore, depends on $M_{N_R}$, decreasing as $M_{N_R}$ increases. The data within each $q^2$ bin are further separated into helicity-angle bins. The integral limits in each part of an asymmetric integral determine the sizes of the helicity-angle bins. For example, let us consider the angular coefficient $S_3$ defined in Eq.~\eqref{eq:S3int}, which involves eight $\chi$ integrals with positive or negative signs. To calculate this observable in each $q^2$ bin, we first divide the $\chi$ range ($0\leq\chi\leq2\pi$) into eight bins of size $\pi/4$ each. The bins are labeled $i = 1, \ldots, 8$, with $N_i$ events in the $i$th bin. $S_3$ is then obtained by appropriately combining the eight $N_i$ with signs corresponding to the integral in Eq.~\eqref{eq:S3int} and normalizing by the total number of events, $\sum\limits_{i=1}^8N_i$.
\begin{figure}[!htbp]
\centering
\includegraphics[width=0.45\textwidth]{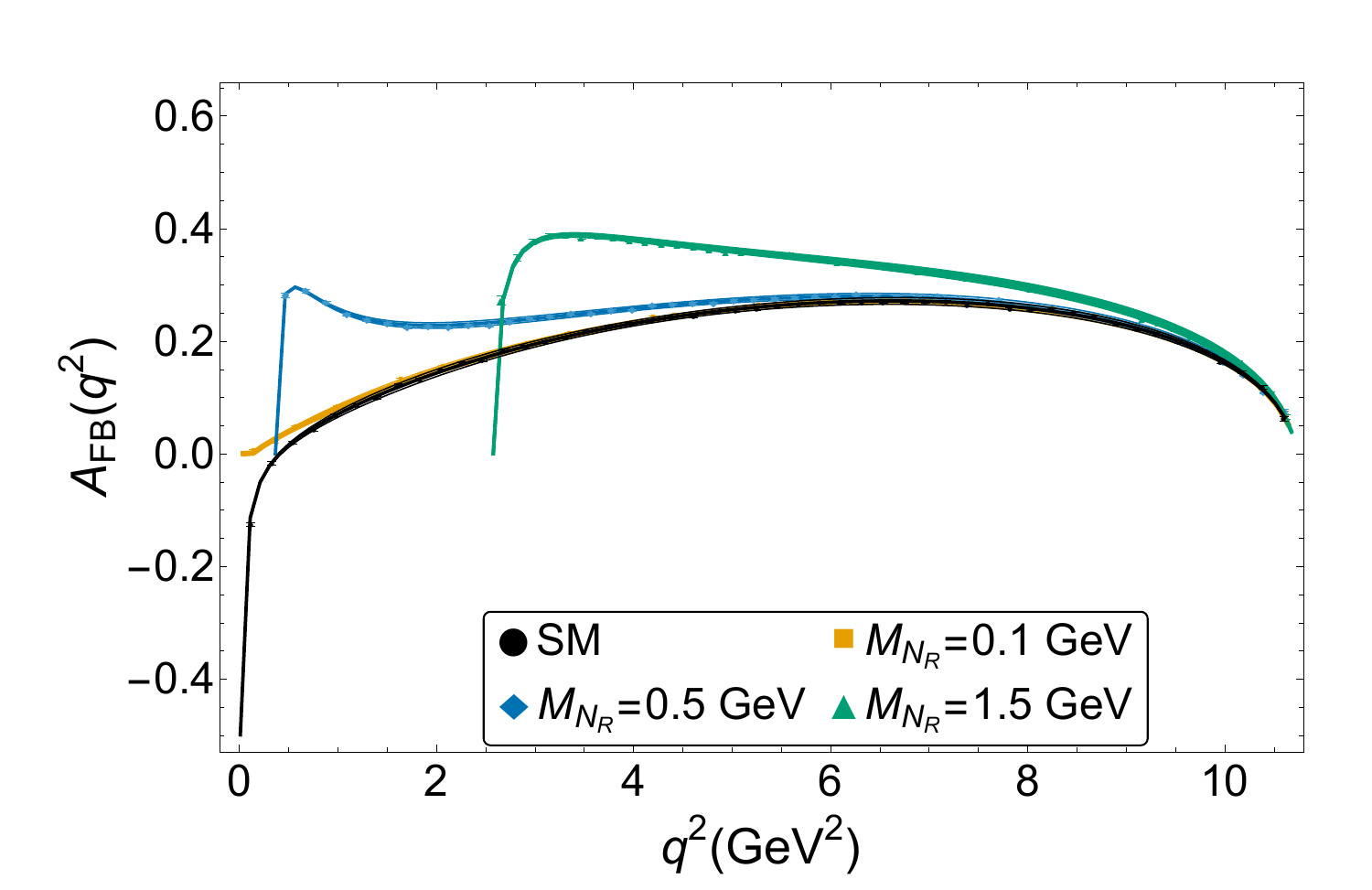}
\includegraphics[width=0.45\textwidth]{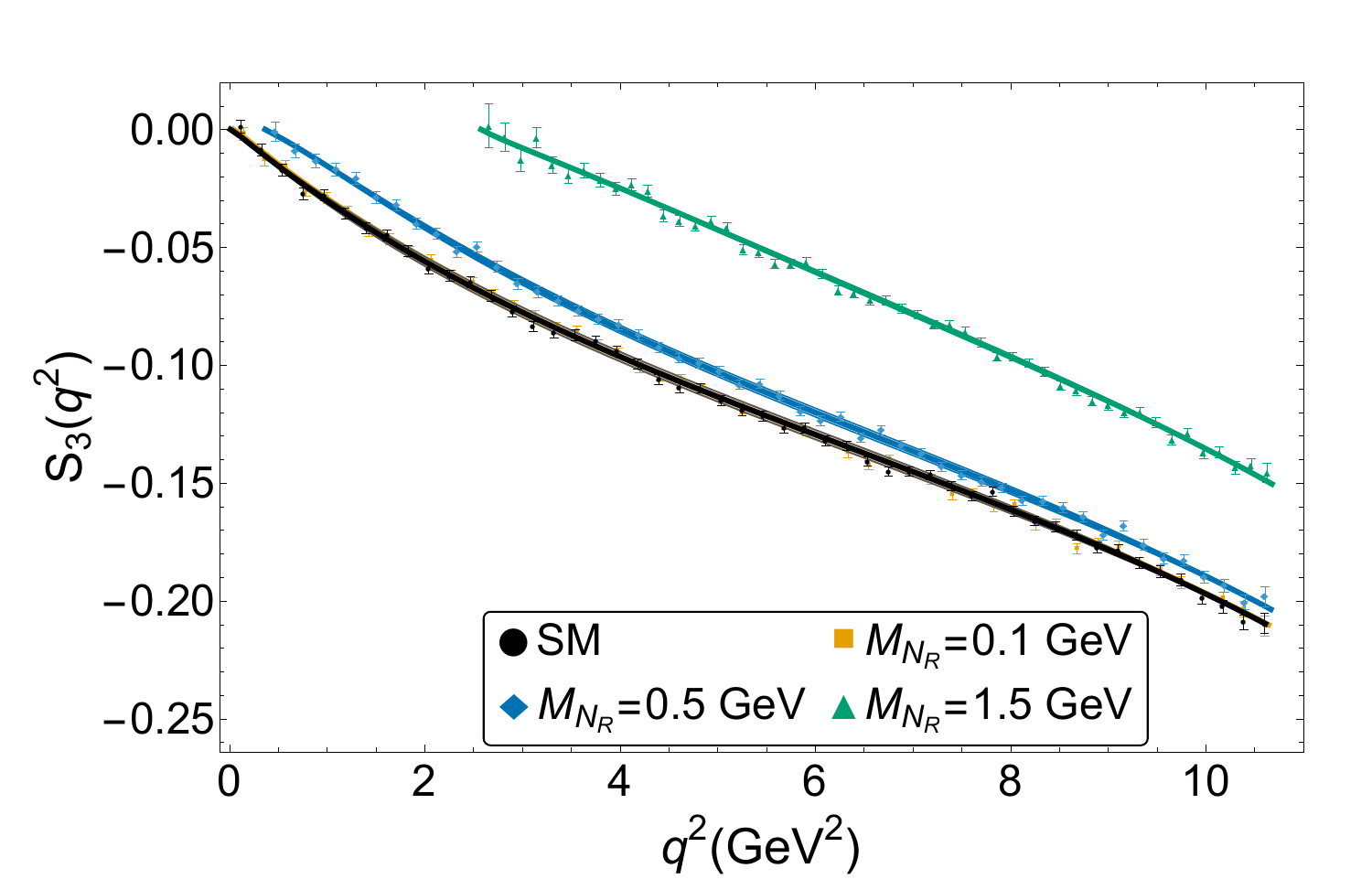}
\includegraphics[width=0.45\textwidth]{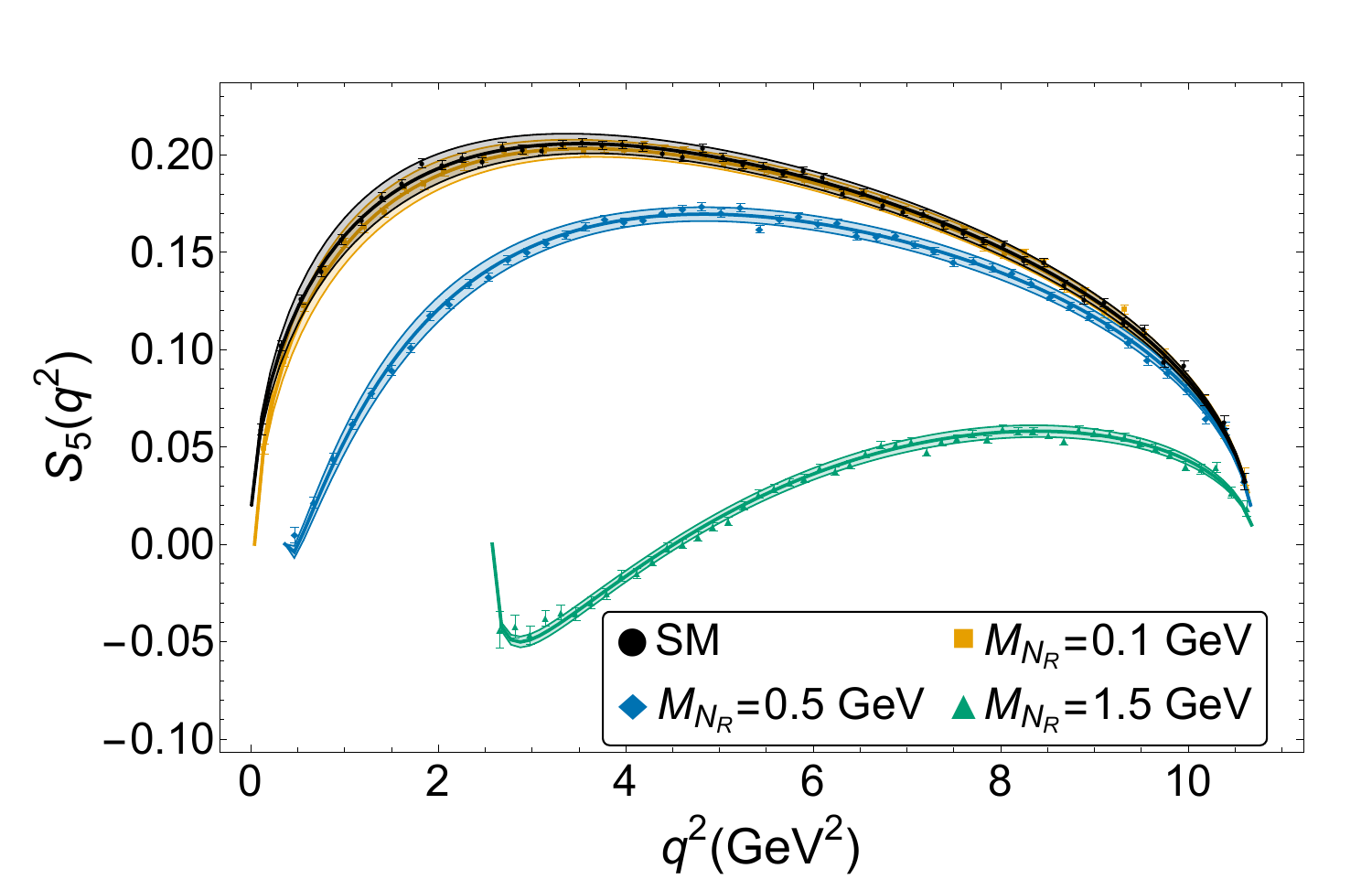}
\caption{Angular observables $A_{FB}$ (top left), $S_3$ (top right), and $S_5$ (bottom) as functions of $q^2$ for the decay $B^+\to{\bar D}^{*0}\mu^{+}N_R$. We use three benchmark RHN masses: $M_{N_R} = 0.1$ GeV (orange), 0.5 GeV (blue), and 1.5 GeV (teal). The EvtGen data are plotted as points with error bars, while solid curves provide the corresponding theoretical predictions; the SM expectation (left-handed neutrino) is shown in black.
\label{fig:AngObsMu}}
\end{figure}

The extracted observables are shown as functions of $q^2$ in Figs.~\ref{fig:AngObsEle} and \ref{fig:AngObsMu} for the electron and muon modes, respectively. The three benchmark RHN masses were used to generate each plot. In each panel, the EvtGen data are compared with the corresponding theoretical predictions, with the SM expectation shown for reference. The theory curves and the MC data agree within their uncertainties. This agreement of the EvtGen implementation of the asymmetries provides a more stringent test of our EvtGen model than the check of the one-dimensional distributions shown in Fig.~\ref{fig:elecdist}.

The accessible $q^2$ range begins at $q^2_{\rm min} = (m_\ell + M_{N_R})^2$ rather than at $m_\ell^2$, so the observables for the heavier benchmark masses turn on at progressively larger $q^2$. The shapes of all three observables deviate from the SM expectation across the full $q^2$ range, with the deviations growing with $M_{N_R}$. $A_{FB}(q^2)$ near the low-$q^2$ threshold reveals a small but notable difference: in the electron channel, both the SM and RHN curves rise smoothly from $A_{FB} = 0$ at threshold, whereas in the muon channel the SM curve dips sharply negative just above threshold.

\subsection{Results III: Correlation Among Observables}
\label{sec:Summarized plots}

\begin{figure}[!htbp]
\centering
\includegraphics[width=0.46\textwidth]{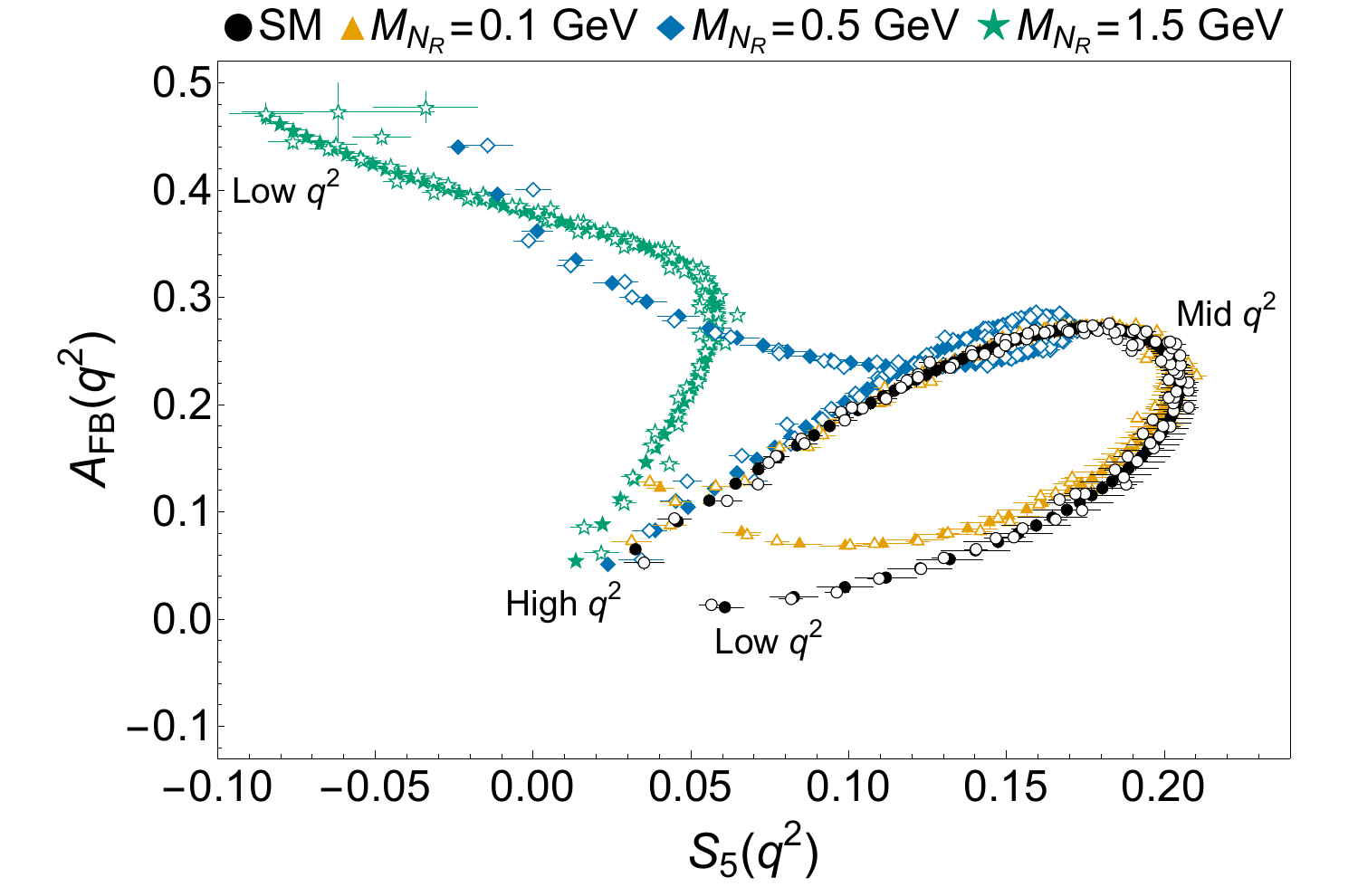}
\includegraphics[width=0.46\textwidth]{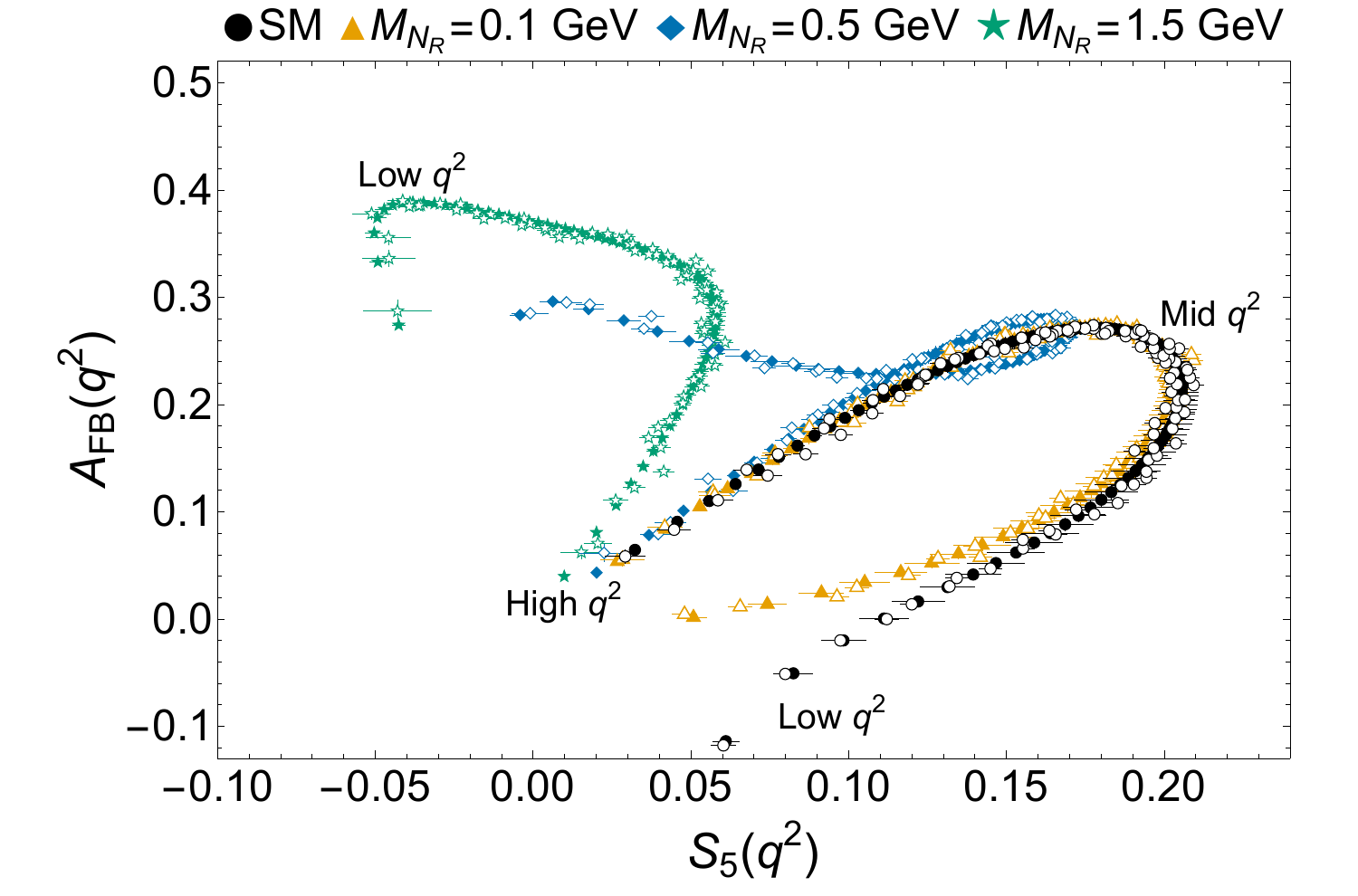}
\caption{Correlation between angular asymmetries $A_{FB}$ and $S_5$. Results for the SM (black) and RHN with mass $M_{N_R} = $ 0.1 GeV (orange), 0.5 GeV (blue), and 1.5 GeV (teal) are shown. Open (closed) points represent results obtained from EvtGen simulations (theoretical expressions). We use 0.1 GeV$^2$-wide $q^2$ bins for both the theory and simulated results. The left panel shows the electron channel, while the right panel shows the muon channel.
\label{fig:afb_s5_corrEvt}}
\end{figure}
Here we examine correlations between angular observables. The first is the $A_{FB}$-$S_5$ correlation, which is sensitive to interference between helicity amplitudes. Fig.~\ref{fig:afb_s5_corrEvt} shows the correlation between $A_{FB}$ and $S_5$ in low-, mid-, and high-$q^2$ ranges for the electron channel (left panel) and muon channel (right panel). In this figure, the open points represent the MC data from EvtGen and the closed points represent theoretical results. For quick comparison between theory and simulation, we have used 0.1 GeV$^2$-wide $q^2$ bins. In both panels, the black points represent the SM, whereas orange, blue, and teal correspond to the NP scenarios with $M_{N_R}$ = 0.1 GeV, 0.5 GeV, and 1.5 GeV, respectively. One can see that for the low-$q^2$ region, the correlation between $A_{FB}$ and $S_5$ is highly sensitive to $M_{N_R}$, so that even for $M_{N_R} = 0.1$ GeV a deviation from the SM curve is observed. For higher RHN masses, the correlation changes significantly. A precise measurement of this correlation will provide a useful and necessary probe of the angular structure of the $B$ decays in the presence of an RHN. Plots showing both MC simulation and theoretical correlations between $A_{FB}$ and $S_3$, and $S_{5}$ and $S_3$ are given in Appendix \ref{app:EvtGen-Theory Correlation}. Even though we show our results separately for the electron and muon channels, the only differences are noticeable in the low-$q^2$ region at threshold.

\section{Experimental Analysis} 
\label{sec:exptanalysis}

The Belle II experiment \cite{Belle-II:2010dht} was designed to search for BSM physics at the SuperKEKB facility \cite{AKAI2018188}, an asymmetric-energy electron–positron collider located at the KEK laboratory in Tsukuba, Japan. The beam energies are chosen to provide a center-of-mass energy of approximately 10.58 GeV, corresponding to the $\Upsilon (4S)$ resonance. At this energy, $B$ pairs are produced through the process $e^+e^-\to\Upsilon(4S)\to B{\bar B}$ with a production cross section of approximately 1.1 nb.

We performed a feasibility study using the Belle II Analysis Software Framework (basf2) \cite{Kuhr2018}, into which the BTODSTARLNUNPRH decay model was incorporated. The detector response was simulated using GEANT4 \cite{GEANT4:2002zbu}, after which the simulated events were processed and reconstructed using the standard Belle II reconstruction software.  The decay is reconstructed in the channel $B^0\to D^{*-}\ell^{+} N_R$, followed by $D^{*-}\to\bar{D}^0 \pi^-$ and $\bar{D}^0 \to K^+\pi^-$. The charge-conjugate decay chain is also included. We generated 100,000 events for the SM decay and 30,000 for each of the following mass hypotheses: 0.1, 0.5, 1.0, 1.5, 2.0, 2.5 and 3.0 GeV. In our samples, one $B$ meson was forced to decay as indicated previously, while the companion $B$ meson decayed generically. 

\begin{figure}[!htbp]
\centering
\includegraphics[width=0.48\textwidth]{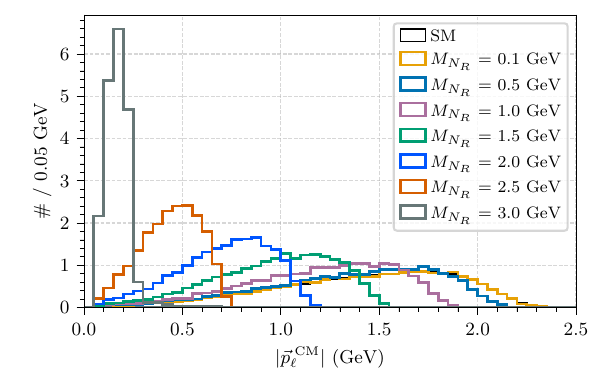}
\includegraphics[width=0.48\textwidth]{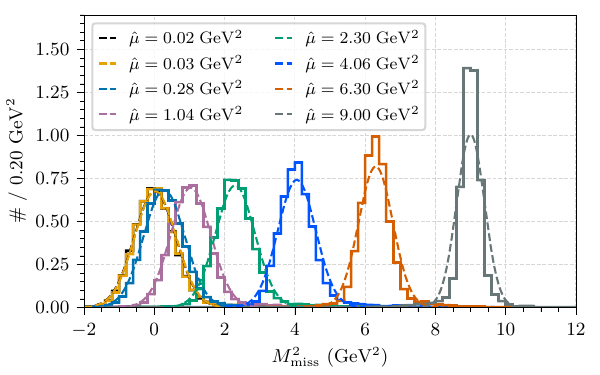}
\caption{Reconstructed distributions for $B^0\to D^{*-}\ell^{+} N_R$ events obtained using basf2, the Belle II Software. The left panel shows the lepton-momentum spectrum in the CM frame. As $M_{N_R}$ increases, the spectrum becomes softer. The right panel shows the missing-mass-squared ($M_{\rm miss}^2$) distribution calculated under the approximation that the $B$ meson is at rest in the CM frame. As $M_{N_R}$ increases, the lepton momentum distribution becomes narrower and shifts to the right, while the missing-mass-squared distribution increases as expected. The Gaussian distribution (dashed lines) was fitted using an unbinned maximum-likelihood analysis. The uncertainty on all $\hat{\mu}$ estimates is less than 0.01 GeV$^2$. The distribution for the SM channel largely overlaps with the $M_{N_R}$ = 0.1 GeV distribution. \label{fig:reco_kinematics}}
\end{figure}

\begin{figure}[!htbp]
\centering
\includegraphics[width=0.48\textwidth]{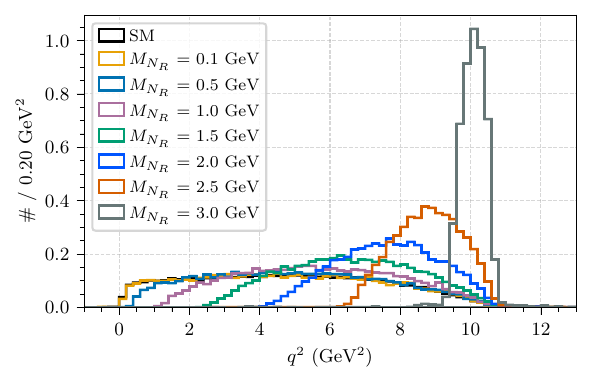}
\includegraphics[width=0.48\textwidth]{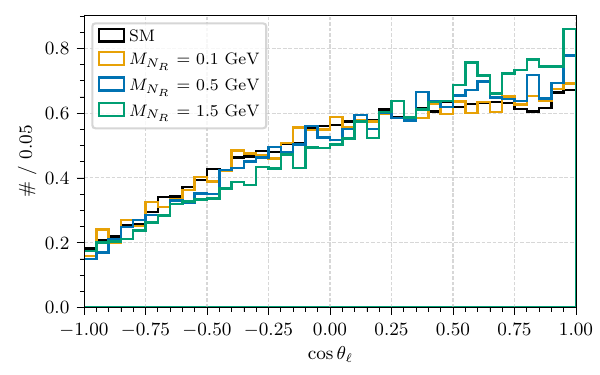}
\includegraphics[width=0.48\textwidth]{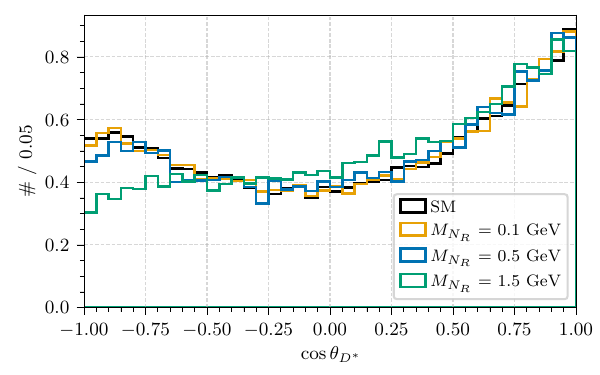}
\includegraphics[width=0.48\textwidth]{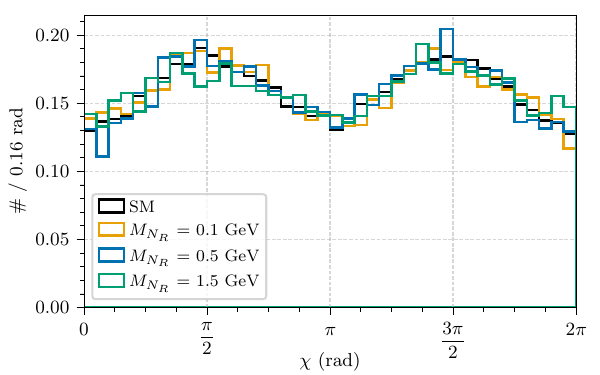}
\caption{Reconstructed event distributions as functions of $q^2$ and helicity angles obtained using the Diamond Frame estimate of the $B$-meson momentum. The top-left panel shows the $q^2$ distribution for all simulated mass hypotheses. As $M_{N_R}$ increases, the lower threshold increases while the upper remains fixed, shifting the distribution toward larger $q^2$ and narrowing the allowed interval.  The remaining panels show $\cos\theta_\ell$ (top right), $\cos\theta_{D^*}$ (bottom left), and $\chi$ (bottom right). Only the SM sample and the representative hypotheses $M_{N_R}$ = 0.1, 0.5, and 1.5 GeV are shown in the angular panels since the remaining distributions largely overlap. The distortion near $\cos\theta_{D^*} = -1$ is associated with reduced reconstruction efficiency for the low-momentum pions in the $D^*$ decay.}
\label{fig:reco_diamond}
\end{figure}

\begin{figure}[!htbp]
\centering
\includegraphics[width=0.48\textwidth]{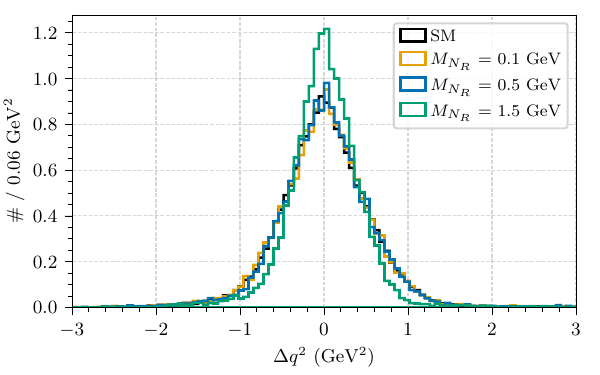}
\includegraphics[width=0.48\textwidth]{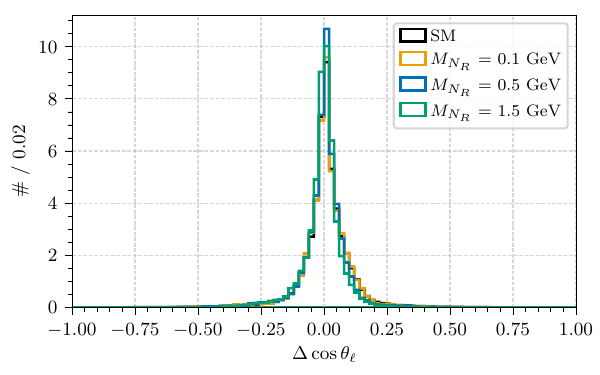}
\includegraphics[width=0.48\textwidth]{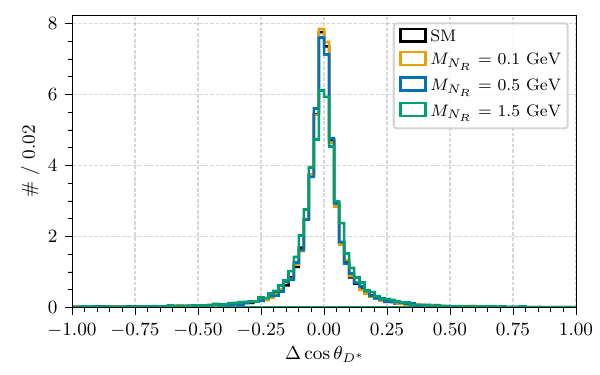}
\includegraphics[width=0.48\textwidth]{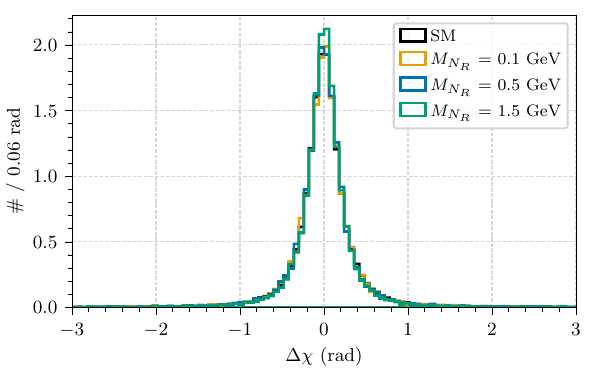}
\caption{Distributions of $q^2$ and helicity angle residuals in reconstructed $B^0\to D^{*-}\ell^{+}N_R$ events. The residual for $x$ is defined as $\Delta x = x_{\rm DF} - x_{\rm gen}$, where $x_{\rm DF}$ is calculated using reconstructed particles and the Diamond Frame estimate of the $B$-meson momentum, while $x_{\rm gen}$ is calculated from generator-level quantities. The panels show $\Delta q^2$ (top left), $\Delta\cos\theta_\ell$ (top right), $\Delta\cos\theta_{D^*}$ (bottom left), and $\Delta\chi$ (bottom right). Distributions for the SM sample (black) and the representative hypotheses $M_{N_R} =$ 0.1 GeV (orange), 0.5 GeV (blue), and 1.5 GeV (teal) are shown. \label{fig:resolution_angular}}
\end{figure}

\begin{figure}[!htbp]
\centering
    \includegraphics[width=0.48\textwidth]{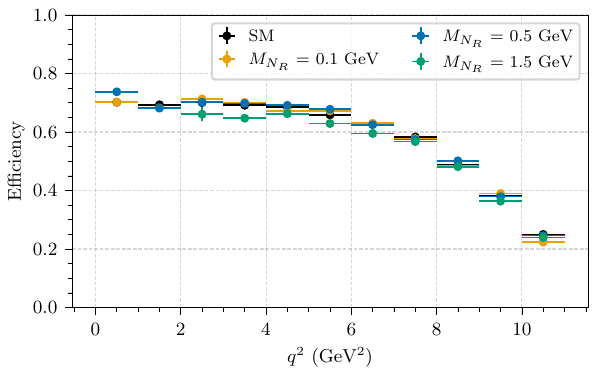}
    \includegraphics[width=0.48\textwidth]{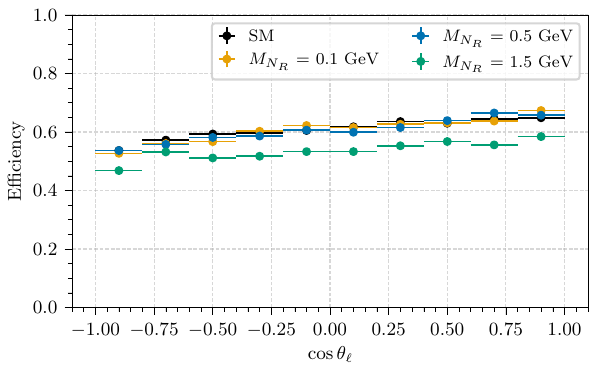}
    \includegraphics[width=0.48\textwidth]{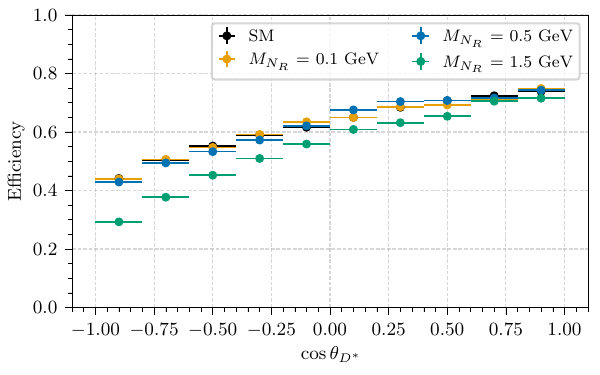}
    \includegraphics[width=0.48\textwidth]{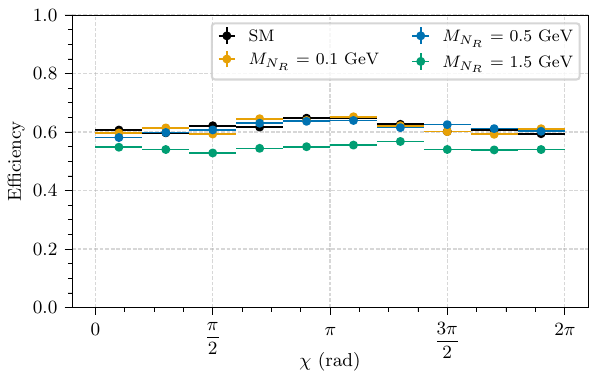}
    \caption{Reconstruction efficiency for $B^0\to D^{*-}\ell^{+} N_R$ as a function of the generator-level variables $q^2$ (top left), $\cos\theta_\ell$ (top right), $\cos\theta_{D^*}$ (bottom left), and $\chi$ (bottom right). Only the SM sample and the representative hypotheses $M_{N_R} =$ 0.1, 0.5, and 1.5 GeV are shown because the remaining results largely overlap. The reduced efficiency near $\cos\theta_{D^*} = -1$ is associated with the lower reconstruction efficiency for low-momentum pions from the $D^{*+}$ decay. The overall reconstruction efficiency decreases as $M_{N_R}$ increases due to a reduction in the available phase space.}
    \label{fig:refficiency_angular}
\end{figure}

Plots corresponding to all mass hypotheses are shown in Fig.~\ref{fig:reco_kinematics} and in the $q^2$ panel of Fig.~\ref{fig:reco_diamond}. In the angular panels of Fig.~\ref{fig:reco_diamond} and throughout Figs.~\ref{fig:resolution_angular} and \ref{fig:refficiency_angular}, only the SM sample and the representative hypotheses $M_{N_R}$ = 0.1, 0.5, and 1.5 GeV are shown because the remaining distributions largely overlap. Reconstructed distributions in Figs.~\ref{fig:reco_kinematics} -- \ref{fig:resolution_angular} are normalized so that the area under the curve is unity. Although we generated plots for both the electron and muon channels, the distributions were compatible with each other. Here we have chosen to show plots only for the electron channel, which has a wider phase space. In general, the distributions for the SM channel largely overlap with the $M_{N_R}$ = 0.1 GeV sample.

Reconstructed charged-particle tracks consistent with pion-, kaon-, and lepton-mass hypotheses are used as final-state candidates. These are subsequently combined to create candidates of the decayed mesons, $D^0$ and $D^{*+}$. A detailed description of the track reconstruction algorithms is given in Ref.~\cite{BelleIITrackingGroup:2020hpx}. 

No analysis-level selection criteria were imposed on the reconstructed events. Instead, we considered only reconstructed objects matched to the corresponding generated particles using MC truth information \cite{SatoMCMatching}. The left panel of Fig.~\ref{fig:reco_kinematics} shows the distribution of the reconstructed lepton momentum after MC matching. In this figure, we see that the lepton momentum becomes softer as the RHN mass increases. For $M_{N_R} = 3.0$ GeV, all leptons in the distribution have momenta less than 300 MeV.

Neutrinos in the final state escape detection, resulting in incomplete kinematic information for the decay. This presents an experimental challenge. This challenge is addressed by exploiting the production of $B\bar{B}$ pairs at the $\Upsilon (4S)$ threshold. In a tagged or Full Event Interpretation (FEI) analysis, the partner $B$ meson is fully reconstructed, allowing the four-momentum of the signal $B$ meson to be inferred from the known initial-state four-momentum. Alternatively, in an untagged analysis, the signal $B$-meson momentum can be estimated using kinematic constraints without explicitly reconstructing the companion $B$.

Here we adopt an untagged approach based on the Diamond Frame method \cite{BaBar:2006taf}, modified to account for a nonzero $N_R$ mass. This method was originally introduced by the \babar collaboration \cite{BaBar:2006taf} as a refinement of the Y-frame method developed by the CLEO Collaboration \cite{CLEO:1995zse}. The Diamond Frame method has been recently used by the Belle II collaboration in Ref.~\cite{Belle-II:2022ffa}.

Let $Y = D^* + \ell$ denote the system formed by the $D^*$ meson and the charged lepton $\ell$. Assigning four-momentum $p_X$ to particle $X$ (here $X = B, Y, N_R$), conservation of momentum implies $p_{N_R} = p_B - p_Y$. In the center-of-mass (CM) frame, the angle $\theta_{BY}$ between ${\vec p}_Y$ and ${\vec p}_B$ is given as,
\beq
    \cos\theta_{BY} = \dfrac{2 E_B E_Y - M_B^2 - M_Y^2 + M_{N_R}^2}{2 |\vec p_B| |\vec p_Y|} ,
\label{eq:costhetaBY}
\eeq
where $M_X$ represents the rest mass of particle $X$, $E_{X}$ and ${\vec p}_X$ respectively represent the energy and three-momentum of particle $X$ in the CM frame. We take $E_B = E_{\rm CM}/2$, and $|\vec p_B| = \sqrt{E_B^2 - M_B^2 }$. The kinematically allowed $B$-meson directions lie on a cone about $\vec p_Y$ with opening angle $\theta_{BY}$.

The Y-frame averages two candidate directions given by the intersection of this cone with the plane defined by the $Y$ and $D^*$ momenta. The Diamond Frame approach additionally exploits the expected $\sin^2 \theta_B$ angular distribution of the $B$ mesons, where $\theta_B$ is the angle between the $B$-meson direction and the beam axis. Four candidate directions are considered: the two used in the Y-frame and two additional orthogonal directions forming a diamond-shaped configuration. Each direction is weighted by its corresponding $\sin^2 \theta_B$ factor. Increasing the number of candidate directions only yields a negligible improvement.

The Diamond Frame method requires an estimation of the neutrino mass. In the SM, the neutrino mass can be neglected when calculating $\cos\theta_{BY}$ from Eq.~\ref{eq:costhetaBY}. For a nonzero RHN mass, we estimate $M_{N_R}^2$ from the missing-mass-squared variable, $M_{\rm miss}^2$. To calculate this variable, we approximate the $B$-meson four-momentum in the CM frame as $p_B^{(0)} = (E_{\rm CM}/2,\vec 0)$ to obtain $p_{\rm miss} = p_B^{(0)} - p_{D^*} - p_{\ell} \approx p_{N_R}$. Thus,
\beq
    M_{\rm miss}^2 = (E_{\rm CM}/2-E_Y)^2 - |\vec p_{D^*} + \vec{p_\ell}|^2,
\eeq
At the reconstruction level, $M_{\rm miss}^2$ is distributed around $M_{N_R}^2$, as shown in the right panel of Fig.~\ref{fig:reco_kinematics}. We performed a maximum-likelihood fit of a Gaussian distribution to $M_{\rm miss}^2$ and took the fitted location parameter, $\hat{\mu}$, as an estimator of $M_{N_R}^2$. The bias of this estimator, measured as $|\hat{\mu}-M_{N_R}^2|$, was less than 0.06 GeV$^2$ for the considered mass hypotheses.

Fig.~\ref{fig:reco_kinematics} shows the missing-mass-squared and the lepton-momentum distributions in the CM frame for different mass hypotheses. The initial assumption of the $B$ meson at rest in the CM frame, which can be regarded as a first-order approximation, was used only to obtain the $M_{N_R}^2$ estimate. The generated value of $M_{N_R}^2$ was used as the nominal input to the Diamond Frame reconstruction. We varied this value by 0.06 GeV$^2$, obtained from the maximum bias of the estimator, to assess its effect on bias and resolution.

Fig.~\ref{fig:reco_diamond} shows the reconstructed event distributions as functions of $q^2$ and helicity angles. These variables were calculated using the resulting Diamond Frame ${\bar B}^0$ momentum and the generated RHN mass. As the RHN mass increases, the $q^2$ distribution becomes narrower. This behavior is consistent with the description in earlier sections of this paper.

For each variable $x$, we distinguish between its Diamond Frame reconstructed value, $x_{\mathrm{DF}}$, and its generator-level value, $x_{\mathrm{gen}}$. The former is calculated using the reconstructed $D^*$ and lepton candidates together with the Diamond Frame estimate of the $B$-meson momentum, whereas the latter is calculated from the generated four-momenta. The residual is defined as $\Delta x = x_{\mathrm{DF}} - x_{\mathrm{gen}}$. We characterize the reconstruction bias of the residual distribution using its median. The resolution of the residual distribution is characterized using half of the central 68\% interval, i.e., the difference between its 84th and 16th percentiles. For a Gaussian distribution, this resolution corresponds approximately to one standard deviation. 

The angular biases remain close to zero across the simulated mass hypotheses: their absolute values do not exceed 0.030 for the cosine variables or 0.003 rad for $\chi$. The resolutions of $\cos\theta_\ell$ and $\chi$ improve modestly up to $M_{N_R}$ = 1.5 GeV and then deteriorate, whereas the resolution of $\cos\theta_{D^*}$ worsens progressively with increasing $M_{N_R}$. At $M_{N_R}$ = 3.0 GeV, the resolutions of $\cos\theta_\ell$, $\cos\theta_{D^*}$, and $\chi$ are 0.16, 0.19, and 0.39 rad, respectively. This is to be compared with 0.08, 0.08, and 0.30 rad for the SM sample. In contrast, the $q^2$ resolution improves as the allowed kinematic range narrows, decreasing from 0.54 GeV$^2$ for the SM sample to 0.14 GeV$^2$ for $M_{N_R}$ = 3.0 GeV. The $q^2$-reconstruction bias increases from approximately 0.06 to 0.11 GeV$^2$ over the same range.

To assess the sensitivity to the assumed RHN mass, we repeated the Diamond Frame reconstruction after shifting the nominal mass-squared input by $\pm$0.06 GeV$^2$. The resolutions of $\cos\theta_{D^*}$ and $\chi$ changed by at most 1.2\% and 1.6\%, respectively. Larger effects were observed for $\cos\theta_\ell$ and $q^2$. The largest change in the $\cos\theta_\ell$ resolution was 10.5\%, occurring for the SM sample; among the RHN samples, the maximum was 10.4\% at $M_{N_R}$ = 0.1 GeV. The $q^2$ resolution changed by less than 1.8\% for $M_{N_R}\leq$ 2.5 GeV, but increased by 10.2\%, from 0.124 to 0.137 GeV$^2$, for $M_{N_R}$ = 3.0 GeV.

For the reconstruction bias, we instead consider the absolute difference between the varied and nominal median residuals. The largest differences were 0.020 for $\cos\theta_\ell$, occurring at $M_{N_R}$ = 3.0~GeV, and 0.047~GeV$^2$ for $q^2$, occurring at $M_{N_R}$ = 2.5~GeV. The corresponding maximum differences for $\cos\theta_{D^*}$ and $\chi$ were 0.0009 and 0.0010~rad, respectively.

The reconstruction efficiency is evaluated as a function of the generator-level variables. For each bin $i$, $n_i^{\rm gen}$ denotes the number of generated signal decays with generator-level values in that bin. On the other hand, $n_i^{\rm matched}$ denotes the number of such decays with a correctly reconstructed, MC-matched signal candidate. The efficiency is, therefore,
\begin{equation}
    \epsilon_i=\frac{n_i{\rm{(matched)}}} {n_i{\rm{(gen)}}}.
\end{equation}
For the RHN samples, each generated event contains one $B^0\to D^{*-}\ell^{+}N_R$ decay. The companion $B^0$ always decays according to its SM branching fractions. In the SM sample, therefore, the companion $B^0$ decay will produce a small number of events with two $B\to D^{*}\ell\nu_L$ decays. In these events, each generated signal decay is counted separately, along with its corresponding matched candidate, and the uncertainties are evaluated using the binomial distribution.

Both the numerator and denominator are binned using the generator-level value of the observable. Reconstructed values are not used in this efficiency definition; the effects of reconstruction resolution and bin-to-bin migration are studied separately through the residual distributions in Fig.~\ref{fig:resolution_angular}.

Fig.~\ref{fig:refficiency_angular} shows the reconstruction efficiency as a function of $q^2$ and helicity angles. In particular, the efficiency decreases toward $\cos\theta_{D^*} = -1$, where the pion from the $D^*$ decay tends to have lower momentum and is therefore less likely to be reconstructed. The overall reconstruction efficiency also decreases as $M_{N_R}$ increases because the available phase space is reduced.

Our feasibility study depends on MC truth-table matching, as no reconstruction-level selection has been applied. We evaluated the detector-level reconstruction acceptance and kinematic resolutions for correctly reconstructed signal candidates. The missing-mass-squared distribution will have a negligible bias in the presence of an RHN. Therefore, the resolutions and reconstruction biases for other variables can be determined using the RHN mass estimation. The efficiencies given here provide upper limits on the final efficiency; With the inclusion of background and further selections, the efficiencies may be lower. An untagged approach was used here. Future work will investigate reconstruction of the companion $B$ meson using Full Event Interpretation (FEI), which is expected to improve the kinematic resolutions relative to the untagged approach, albeit at the cost of much reduced efficiency.

\section{Conclusion}
\label{sec:conclusion}

We have calculated the kinematic and angular distributions of ${\bar B}\to D^*\ell^-{\bar N}_R$, where $N_R$ is a sterile right-handed neutrino (RHN). Three benchmark values were chosen for the RHN mass in the MeV--GeV range. This calculation has been implemented in an EvtGen-based Monte Carlo (MC) generator, which should enable future experimental searches for RHNs. We have examined event distributions as functions of $q^2, \cos\theta_{l}, \cos\theta_{D^*}$, and $\chi$. As we increase the RHN mass, the lower limit for $q^2$ also increases, and the lepton-momentum spectrum becomes softer -- a feature that should be noticed in future experimental searches for RHNs. We find an interesting and unexpected correlation between the angular asymmetries $A_{FB}$ and $S_5$ as a function of $q^2$; the shape of this correlation depends on the RHN mass and can be tested in future experiments.

We also performed an initial experimental feasibility study with the RHN MC generator and the Belle II detector. Our simulation suggested that a massive sterile RHN in the GeV mass range could be detected in semileptonic $B$ decays with future Belle II data. Future work will examine the sensitivity of Belle II, including backgrounds, tagging strategies, and experimental selections.

\acknowledgments
We thank L.~Mukherjee for helpful communications, K. Yoshihara for useful discussions, and K. Flood for editorial comments. B.B.~is supported by the U.S.~National Science Foundation through Grant No.~PHY-2609773. A.D.~and K.P.~are supported in part by the U.S.~National Science Foundation under Grant No.~PHY-2309937 and PHY-2607841. T.E.B., H.C.L., and A.S.~thank the DOE Office of High Energy Physics for support through DOE Grant No.~DE-SC0010504. N.D.~acknowledges support from the SERB Grant No.~SPG/2022/001238.

\appendix

\section{Canonical form factor parametrization and helicity amplitudes}
\label{app:tradformfactors}

The helicity amplitudes are projections of the ${\bar B}\to D^*$ matrix elements in the basis of $W$-boson polarization vectors. To write expressions for the helicity amplitudes, we first express the matrix elements in terms of form factors. The vector and axial-vector operator matrix elements are given by 
\begin{align}
\langle D^* (p_{D^*},\eds) | \bar{c} \gamma_\mu b | \bar{B}(p_B) \rangle &= -i \epsilon_{\mu\nu\rho\sigma} \eds^{*\nu}p_B^{\rho}p_{D^*}^{\sigma} \frac{2 V(q^2)}{m_B + m_{D^*}}\,, \label{eq:vectormatrixcanonical} \\
\langle D^* (p_{D^*}\eds) | \bar{c} \gamma_\mu \gamma_5 b | \bar{B}(p_B) \rangle &= \epsilon^{*}_{D^*,\mu} (m_B + m_{D^*}) A_1(q^2) \nonumber \\
&- (p_B+p_{D^*})_{\mu}(\eds^*.q) \frac{A_2(q^2)}{m_B + m_{D^*}} \nonumber \\
&- q_\mu (\eds^*.q) \frac{2 m_{D^*}}{q^2} [A_3(q^2) - A_0(q^2)] \,,\label{eq:axialvectormatrixcanonical}
\end{align}
where
\begin{align}
A_3(q^2) = \frac{m_B + m_{D^*}}{2m_{D^*}} A_1(q^2) - \frac{m_B - m_{D^*}}{2m_{D^*}} A_2(q^2)\,.
\end{align}
Here, $q^\mu = p_B^\mu - p_{D^*}^\mu$, $\eds$ is the polarization vector of $D^*$, and $p_{B(D^*)}$ is the momentum of $B(D^*)$. $V$ is the canonical vector form factor, while $A_{0,1,2,3}$ are the canonical axial-vector form factors. Since $A_3(0) = A_0(0)$, only four form factors are independent.

The tensor matrix element can be parametrized as
\begin{align}
\langle D^* (p_{D^*},\eds) | \bar{c} \sigma_{\mu\nu}  b | \bar{B}(p_B) \rangle &= \epsilon_{\mu\nu\rho\sigma} \Bigg\{ -\eds^{*\rho} (p_B + p_{D^*})^{\sigma} T_1(q^2) \nonumber \\
&+ \eds^{*\rho}q^{\sigma} \frac{m_B^2 - m_{D^*}^2}{q^2} [T_1(q^2) - T_2(q^2)] \label{eq:tensormatrixcanonical} \\
&+ 2 \frac{(\eds^* . q)}{q^2} p_B^{\rho} p_{D^*}^{\sigma} \left[T_1(q^2) - T_2(q^2) - \frac{q^2}{m_B^2 - m_{D^*}^2}T_3(q^2) \right] \Bigg\}\,, \nonumber
\end{align}
where $T_{1,2,3}$ are the canonical tensor form factors. The pseudotensor matrix elements can be related to the tensor matrix elements by the relation $\bar{c}\sigma_{\mu\nu} \gamma^5 b = -\frac{i}{2} \epsilon_{\mu\nu\alpha\beta} \bar{c}\sigma^{\alpha\beta}b$, with the convention $\epsilon_{0123} = 1$. Similarly, the pseudoscalar matrix element can be expressed in terms of $A_0$ as
\begin{align}\label{eq:pseudoscalarmatrixcanonical}
\langle D^* (p_{D^*},\eds) | \bar{c} \gamma_5 b | \bar{B} (p_B) \rangle = -(\eds^*.q) \frac{2 m_{D^*}}{m_b + m_c} A_0(q^2)\,,
\end{align}
where $m_{b(c)}$ is the $b(c)$-quark mass. 
\par 
Having defined the scalar, vector, and tensor matrix elements, using Eq.~\ref{hadhelamp1}, we define the hadronic helicity amplitudes as follows.
\begin{align}\label{eq:helamp}
\begin{aligned}
H_{ V}^\pm(q^2)  &\equiv H_{ V_L,\pm}^{\pm}(q^2) = -H_{ V_R,\mp}^{\mp}(q^2) \\
&= (m_B + m_{D^*}) A_1(q^2) \mp \frac{\sqrt{\lambda_{D^*}(q^2)}}{m_B + m_{D^*}}V(q^2)\,, \\
H_{ V}^0(q^2) &\equiv H_{ V_L,0}^{0}(q^2) = - H_{ V_R,0}^{0}(q^2) \\
&= \frac{m_B + m_{D^*}}{2 m_{D^*}\sqrt{q^2}} \left[ -(m_B^2 - m_{D^*}^2 - q^2) A_1(q^2) + \frac{\lambda_{D^*}(q^2)}{(m_B + m_{D^*})^2}A_2(q^2) \right]\,, \\
H_{P}(q^2) &\equiv H_{P}^0(q^2) = H_{S_R}^0 = - H_{S_L}^0 \\
&= -\frac{\sqrt{\lambda_{D^*}(q^2)}}{m_B + m_c} A_0(q^2)\,, \\
H_{T}^\pm(q^2) &\equiv \pm H_{\rm{T},\pm t}^{\pm}(q^2) \\
&= \frac{1}{\sqrt{q^2}} \bigg[ \sqrt{\lambda_{D^*}(q^2)} T_1(q^2) \pm (m_B^2 - m_{D^*}^2) T_2(q^2) \bigg]\,, \\
H_{T}^0(q^2) &\equiv H^0_{T,+-}(q^2) = H^0_{T,0t}(q^2) \\
&= \frac{1}{2 m_{D^*}} \bigg[ (m_B^2 + 3 m_{D^*}^2 - q^2) T_2(q^2) - \frac{\lambda_{D^*}(q^2) }{(m_B^2 - m_{D^*}^2)}T_3(q^2) \bigg]\,. \\
\end{aligned}
\end{align}
These expressions agree with the results in Refs.~\cite{Bhattacharya:2019olg} and \cite{helampref2}, and, up to an overall sign difference in $H_{T}^0$, with Ref.~\cite{tensorff}.

\section{BGL parametrization of hadronic form factors}
\label{app:bglformfactors}

The Boyd-Grinstein-Lebed (BGL) parametrization of hadronic form factors uses a series expansion in terms of a small parameter $z$ derived from the conformal mapping of the kinematic variable $w$ \cite{Boyd:1997kz}. This mapping allows $z$ to map the kinematic range of $q^2$ (or $w$) to a small interval, improving the convergence properties of the series.  In addition, with more parameters, the BGL parametrization can accommodate a wider range of possible form-factor behaviors, potentially leading to better fits to experimental data.  The canonical form factors, $V, A_{0,1,2}$ are related to the BGL form factors, $g, f, {\cal F}_{1,2}$, as follows: 
\begin{align}\label{eq:bgltostdff}
\begin{aligned}
g&=\frac{2}{m_B+m_{D^*}}V\,, \\
f&=(m_B+m_{D^*})A_1\,, \\
\mathcal{F}_1 &= \frac{1}{2m_{D^*}}\left[ (m_B^2-m_{D^*}^2-q^2)(m_B+m_{D^*}) A_1  -  
\frac{4m_B^2|{ \vec{p}}_{D^*}|^2}{ m_B+m_{D^*}} A_2 \right]\,,  \\
\mathcal{F}_2 &= 2A_0\,,
\end{aligned}
\end{align}
where the $q^2$ dependence of the form factors is implicit. 
With these definitions, the helicity amplitudes in Eq.~\eqref{eq:helamp} can be expressed as follows. 
\begin{align} \label{eq:helampbgl}
\begin{aligned}
H_{V}^\pm(w) &= f\mp g m_B |{\vec{p}_{D^*}}|\,, \\
H_{V}^0(w) &=-\frac{\mathcal{F}_1 }{M_{B}\sqrt{1-2wr+r^2}}\,, \\
H_{P}(w)&=  -\frac{\mathcal{F}_2\,m_B\,|{ \vec{p}_{D^*}}|}{m_b + m_c}\,, \\
H_{T}^\pm(w)&=\frac{\pm f (m_b - m_c) + g m_B |{ \vec{p}_{D^*}}|(m_b + m_c)}{ M_{B}\sqrt{1-2wr+r^2} }\,, \\
H_{T}^0(w) &= \frac{-(m_b-m_c)(-\mathcal{F}_1 (m_B^2 - m_{D^*}^2) + 2 \mathcal{F}_2 m_B^2 |{ \vec{p}_{D^*}}|^2 ) }{(m_B^2 - m_{D^*}^2) {(M_{B}^2 + m_{D^*}^2 - 2M_{B}m_{D^*} w  )}}\,,
\end{aligned}
\end{align}
where we have changed variables from $q^2$ to $w = (m_B^2 + m_{D^*}^2 - q^2)/(2 m_B m_{D^*})$.  Note that the pseudoscalar and tensor form factors are reduced to the above form factors via the relations given in Appendix A of Ref.~\cite{Kapoor:2024ufg}.

The momentum dependence of these form factors is given by a 
$z$-expansion with
\begin{equation}
z \equiv \frac{\sqrt{w+1}-\sqrt{2}}{\sqrt{w+1}+\sqrt{2}}\,.
\end{equation}
As functions of $z$, the BGL form factors can be expressed as,
\begin{align} \label{eq:bglff}
\begin{aligned}
g(z) &= \frac{1}{P_{1^-}(z)\phi_g(z)} \sum_{n=0}^{\infty} a^g_n z_n\,,  \quad \quad \;
f(z) = \frac{1}{P_{1^+}(z)\phi_f(z)} \sum_{n=0}^{\infty} a^f_n z^n\,, \\
\mathcal{F}_1(z) &= \frac{1}{P_{1^+}(z)\phi_{\mathcal{F}_1}(z)} \sum_{n=0}^{\infty} a^{\mathcal{F}_1}_n z_n\,,  \quad
\mathcal{F}_2(z) = \frac{1}{P_{0^-}(z)\phi_{\mathcal{F}_2}(z)} \sum_{n=0}^{\infty} a^{\mathcal{F}_2}_n z_n\,.
\end{aligned}
\end{align}
The Blaschke factors, $P_{J^P}(z)$, appearing in Eq.~\eqref{eq:bglff} are given by
\begin{equation}\label{eq:blaschke}
P_{J^P}(z)=\prod_{p}\frac{z-z_p}{1-z\,z_p}\,,
\qquad
z_p=\frac{\sqrt{t_+-M_p^2}-\sqrt{t_+-t_-}}{\sqrt{t_+-M_p^2}+\sqrt{t_+-t_-}}\,,
\end{equation}
where $t_\pm=(m_B\pm m_{D^*})^2$, so that $t_+-t_-=4m_Bm_{D^*}$. The pole masses $M_p$ of
the relevant $B_c$ resonances are listed in Table~\ref{inp_polemass}.
\begin{table}[!htbp]
\centering
\setlength{\tabcolsep}{10pt}
\renewcommand{\arraystretch}{1.25}
\begin{tabular}{@{}ccl@{}}
\toprule
Form Factor & Type & Pole Masses, $M_p$ (GeV) \\
\midrule
$g$                  & $1^{-}$ & $6.329,\; 6.920,\; 7.020$          \\
$f,\; \mathcal{F}_1$ & $1^{+}$ & $6.739,\; 6.750,\; 7.145,\; 7.150$ \\
$\mathcal{F}_2$      & $0^{-}$ & $6.275,\; 6.842,\; 7.250$          \\
\bottomrule
\end{tabular}
\caption{The pole masses corresponding to the different types of $B_c$ resonances,
taken from Ref.~\cite{gambino}}
\label{inp_polemass}
\end{table}

The outer functions, $\phi_{g,f,{\cal F}_1,{\cal F}_2}$, also appearing in Eq.~\eqref{eq:bglff}, are \cite{Boyd:1997kz,gambino}
\begin{align}\label{eq:outerfns}
\begin{aligned}
\phi_g(z) &= 16\,r^2\sqrt{\frac{n_I}{3\pi\,\tilde\chi^{T}_{1^-}(0)}}\;
 \frac{(1+z)^2(1-z)^{-1/2}}{\big[(1+r)(1-z)+2\sqrt{r}\,(1+z)\big]^{4}}\,, \\[4pt]
\phi_f(z) &= \frac{4r}{m_B^2}\sqrt{\frac{n_I}{3\pi\,\chi^{T}_{1^+}(0)}}\;
 \frac{(1+z)(1-z)^{3/2}}{\big[(1+r)(1-z)+2\sqrt{r}\,(1+z)\big]^{4}}\,, \\[4pt]
\phi_{\mathcal{F}_1}(z) &= \frac{4r}{m_B^3}\sqrt{\frac{n_I}{6\pi\,\chi^{T}_{1^+}(0)}}\;
 \frac{(1+z)(1-z)^{5/2}}{\big[(1+r)(1-z)+2\sqrt{r}\,(1+z)\big]^{5}}\,, \\[4pt]
\phi_{\mathcal{F}_2}(z) &= 8\sqrt{2}\,r^2\sqrt{\frac{n_I}{\pi\,\tilde\chi^{L}_{1^+}(0)}}\;
 \frac{(1+z)^2(1-z)^{-1/2}}{\big[(1+r)(1-z)+2\sqrt{r}\,(1+z)\big]^{4}}\,,
\end{aligned}
\end{align}
where $n_I$ counts the spectator-quark channels, including $SU(3)$ breaking, for which we
take $n_I=2.6$. For the susceptibilities $\chi^{T}_{1^+}(0)$, $\tilde\chi^{T}_{1^-}(0)$
and $\tilde\chi^{L}_{1^+}(0)$ we use the values given in Ref.~\cite{fermilab}, which were originally obtained in Ref.~\cite{gambino}; these, together with the meson and quark masses, are collected in Table \ref{inp_formfactor}.
\begin{table}[!htbp]
\centering
\setlength{\tabcolsep}{10pt}
\renewcommand{\arraystretch}{1.25}
\begin{tabular}{llll} \hline
Mass & Value (MeV) & Input & Value \\ \hline
$m_B$ & 5279.63(20) & $|V_{cb}|$ & 0.041 \\
$m_{D^*}$ & 2010.26(05) & $n_I$  & 2.6 \\
$m_e$ & 0.000(00) & $\chi_{1^+}^{T}(0)$ & $3.894 \times 10^{-4}$ GeV$^{-2}$ \\
$m_\mu$ & 105.6583745(24) & $\tilde{\chi}_{1^-}^{T}(0)$ & $5.131 \times 10^{-4}$ GeV$^{-2}$ \\
$m_b$ & 4180 & $\tilde{\chi}_{1^+}^{L}(0)$ & $1.9421 \times 10^{-2}$ \\
$m_c$ & 1270 &  &       \\ \hline
\end{tabular}
\caption{Input values for $|V_{cb}|$ and particle masses taken from the Particle Data Group \cite{ParticleDataGroup:2020ssz}, together with relevant input values for parameters in the outer functions taken from Ref.~\cite{gambino}.
\label{inp_formfactor}}
\end{table}

In the limit of zero-recoil, i.e., $|{\vec p}_{D^*}|=0$ $(w_{\rm min}=1)$, we find that both $f$ and $\mathcal{F}_1$ depend only on the form factor $A_1$ and are related by 
\begin{align}\label{eq:relF1}
\mathcal{F}_1(0)=(m_B-m_{D^*})f(0) \,.
\end{align}
On the other hand, for maximum recoil, $q^2=0$ [$w_{\rm max} = (m_B^2 + m_{D^*}^2)/(2 m_B m_{D^*})$], we require  $A_0 = \{(m_B+m_{D^*})A_1 - (m_B-m_{D^*})A_2\}/(2 m_{D^*})$ to avoid the $q^2$ pole in the axialvector operator matrix element \cite{tensorff}. This leads to the following relationship.
\begin{align}\label{eq:relF2}
\mathcal{F}_1(z(w_{\rm max}))=\frac{(m_B^2-m_{D^*}^2)}{2}\mathcal{F}_2(z(w_{\rm max})) \,.
\end{align}

In the present analysis, the form-factor parameters $a_n^{g, f, {\cal F}_1, {\cal F}_2}$ are taken from the fitted results reported in Ref.~\cite{Biswas:2021pic}. These parameters appear in Eq.~\eqref{eq:bglff}; their numerical values are given in Table \ref{inp_bglcoeff}.
\begin{table}[!htbp]
\centering
\setlength{\tabcolsep}{10pt}
\renewcommand{\arraystretch}{1.25}
\begin{tabular}{llll} \hline
Coefficient & Value & Coefficient & Value \\ \hline
${a_0^{f}}$             & ${0.0123(1)}$   & ${a_0^{g}}$             & ${0.0318(10)}$ \\
${a_1^{f}}$             & ${0.0222(96)}$  & ${a_1^{g}}$             & ${-0.133(63)}$ \\
${a_2^{f}}$             & ${-0.522(196)}$ & ${a_2^{g}}$             & ${-0.62(146)}$ \\
${a_1^{\mathcal{F}_1}}$ & ${0.0021(15)}$  & ${a_0^{\mathcal{F}_2}}$ & ${0.0515(12)}$ \\
${a_1^{\mathcal{F}_2}}$ & ${-0.149(59)}$  & ${a_2^{\mathcal{F}_2}}$ & ${0.987(932)}$ \\ \hline
\end{tabular}
\caption{$z$-expansion coefficients needed to compute BGL form factors used in this analysis.
These coefficients appear in Eq.~\eqref{eq:bglff}.
\label{inp_bglcoeff}}
\end{table}

\section{Angular distribution for \boldmath${\bar B}\to D^*\ell^-{\bar N}_R$}

We start with the general differential decay rate given as
\beq
d\Gamma = \frac{1}{2m_B} \int d\Pi_4\,|\mathcal{M}|^2 \,, \label{eq:decayrategeneralmuN}
\eeq
where $\mathcal{M}$ denotes the full decay amplitude and $d\Pi_4$ is the four-body phase-space element. The decay amplitude is the sum of SM and NP contributions, i.e. $\mathcal{M} = \mathcal{M^{\rm SM}} + \mathcal{M^{\rm NP}}$. The SM calculations can be found in~\cite{Bhattacharya:2019olg}; here, we present the NP calculations. 

\subsection{Decay amplitude}
\label{sec:amplitudecalculationmu}

To calculate the amplitude, we use the process $\bar B^0\to D^{*+}W^{*-}(\to \ell^-\bar N_R)$ with
$D^{*+}\to D\pi$, shown in Fig.~\ref{fig:angN-bar}. The corresponding Feynman diagrams are shown in Fig.~\ref{fig:four_diagrams}. The decay amplitude can be expressed as,
\begin{align}
\label{eq:ampNPfullmuN}
\begin{aligned}
\Amp^{\rm NP}
&=
\sum_{\lambda_{D^*}}
\frac{G_F V_{cb}}{\sqrt{2}}
\left\langle D\pi\middle|D^*(\lambda_{D^*})\right\rangle
\Bigg[
C_S^{LR}
\left\langle D^*(\lambda_{D^*})\middle|
\bar c(1-\gamma^5)b
\middle|\bar B\right\rangle
\left(\bar u_\mu(1+\gamma^5)v_{\bar N_R}\right)
\\
&\quad
+
C_S^{RR}
\left\langle D^*(\lambda_{D^*})\middle|
\bar c(1+\gamma^5)b
\middle|\bar B\right\rangle
\left(\bar u_\mu(1+\gamma^5)v_{\bar N_R}\right)
\\
&\quad
+
C_V^{RR}
\left\langle D^*(\lambda_{D^*})\middle|
\bar c\gamma^\alpha(1+\gamma^5)b
\middle|\bar B\right\rangle
\left(\bar u_\mu\gamma_\alpha(1+\gamma^5)v_{\bar N_R}\right)
\\
&\quad
+
C_T^{RR}
\left\langle D^*(\lambda_{D^*})\middle|
\bar c\sigma^{\alpha\beta}(1+\gamma^5)b
\middle|\bar B\right\rangle
\left(\bar u_\mu\sigma_{\alpha\beta}(1+\gamma^5)v_{\bar N_R}\right)
\Bigg]\,.
\end{aligned}
\end{align}
With the definitions
\begin{align}
\label{eq:bdmunampdefs}
P(\lambda_{D^*})
&\equiv
\eds^\alpha(\lambda_{D^*})(p_D)_\alpha
=
\frac{1}{2g_{D^*D\pi}}
\left\langle D\pi\middle|D^*(\lambda_{D^*})\right\rangle ,
\\
\eds^{*\beta}(\lambda_{D^*})T_{\beta\alpha}^{V_R}
&\equiv
\left\langle D^*(\lambda_{D^*})\middle|
\bar c\gamma_\alpha(1+\gamma^5)b
\middle|\bar B\right\rangle ,
\end{align}
and writing the $D^*$ propagator explicitly,
\begin{align}
\sum_{\lambda_{D^*}}
\eds^\alpha(\lambda_{D^*})
\eds^{*\beta}(\lambda_{D^*})
\rightarrow
\frac{
\sum_{\lambda_{D^*}}
\eds^\alpha(\lambda_{D^*})
\eds^{*\beta}(\lambda_{D^*})
}{
p_{D^*}^2-m_{D^*}^2+i m_{D^*}\Gamma_{D^*}
}\,,
\end{align}
The amplitude can be written as
\begin{align}
\label{eq:ampNPmuNprop}
\begin{aligned}
\Amp^{\rm NP}
&=
\sqrt{2}\,G_F V_{cb}\,g_{D^*D\pi}
\sum_{\lambda_{D^*}}
P(\lambda_{D^*})
\frac{1}{
p_{D^*}^2-m_{D^*}^2+i m_{D^*}\Gamma_{D^*}
}
\\
&\quad\times
\Bigg[
C_S^{LR}H_{S_L}^{\lambda_{D^*}}L^S
+
C_S^{RR}H_{S_R}^{\lambda_{D^*}}L^S
+
C_V^{RR}\sum_m g_{mm}H_{V_R,m}^{\lambda_{D^*}}L_m^V
\\
&\qquad\qquad
+
C_T^{RR}\sum_{m,n}g_{mm}g_{nn}
H_{T,mn}^{\lambda_{D^*}}L_{mn}^T
\Bigg].
\end{aligned}
\end{align}
The product of the hadronic and leptonic parts can be written in terms of Lorentz-invariant helicity amplitudes by using the completeness relation for the $W$ polarization vectors,
\begin{align}
\label{eq:completenessrelationmu}
\sum_{mn}\ew^{*\alpha}(m)\ew^\beta(n)g_{mn}=g^{\alpha\beta}\,,
\end{align}
where $g_{mn}$ is the metric tensor and $m,n$ label the $W$ polarizations. For a $W$ moving in the $-z$ direction in the $B$ rest frame, we define:
\begin{align}
\label{eq:Wpolarizationvectorsmu}
\begin{aligned}
\ew^\alpha(\pm)
&=
\frac{1}{\sqrt{2}}(0,\pm 1,-i,0),
\\
\ew^\alpha(0)
&=
\frac{1}{\sqrt{q^2}}(|\vec q|,0,0,-q_0),
\\
\ew^\alpha(t)
&=
\frac{1}{\sqrt{q^2}}(q_0,0,0,-|\vec q|),
\end{aligned}
\end{align}
where $q_0$ and $\vec q$ are the energy and three-momentum of the $W$ in the $B$ rest frame. The leptonic helicity amplitudes are defined as,
\begin{align}
\label{eq:lephelampmu}
\begin{aligned}
L_m^V
&\equiv
\ew^\alpha(m)
\left(\bar u_\mu\gamma_\alpha(1+\gamma^5)v_{\bar N_R}\right),
\\
L^S
&\equiv
\left(\bar u_\mu(1+\gamma^5)v_{\bar N_R}\right),
\\
L_{mn}^T
&\equiv
-i\ew^\alpha(m)\ew^\beta(n)
\left(\bar u_\mu\sigma_{\alpha\beta}(1+\gamma^5)v_{\bar N_R}\right).
\end{aligned}
\end{align}
The corresponding hadronic helicity amplitudes are,
\begin{align}
\label{hadhelamp1}
\begin{aligned}
H_{V_L,m}^{\lambda_{D^*}}(q^2)
&\equiv
\ew^{*\alpha}(m)
\left\langle
D^*(p_{D^*},\eds(\lambda_{D^*}))
\middle|
\bar c\gamma_\alpha(1-\gamma^5)b
\middle|
\bar B(p_B)
\right\rangle ,
\\
H_{V_R,m}^{\lambda_{D^*}}(q^2)
&\equiv
\ew^{*\alpha}(m)
\left\langle
D^*(p_{D^*},\eds(\lambda_{D^*}))
\middle|
\bar c\gamma_\alpha(1+\gamma^5)b
\middle|
\bar B(p_B)
\right\rangle ,
\\
H_{S_L}^{\lambda_{D^*}}(q^2)
&\equiv
\left\langle
D^*(p_{D^*},\eds(\lambda_{D^*}))
\middle|
\bar c(1-\gamma^5)b
\middle|
\bar B(p_B)
\right\rangle ,
\\
H_{S_R}^{\lambda_{D^*}}(q^2)
&\equiv
\left\langle
D^*(p_{D^*},\eds(\lambda_{D^*}))
\middle|
\bar c(1+\gamma^5)b
\middle|
\bar B(p_B)
\right\rangle ,
\\
H_{T,mn}^{\lambda_{D^*}}(q^2)
&\equiv
i\,\ew^{*\alpha}(m)\ew^{*\beta}(n)
\left\langle
D^*(p_{D^*},\eds(\lambda_{D^*}))
\middle|
\bar c\sigma_{\alpha\beta}(1+\gamma^5)b
\middle|
\bar B(p_B)
\right\rangle .
\end{aligned}
\end{align}

The factor $i$ in the hadronic tensor helicity amplitude is a conventional choice that makes the amplitude real. To compensate for this, a factor $-i$ is introduced in the leptonic tensor amplitude. Expressions for the hadronic helicity amplitudes in terms of form factors are provided in Appendix \ref{app:tradformfactors}. The scalar parts of $H_{S_L}^{\lambda_{D^*}}$ and $H_{S_R}^{\lambda_{D^*}}$ vanish \cite{tensorff}, so that $H_{S_L}^{\lambda_{D^*}}=-H_{S_R}^{\lambda_{D^*}}$.

Using the narrow-width approximation for the $D^*$ propagator, the squared amplitude becomes
\begin{align}
\label{eq:ampsqmuNv1}
\begin{aligned}
|\Amp^{\rm NP}|^2
&=
2G_F^2|V_{cb}|^2g_{D^*D\pi}^2
\frac{\pi}{m_{D^*}\Gamma_{D^*}}
\delta(p_{D^*}^2-m_{D^*}^2)
\\
&\quad\times
\Bigg|
\sum_{\lambda_{D^*}}
P(\lambda_{D^*})
\bigg[
C_S^{LR}H_{S_L}^{\lambda_{D^*}}L^S
+
C_S^{RR}H_{S_R}^{\lambda_{D^*}}L^S
+
C_V^{RR}\sum_m g_{mm}
H_{V_R,m}^{\lambda_{D^*}}L_m^V
\\
&\qquad\qquad
+
C_T^{RR}\sum_{m,n}g_{mm}g_{nn}
H_{T,mn}^{\lambda_{D^*}}L_{mn}^T
\bigg]
\Bigg|^2 .
\end{aligned}
\end{align}
Since the two-body decay branching ratio for $D^*\to D\pi$ is,
\begin{align}
\label{eq:DstarBRmu}
\mathcal B(D^*\to D\pi)
=
\frac{
g_{D^*D\pi}^2|\vec p_D|^3
}{
6\pi m_{D^*}^2\Gamma_{D^*}
}\,,
\end{align}
Eq.~\eqref{eq:ampsqmuNv1} can be expressed as
\begin{align}
\label{eq:ampsquaredmuN}
\begin{aligned}
|\Amp^{\rm NP}|^2
&=
\frac{
12\pi^2G_F^2|V_{cb}|^2m_{D^*}
}{
|\vec p_D|^3
}
\mathcal B(D^*\to D\pi)
\delta(p_{D^*}^2-m_{D^*}^2)
\\
&\quad\times
\Bigg|
\sum_{\lambda_{D^*}}
P(\lambda_{D^*})
\bigg[
C_S^{LR}H_{S_L}^{\lambda_{D^*}}L^S
+
C_S^{RR}H_{S_R}^{\lambda_{D^*}}L^S
+
C_V^{RR}\sum_m g_{mm}
H_{V_R,m}^{\lambda_{D^*}}L_m^V
\\
&\qquad\qquad
+
C_T^{RR}\sum_{m,n}g_{mm}g_{nn}
H_{T,mn}^{\lambda_{D^*}}L_{mn}^T
\bigg]
\Bigg|^2 .
\end{aligned}
\end{align}

\subsection{Phase space}\label{sec:phase-space-muN}

The decay chain ${\bar B}\to D^*(\to D\pi)\ell^-{\bar N}_R$ leads to a four-particle final state. The relevant kinematic variables are $q^2$, the polar angle $\theta_{D^*}$ of the $D$ meson three-momentum in the $D^*$ rest frame, the lepton angle $\theta_\ell$ in the $q^2$ rest frame, and the azimuthal angle $\chi$ between the hadronic and leptonic decay planes, as defined in Fig.~\ref{fig:angN-bar}.

Using the recursive decomposition of the phase space, we write
\begin{align}
\label{eq:phasespacerecursionrelationmuN}
d\Pi_4
=
\frac{dq^2}{2\pi}
\frac{dp_{D^*}^2}{2\pi}
d\Pi_2^B(B\to D^* W)
d\Pi_2^{D^*}(D^*\to D\pi)
d\Pi_2^{W}(W\to \ell\bar N_R),
\end{align}
where $q^2=p_W^2$ denotes the invariant mass squared of the fictitious $W$. The phase-space element for the primary $B$ decay, evaluated in the $B$ rest frame, is
\begin{align}
\label{eq:bphasespacemuN}
d\Pi_2^B
=
\frac{1}{2(2\pi)}
\int
\frac{|\vec p_{D^*}|\,d|\vec p_{D^*}|}{m_B}
\delta\left(
|\vec p_{D^*}|-
\frac{\sqrt{\lambda(m_B^2,m_{D^*}^2,q^2)}}{2m_B}
\right).
\end{align} 
For the secondary decay $D^*\to D\pi$, working in the $D^*$ rest frame, we obtain:
\begin{align}
\label{eq:dstarphasespacemuN}
d\Pi_2^{D^*}
=
\frac{1}{4(2\pi)}
\int
\frac{|\vec p_D|\,d|\vec p_D|\,d\cos\theta_{D^*}}{m_{D^*}}
\delta\left(
|\vec p_D|
-
\frac{\sqrt{\lambda(m_{D^*}^2,m_D^2,m_\pi^2)}}{2m_{D^*}}
\right).
\end{align}
The leptonic phase space for the secondary decay $W\to \ell\bar N_R$ is evaluated in the $q^2$ rest frame as follows.
\begin{align}
\label{eq:wphasespacemuN}
d\Pi_2^W
=
\frac{1}{4(2\pi)^2}
\int
\frac{|\vec p_\ell|\,d|\vec p_\ell|\,
d\cos\theta_\ell\,d\chi}{\sqrt{q^2}}
\delta\left(
|\vec p_\ell|-
\frac{\sqrt{\lambda(q^2,m_\ell^2,M_{N_R}^2)}}{2\sqrt{q^2}}
\right).
\end{align}

Combining Eqs.~\eqref{eq:bphasespacemuN}--\eqref{eq:wphasespacemuN} with Eq.~\eqref{eq:phasespacerecursionrelationmuN}, and putting it in Eq.~\eqref{eq:decayrategeneralmuN}, the differential decay rate becomes
\begin{align}
\label{eq:decayratewithphasespacemuN}
\frac{d\Gamma}
{dq^2\,d\cos\theta_{D^*}\,d\cos\theta_\ell\,d\chi}
=
\int
\frac{
dp_{D^*}^2\,
|\vec p_{D^*}|\,
|\vec p_D|\,
|\vec p_\ell|
}
{2^{12}\pi^6 m_B^2 m_{D^*}\sqrt{q^2}}
|\mathcal{M}|^2\,,
\end{align}
where, using the K\"all\'en function $\lambda(a,b,c) = a^2 + b^2 + c^2 - 2ab - 2bc - 2ca$,
\beq
|{\vec p}_{D^*}| = \frac{\sqrt{\lambda(m_B^2,m_{D^*}^2,q^2)}}{2m_B}\,,~~
|{\vec p}_D| = \frac{\sqrt{\lambda(m_{D^*}^2,m_D^2,m_\pi^2)}}{2m_{D^*}}\,,~~
|{\vec p}_\ell| = \frac{\sqrt{\lambda(q^2,m_\ell^2,M_{N_R}^2)}}{2\sqrt{q^2}}\,. \label{eq:momentums}
\eeq
In the $q^2$ rest frame, the lepton energy is $E_\ell = (q^2+m_\ell^2-M_{N_R}^2)/(2\sqrt{q^2})$, while the kinemtically allowed integration range for $q^2$ is $(m_\ell+M_{N_R})^2
\leq q^2 \leq (m_B-m_{D^*})^2$.

Using the amplitude squared obtained from the effective Hamiltonian (given in Eq.~\eqref{eq:ampsquaredmuN}), Eq.~\eqref{eq:decayratewithphasespacemuN} can be expressed in terms of angular functions as
\begin{align}
\label{eq:rateJmuN}
\frac{d\Gamma}
{dq^2\,d\cos\theta_{D^*}\,d\cos\theta_\ell\,d\chi}=
\sum\limits_{i = 1}^9 J_i(q^2)\,
f_i(\theta_{D^*}\, \theta_\ell\,\chi),
\end{align}
where the functions $f_i$, listed earlier in Eq.~\eqref{eq:rateJ}, contain the full angular dependence. The functions $J_i(q^2)$, referred to as the ``$J$ functions,'' are given in Appendix \ref{app:J functions}.

\section{\boldmath$J$ functions}
\label{app:J functions}

The $J$ functions in the angular distribution of $\bar B\to D^*(\to D\pi)\ell\bar N_R$ (Eqs.~\eqref{eq:rateJ} and \eqref{eq:rateJmuN}) are given below.
The SM angular coefficients are,
\bea
J_{1s}^{\text{(SM)}} &=& \frac{k}{2}\left(m_\ell^2 + 3q^2\right)\left(|H_V^-|^2 + |H_V^+|^2\right)\,, \\
J_{1c}^{\text{(SM)}} &=& k \left[2\left(m_\ell^2+q^2\right) |H_V^0|^2 + 4m_\ell^2 |H_t|^2\right]\,, \\
J_{2s}^{\text{(SM)}} &=& -\frac{k}{2}\left(m_\ell^2-q^2\right)\left(|H_V^-|^2 + |H_V^+|^2\right)\,, \\
J_{2c}^{\text{(SM)}} &=& 2k\left(m_\ell^2 - q^2\right)|H_V^0|^2\,, \\
J_{3}^{\text{(SM)}} &=& k\left(m_\ell^2 - q^2\right)\left(H_V^+ H_V^{-*} + H_V^- H_V^{+*}\right)\,, \\
J_{4}^{\text{(SM)}} &=& -\frac{k}{2} \left(m_\ell^2 - q^2\right)
\left[H_V^{0*}\left(H_V^- + H_V^+\right) + H_V^0\left(H_V^{-*} + H_V^{+*}\right) \right]\,, \\
J_{5}^{\text{(SM)}} &=& k\Big[q^2 \left\{ H_V^{0*}\left(H_V^- - H_V^+\right) + H_V^0 \left(H_V^{-*} - H_V^{+*}\right)\right\} \nonumber\\
&& ~~~~~+~m_\ell^2\left\{H_t^*\left(H_V^- + H_V^+\right) + H_t\left(H_V^{-*} + H_V^{+*} \right) \right\} \Big], \\
J_{6s}^{\text{(SM)}} &=& 2kq^2\left(|H_V^-|^2 - |H_V^+|^2\right)\, \\
J_{6c}^{\text{(SM)}} &=& -\,4k m_\ell^2 \left(H_t H_V^{0*} + H_V^0 H_t^*\right)\,, \\
J_{7}^{\text{(SM)}} &=& i\,k\,\Big[q^2\left\{ H_V^{0*} \left(H_V^- + H_V^+\right) - H_V^0\left(H_V^{-*} + H_V^{+*}\right)\right\} \nonumber\\
&& ~~~~~+~m_\ell^2\left\{H_t^*\left(H_V^- - H_V^+\right) - H_t\left(H_V^{-*} - H_V^{+*}\right)\right\}\Big]\,, \\
J_{8}^{\text{(SM)}} &=& \frac{ik}{2}\left(m_\ell^2 - q^2\right) \left[H_V^{0*}\left(H_V^+ - H_V^-\right) + H_V^0\left(H_V^{-*} - H_V^{+*}\right)\right]\,, \\
J_{9}^{\text{(SM)}} &=& -\,i\,k\left(m_\ell^2 - q^2\right)\left(H_V^+ H_V^{-*} - H_V^- H_V^{+*}\right)\,,
\eea
where $k = |\vec{p}_D|^2(q^2 - m_\ell^2)/(4\,q^2) = |\vec{p}_D|^2\,|\vec{p}_\ell|/(2\sqrt{q^2})$. The NP angular coefficients with the $C_V^{RR}$ coupling can be written as follows.
\bea
J_{1s}^{(C_V^{RR})} &=&
\frac{|C_V^{RR}|^2 |p_D|^2}{2}
\left(
|H_V^-|^2 + |H_V^+|^2
\right)
\left[
2E_\ell\left(\sqrt{q^2}-E_\ell\right)
+|p_\ell|^2
\right],
\\
J_{1c}^{(C_V^{RR})} &=&
-2|C_V^{RR}|^2 |p_D|^2
\Big[
E_\ell\left(E_\ell - \sqrt{q^2}\right)\left(|H_t|^2 + |H_V^0|^2\right)
+ |p_\ell|^2|H_t|^2\Big], \\
J_{2s}^{(C_V^{RR})} &=&
\frac{|C_V^{RR}|^2 |p_D|^2}{2}
|p_\ell|^2
\left(|H_V^-|^2 + |H_V^+|^2\right)\,, \\
J_{2c}^{(C_V^{RR})} &=& - 2|C_V^{RR}|^2 |p_D|^2 |p_\ell|^2 |H_V^0|^2\,, \\
J_{3}^{(C_V^{RR})} &=&
-|C_V^{RR}|^2 |p_D|^2 |p_\ell|^2
\left(
H_V^+ H_V^{-*}
+
H_V^- H_V^{+*}
\right),
\\
J_{4}^{(C_V^{RR})} &=&
\frac{|C_V^{RR}|^2 |p_D|^2}{2}
|p_\ell|^2
\Big[
H_V^{0*}\left(H_V^-+H_V^+\right)
+
H_V^0 \left(H_V^{-*}
+
H_V^{+*}\right)
\Big],
\\
J_{5}^{(C_V^{RR})} &=&
\frac{|C_V^{RR}|^2 |p_D|^2}{2}
|p_\ell|
\Bigg[
\sqrt{q^2}
\Big[
H_V^{0*}\left(H_V^- - H_V^+\right)
+H_V^0\left(H_V^{-*} - H_V^{+*}\right)
\Big]
\nonumber\\
&&~~~~~~~~~~
+~\left(2E_\ell-\sqrt{q^2}\right)
\Big[
H_t^*\left(H_V^- + H_V^+\right)
+H_t\left(H_V^{-*} + H_V^{+*}\right)
\Big]
\Bigg],
\\
J_{6s}^{(C_V^{RR})} &=& |C_V^{RR}|^2 |p_D|^2 |p_\ell|\sqrt{q^2} \left(|H_V^-|^2 - |H_V^+|^2\right)\,, \\
J_{6c}^{(C_V^{RR})} &=&
-2|C_V^{RR}|^2 |p_D|^2
|p_\ell|
\left(
2E_\ell-\sqrt{q^2}
\right)
\left(
H_t H_V^{0*}
+
H_V^0 H_t^*
\right),
\\
J_{7}^{(C_V^{RR})} &=&
\frac{i|C_V^{RR}|^2 |p_D|^2}{2}
|p_\ell|
\Bigg[
\sqrt{q^2}
\Big[
H_V^0\left(H_V^{-*}+H_V^{+*}\right)
-H_V^{0*}\left(H_V^-+H_V^+\right)
\Big]
\nonumber\\
&&~~~~~~~~~~~~
+\left(2E_\ell-\sqrt{q^2}\right)
\Big[
H_t\left(H_V^{-*}-H_V^{+*}\right)
-H_t^*\left(H_V^- - H_V^+\right)
\Big]
\Bigg],
\\
J_{8}^{(C_V^{RR})} &=&
\frac{i|C_V^{RR}|^2 |p_D|^2}{2}
|p_\ell|^2
\Big[
H_V^{0*}
\left(H_V^+-H_V^-\right)
+
H_V^0\left(H_V^{-*}
-
H_V^{+*}\right)
\Big],
\\
J_{9}^{(C_V^{RR})} &=&
-i|C_V^{RR}|^2 |p_D|^2 |p_\ell|^2
\left(
H_V^+ H_V^{-*}
-
H_V^- H_V^{+*}
\right)\,.
\eea
The NP angular coefficients corresponding to the $C_V^{LR}$ coupling are:
\bea
J_{1s}^{(C_V^{LR})} &=&
\frac{|C_V^{LR}|^2 |p_D|^2}{2}
\left(
|H_V^-|^2 + |H_V^+|^2
\right)
\left[
2E_\ell\left(\sqrt{q^2}-E_\ell\right)
+|p_\ell|^2
\right],
\\
J_{1c}^{(C_V^{LR})} &=&
-2|C_V^{LR}|^2 |p_D|^2
\Big[
E_\ell\left(E_\ell-\sqrt{q^2}\right)
\left(|H_t|^2+|H_V^0|^2\right)
+|p_\ell|^2|H_t|^2
\Big],
\\
J_{2s}^{(C_V^{LR})} &=&
\frac{|C_V^{LR}|^2 |p_D|^2}{2}
|p_\ell|^2
\left(|H_V^-|^2 + |H_V^+|^2\right)\,, \\
J_{2c}^{(C_V^{LR})} &=&
-2|C_V^{LR}|^2 |p_D|^2 |p_\ell|^2
|H_V^0|^2\,, \\
J_{3}^{(C_V^{LR})} &=&
-|C_V^{LR}|^2 |p_D|^2 |p_\ell|^2
\left(
H_V^+ H_V^{-*}
+
H_V^- H_V^{+*}
\right),
\\
J_{4}^{(C_V^{LR})} &=&
\frac{|C_V^{LR}|^2 |p_D|^2}{2}
|p_\ell|^2
\Big[
H_V^{0*}\left(H_V^-+H_V^+\right)
+
H_V^0 \left(H_V^{-*}
+ H_V^{+*} \right)
\Big],
\\
J_{5}^{(C_V^{LR})} &=&
\frac{|C_V^{LR}|^2 |p_D|^2}{2}
|p_\ell|
\Bigg[
\sqrt{q^2}
\Big[
H_V^{0*}\left(H_V^+ - H_V^-\right)
+H_V^0\left(H_V^{+*} - H_V^{-*}\right)
\Big]
\nonumber\\
&&~~~~~~~~~~
+\left(2E_\ell-\sqrt{q^2}\right)
\Big[
H_t^*\left(H_V^- + H_V^+\right)
+H_t\left(H_V^{-*} + H_V^{+*}\right)
\Big]
\Bigg],
\\
J_{6s}^{(C_V^{LR})} &=&
|C_V^{LR}|^2 |p_D|^2
|p_\ell|\sqrt{q^2}
\left(|H_V^+|^2 - |H_V^-|^2\right)\,, \\
J_{6c}^{(C_V^{LR})} &=&
-2|C_V^{LR}|^2 |p_D|^2
|p_\ell|
\left(
2E_\ell-\sqrt{q^2}
\right)
\left(
H_t H_V^{0*}
+
H_V^0 H_t^*
\right),
\\
J_{7}^{(C_V^{LR})} &=&
-\frac{i|C_V^{LR}|^2 |p_D|^2}{2}
|p_\ell|
\Bigg[
\sqrt{q^2}
\Big[
H_V^{0*}\left(H_V^- + H_V^+\right)
-H_V^0\left(H_V^{-*} + H_V^{+*}\right)
\Big]
\nonumber\\
&&~~~~~~~~~~~~
+\left(2E_\ell-\sqrt{q^2}\right)
\Big[
H_t\left(H_V^{-*} - H_V^{+*}\right)
-H_t^*\left(H_V^- - H_V^+\right)
\Big]
\Bigg],
\\
J_{8}^{(C_V^{LR})} &=&
-\frac{i|C_V^{LR}|^2 |p_D|^2}{2}
|p_\ell|^2
\Big[
H_V^{0*}
\left(H_V^+-H_V^-\right)
+
H_V^0\left(H_V^{-*}
- H_V^{+*}\right)
\Big],
\\
J_{9}^{(C_V^{LR})} &=&
i|C_V^{LR}|^2 |p_D|^2 |p_\ell|^2
\left(
H_V^+ H_V^{-*}
-
H_V^- H_V^{+*}
\right).
\eea
The only nonzero NP angular coefficient corresponding to the $C_S^{RR}/C_S^{LR}$ coupling is
\beq
J_{1c}^{C_S^{RR}(C_S^{LR})} ~=~ 2\, |C_S^{RR}(C_S^{LR})|^2\, |H_S^P|^2\, |p_D|^2\, \left(-E_\ell^2 + E_\ell \sqrt{q^2} + |p_\ell|^2\right)\,.
\eeq
Finally, the NP angular coefficients corresponding to the $C_T^{RR}$ coupling are as follows.
\bea
J_{1s}^{(C_T^{RR})} &=&
-8 |C_T^{RR}|^2 |p_D|^2
\left(
|H_T^-|^2 + |H_T^+|^2
\right)
\left[
2E_\ell\left(E_\ell-\sqrt{q^2}\right)
+|p_\ell|^2
\right],
\\
J_{1c}^{(C_T^{RR})} &=&
-~32 |C_T^{RR}|^2 |p_D|^2
E_\ell \left(E_\ell - \sqrt{q^2}\right)
|H_T^0|^2,
\\
J_{2s}^{(C_T^{RR})} &=&
-8 |C_T^{RR}|^2 |p_D|^2 |p_\ell|^2
\left(|H_T^-|^2 + |H_T^+|^2\right)\,, \\
J_{2c}^{(C_T^{RR})} &=&
32 |C_T^{RR}|^2 |p_D|^2 |p_\ell|^2 |H_T^0|^2\,, \\
J_{3}^{(C_T^{RR})} &=&
-16 |C_T^{RR}|^2 |p_D|^2 |p_\ell|^2
\left(
H_T^+ H_T^{-*}
+
H_T^- H_T^{+*}
\right),
\\
J_{4}^{(C_T^{RR})} &=&
8 |C_T^{RR}|^2 |p_D|^2 |p_\ell|^2
\Big[
H_T^{0*}\left(H_T^+ - H_T^-\right)
+H_T^0\left(H_T^{+*} - H_T^{-*}\right)
\Big],
\\
J_{5}^{(C_T^{RR})} &=&
8 |C_T^{RR}|^2 |p_D|^2 |p_\ell|
\left(2E_\ell-\sqrt{q^2}\right)
\Big[
H_T^{0*}\left(H_T^- + H_T^+\right)
+H_T^0\left(H_T^{-*} + H_T^{+*}\right)
\Big],
\\
J_{6s}^{(C_T^{RR})} &=&
16 |C_T^{RR}|^2 |p_D|^2 |p_\ell|
\left(2E_\ell-\sqrt{q^2}\right)
\left(|H_T^-|^2 - |H_T^+|^2\right)\,, \\
J_{6c}^{(C_T^{RR})} &=& 0,
\\
J_{7}^{(C_T^{RR})} &=&
8i |C_T^{RR}|^2 |p_D|^2 |p_\ell|
\left(2E_\ell-\sqrt{q^2}\right)
\Big[
H_T^{0*}\left(H_T^+ - H_T^-\right)
+H_T^0\left(H_T^{-*} - H_T^{+*}\right)
\Big],
\\
J_{8}^{(C_T^{RR})} &=&
8i |C_T^{RR}|^2 |p_D|^2 |p_\ell|^2
\Big[
H_T^{0*}\left(H_T^- + H_T^+\right)
-H_T^0\left(H_T^{-*} + H_T^{+*}\right)
\Big],
\\
J_{9}^{(C_T^{RR})} &=&
-16i |C_T^{RR}|^2 |p_D|^2 |p_\ell|^2
\left(
H_T^+ H_T^{-*}
-
H_T^- H_T^{+*}
\right).
\eea

\section{Implementation and Validation of the RHN model in EvtGen}
\label{app:EvtHistogram}

\begin{figure}[!htbp]
\centering
\includegraphics[width=0.47\textwidth]{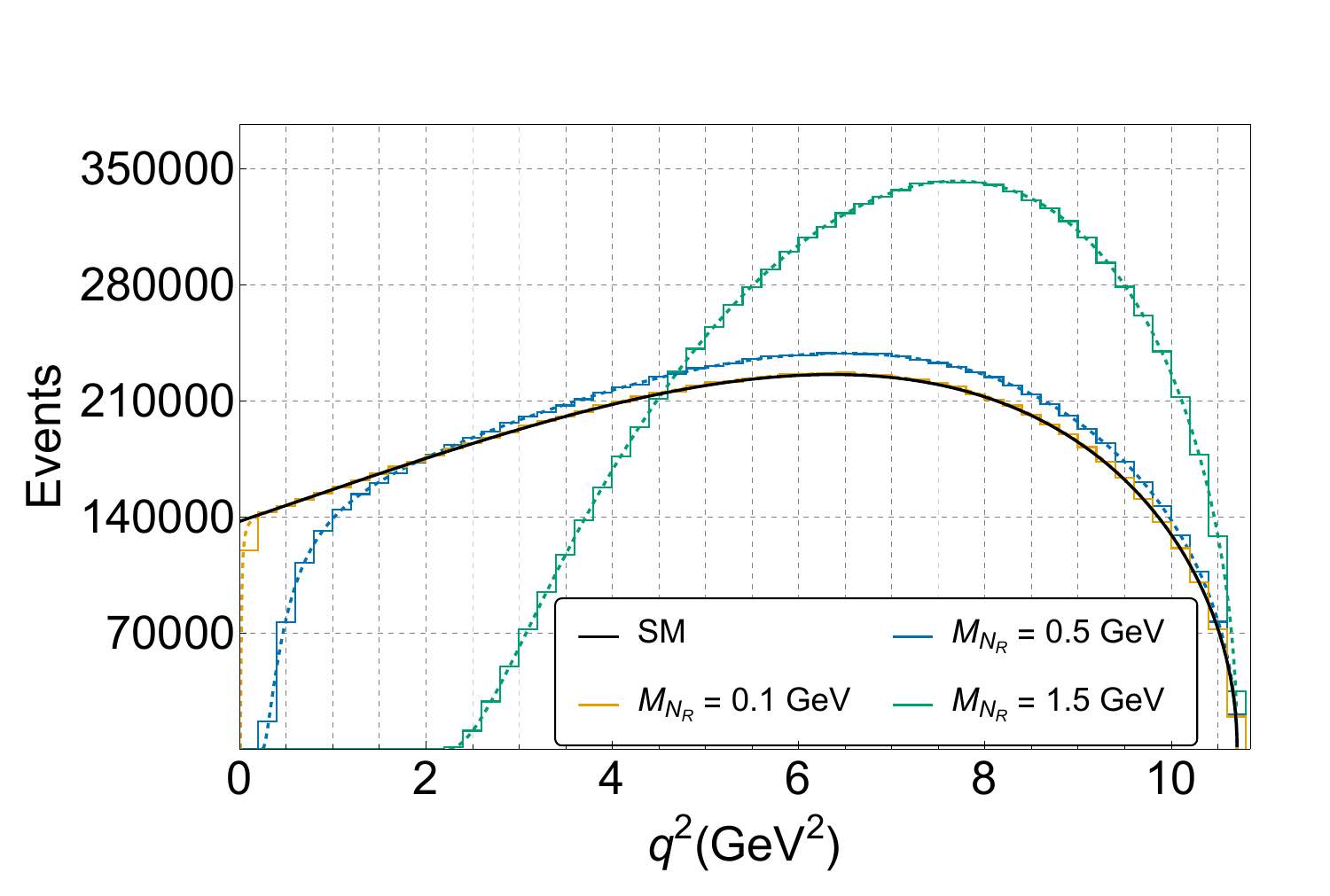}
\includegraphics[width=0.47\textwidth]{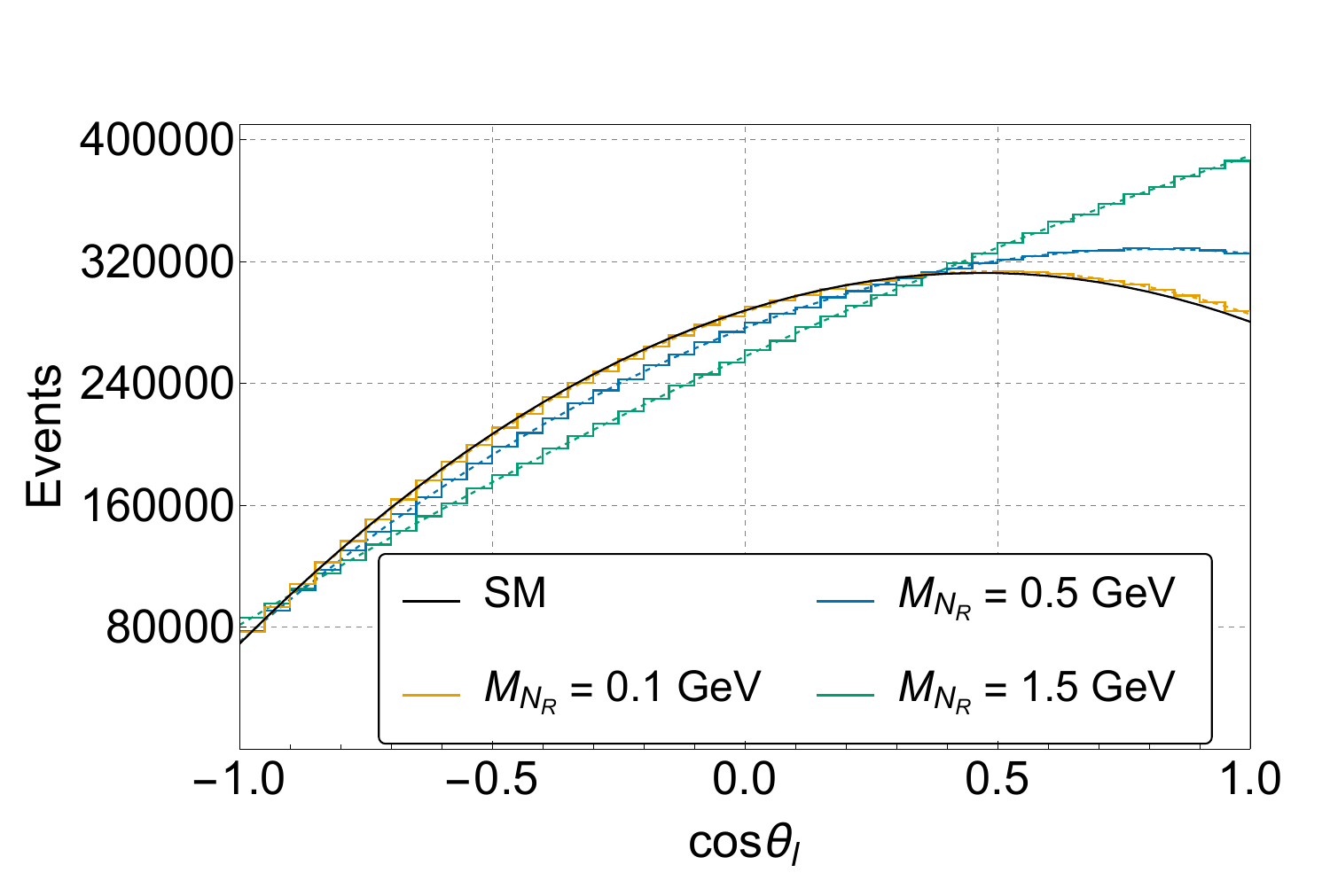}
\includegraphics[width=0.47\textwidth]{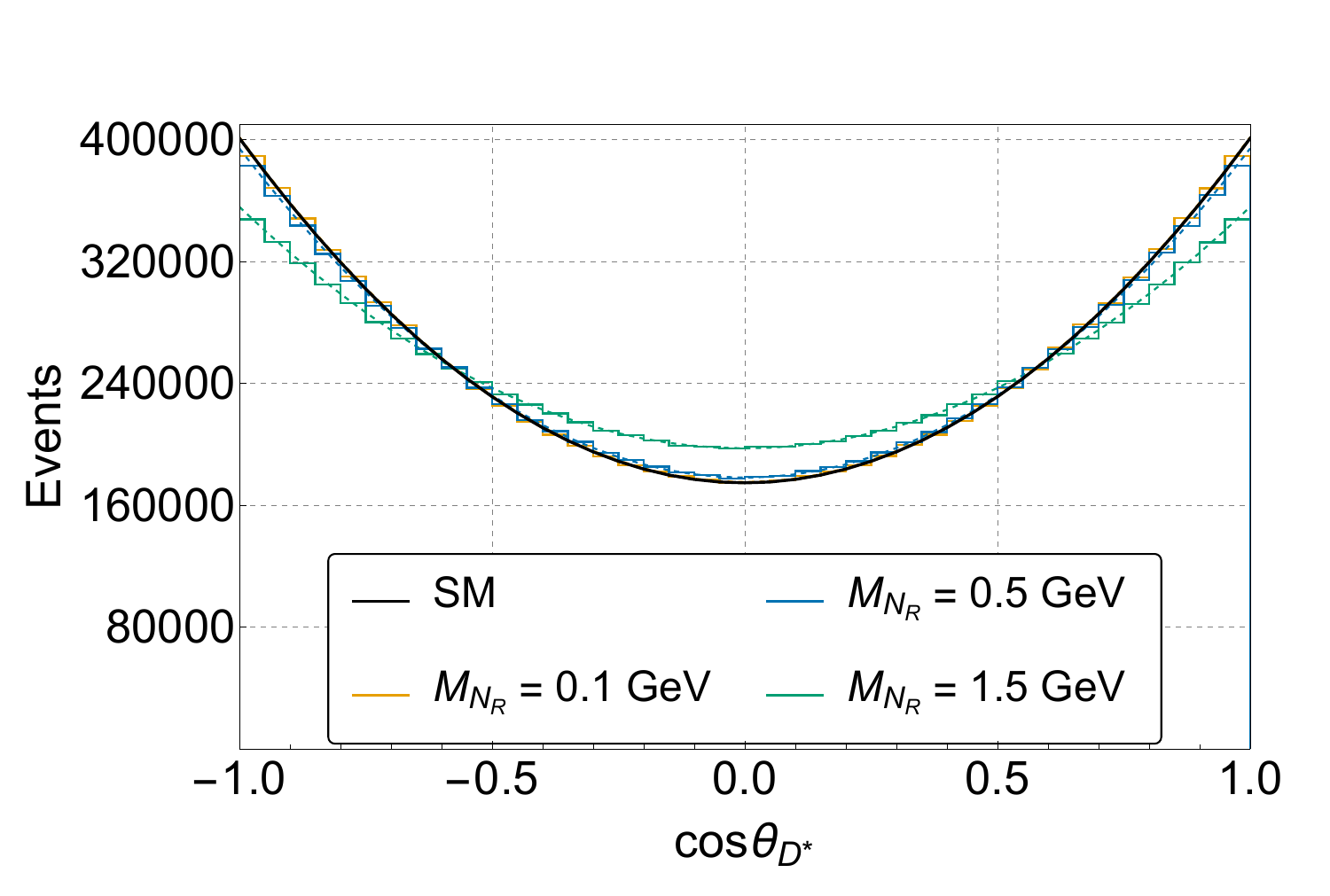}
\includegraphics[width=0.47\textwidth]{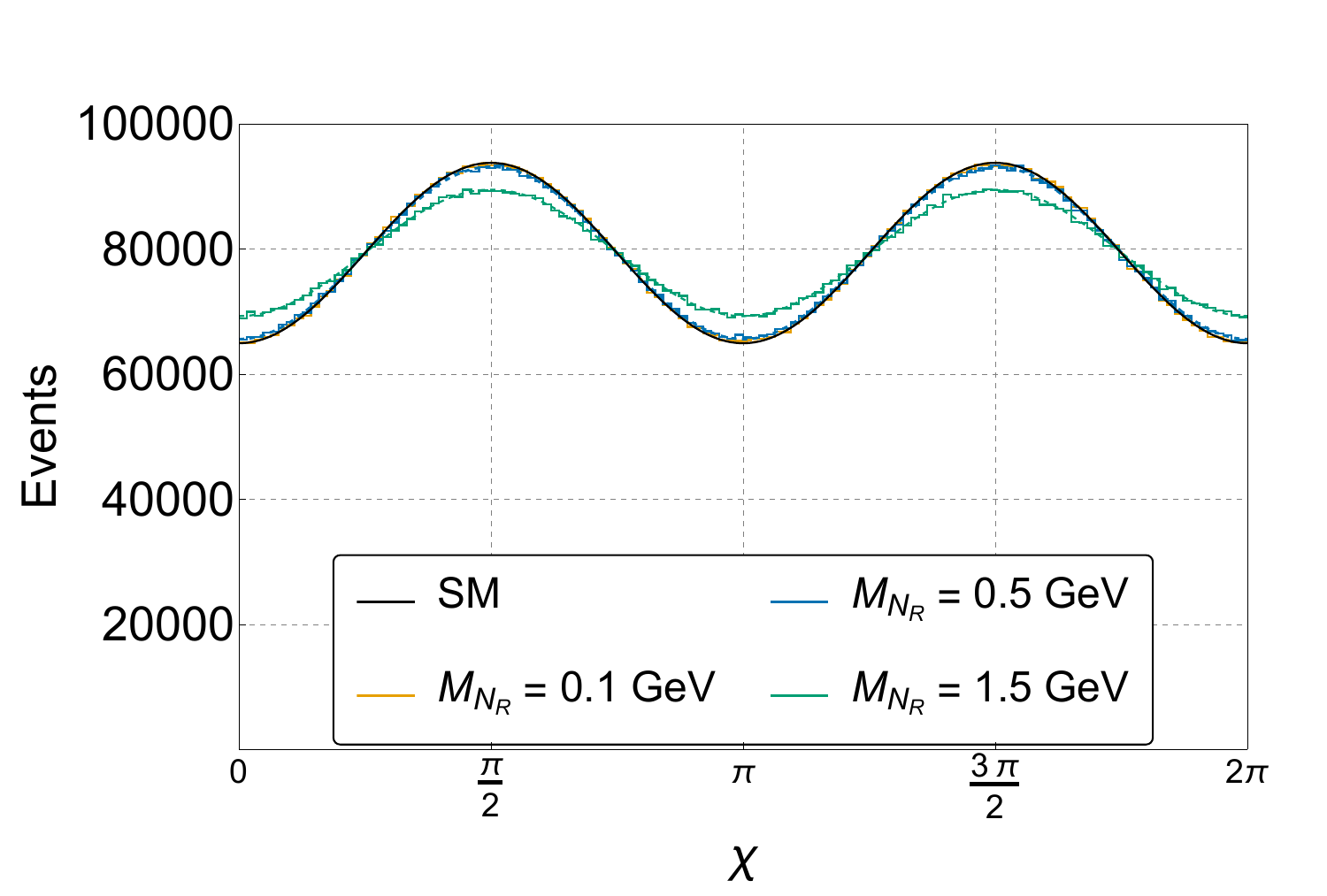}
\caption{Kinematic distributions of $B^+\to{\bar D}^{*0}e^{+}N_R$ events as functions of (clockwise from top left) $q^2$, $\cos\theta_\ell$, $\cos\theta_{D^*}$, and $\chi$ for the three benchmark masses $M_{N_R} = 0.1$, $0.5$, and $1.5$ GeV with $C_{V_{RR}} = 1$. EvtGen data are shown as histograms and the corresponding theoretical predictions as dashed curves; the solid black curve shows the SM expectation. The MC data and theoretical curves agree.}
\label{fig:EvtGenDistEle}
\end{figure}
For each of the three benchmark RHN masses, $M_{N_R} = 0.1$ GeV, 0.5 GeV, and 1.5 GeV, we reconstruct the distributions of the generated events as functions of the four kinematic variables, $q^2$, $\cos\theta_\ell$, $\cos\theta_{D^*}$, and $\chi$. Starting from the differential distribution given in Eq.~\eqref{eq:rateJ}, these kinematic event distributions are obtained by integrating over the remaining three variables. Since the RHN amplitude does not interfere with the SM ($N_R \neq \nu_L$), the NP samples are generated with $C_{V_{RR}} = 1$ and the SM contribution is switched off using $C_{V_{LL}} = -1$\footnote{Within our EvtGen model, setting the NP coefficient $C_{V_{LL}} = -1$ eliminates the SM contribution from the operator $O_V^{LL}$.}. The histograms obtained in this way represent pure RHN distributions.

\begin{figure}[!htbp]
\centering
\includegraphics[width=0.47\textwidth]{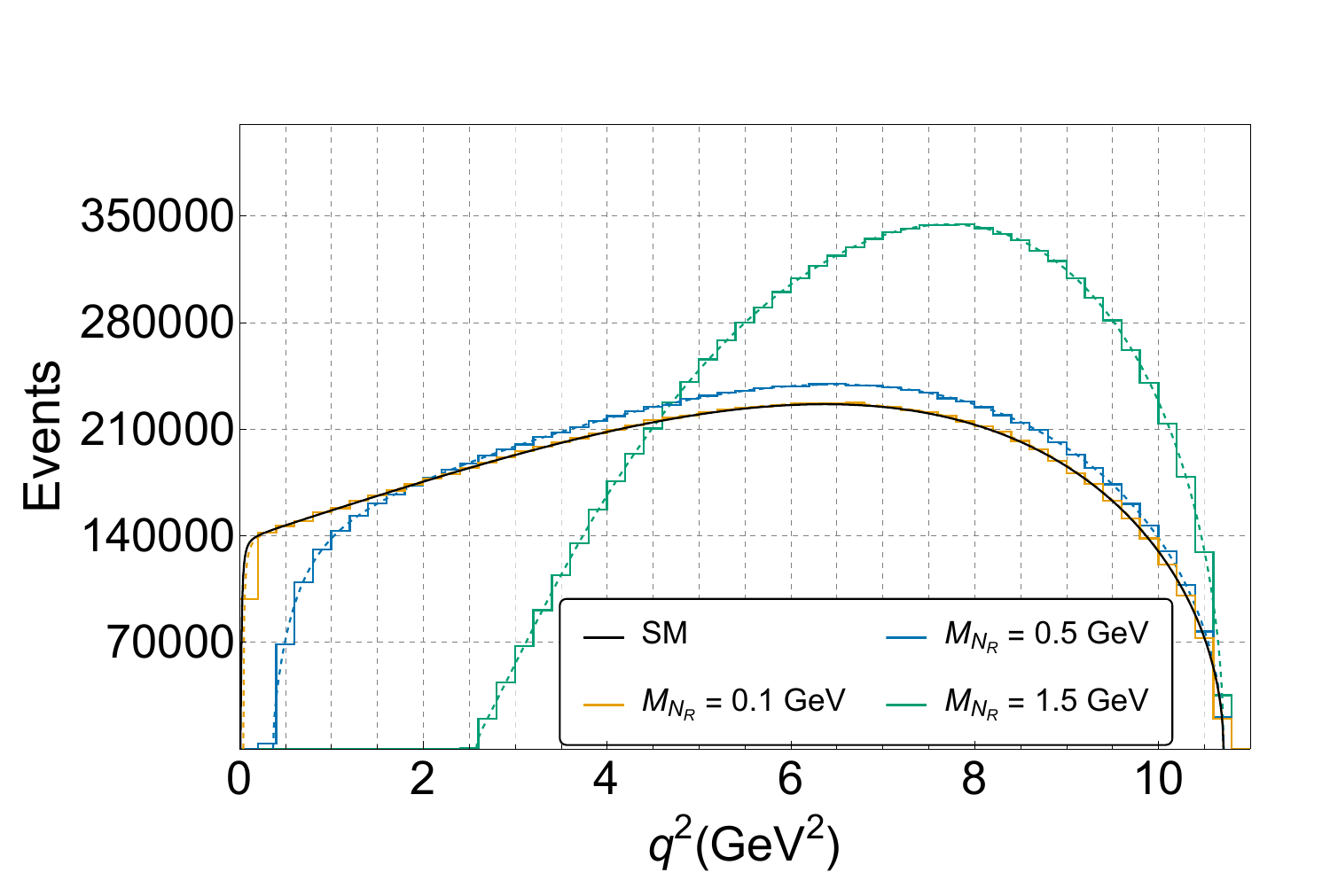}
\includegraphics[width=0.47\textwidth]{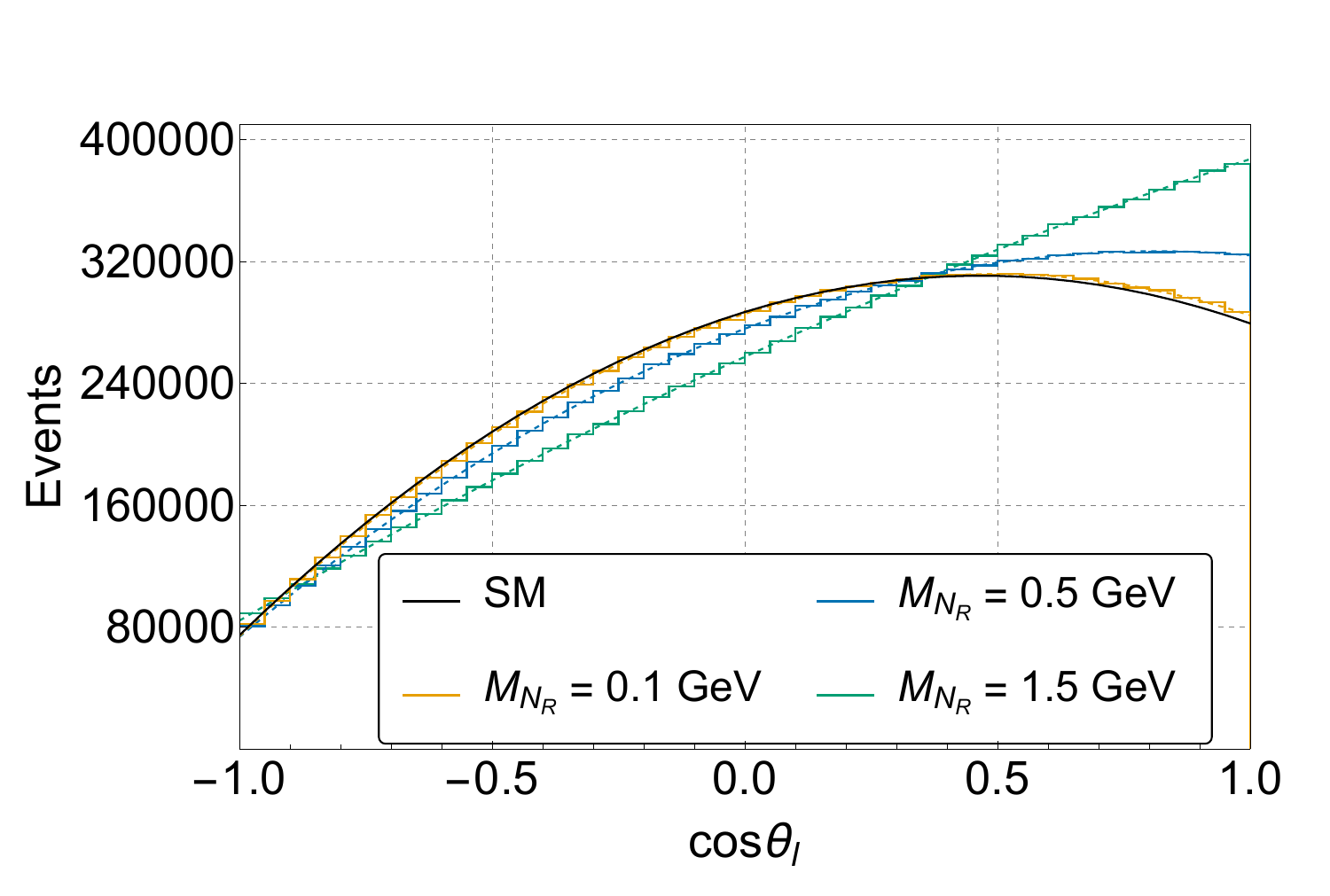}
\includegraphics[width=0.47\textwidth]{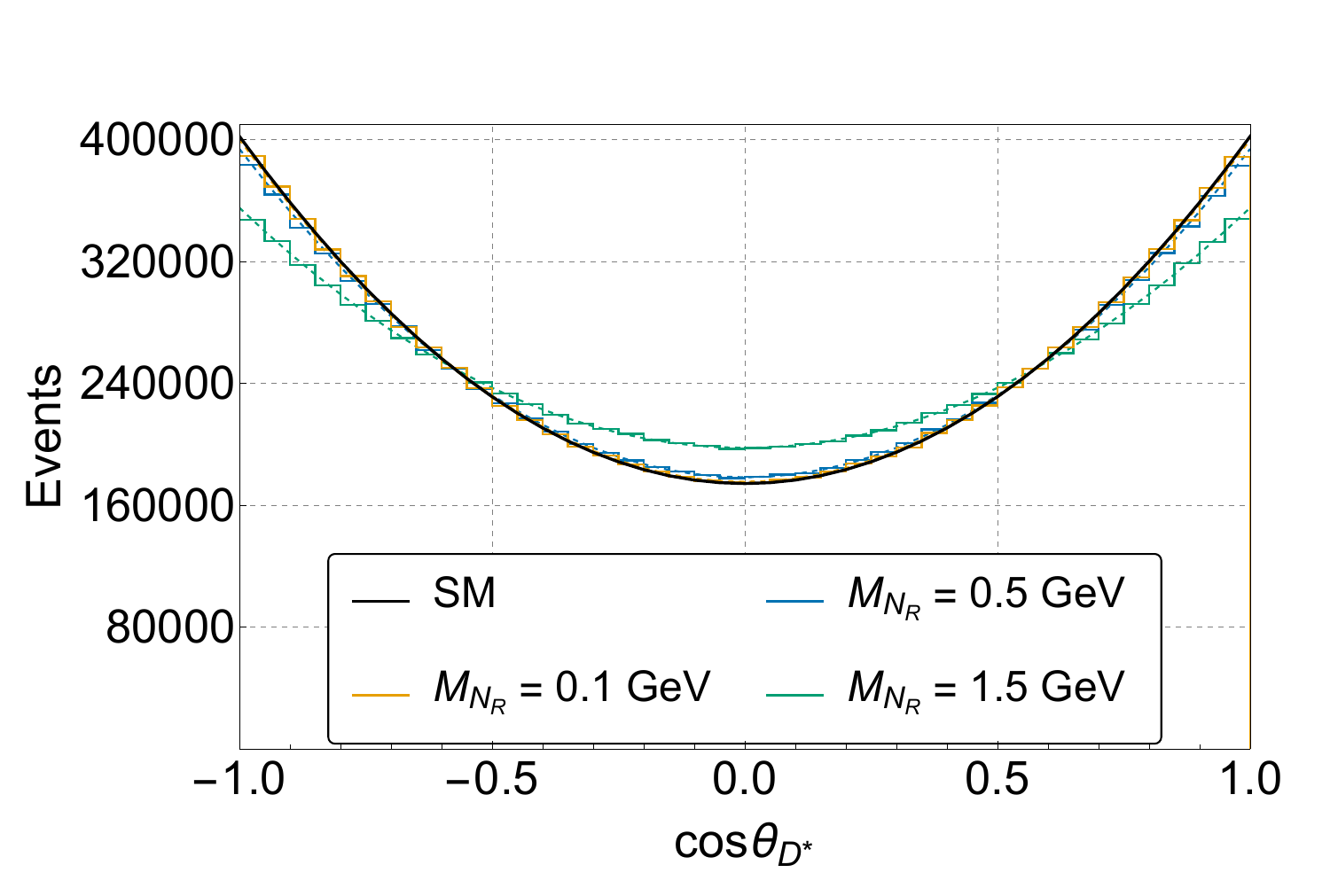}
\includegraphics[width=0.47\textwidth]{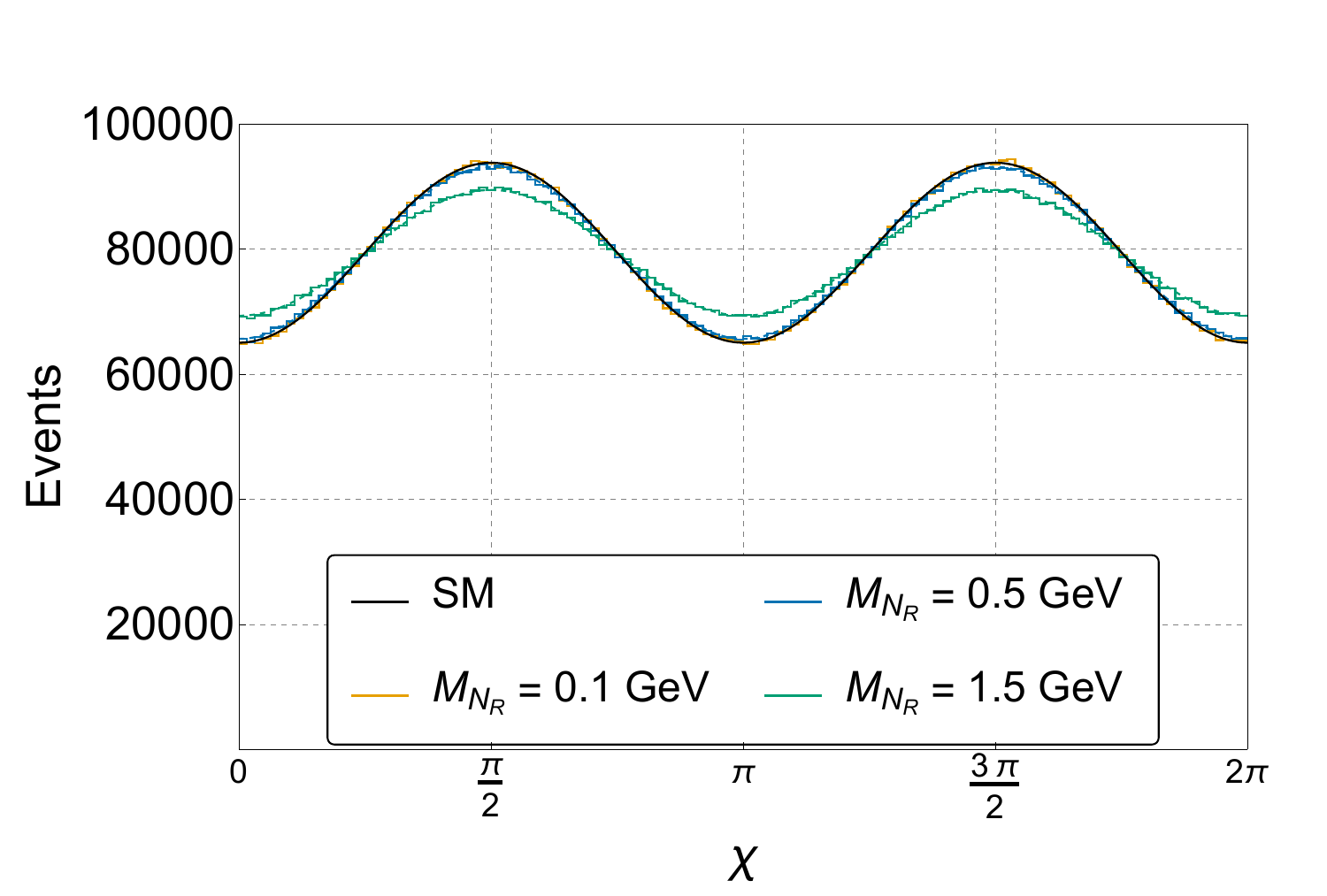}
\caption{Kinematic distributions of $B^+\to{\bar D}^{*0}\mu^{+}N_R$ events as functions of (clockwise from top left) $q^2$, $\cos\theta_\ell$, $\cos\theta_{D^*}$, and $\chi$ for the three benchmark masses $M_{N_R} = 0.1$, $0.5$, and $1.5$ GeV with $C_{V_{RR}} = 1$. EvtGen data are shown as histograms and the corresponding theoretical predictions as dashed curves; the solid black curve shows the SM expectation. The MC data and theoretical curves agree.}
\label{fig:EvtGenDistMu}
\end{figure}
The results for the electron and muon modes are shown in Figs.~\ref{fig:EvtGenDistEle} and \ref{fig:EvtGenDistMu}, respectively. In each panel, the EvtGen data are shown as solid colored histograms. At the same time, the corresponding theoretical predictions, obtained from Eq.~\eqref{eq:rateJ} with the BGL form-factor parametrization of Appendix \ref{app:bglformfactors}, are overlaid as dashed
curves of the same color. The solid black curves show the SM expectation. The agreement between the MC data and the theoretical curves across all four kinematic variables, all three benchmark masses, and both lepton channels validates the implementation of the EvtGen model.

\section{ EvtGen-Theory Correlation Plots}
\label{app:EvtGen-Theory Correlation}

\begin{figure}[!ht]
\centering
\includegraphics[width=0.45\textwidth]{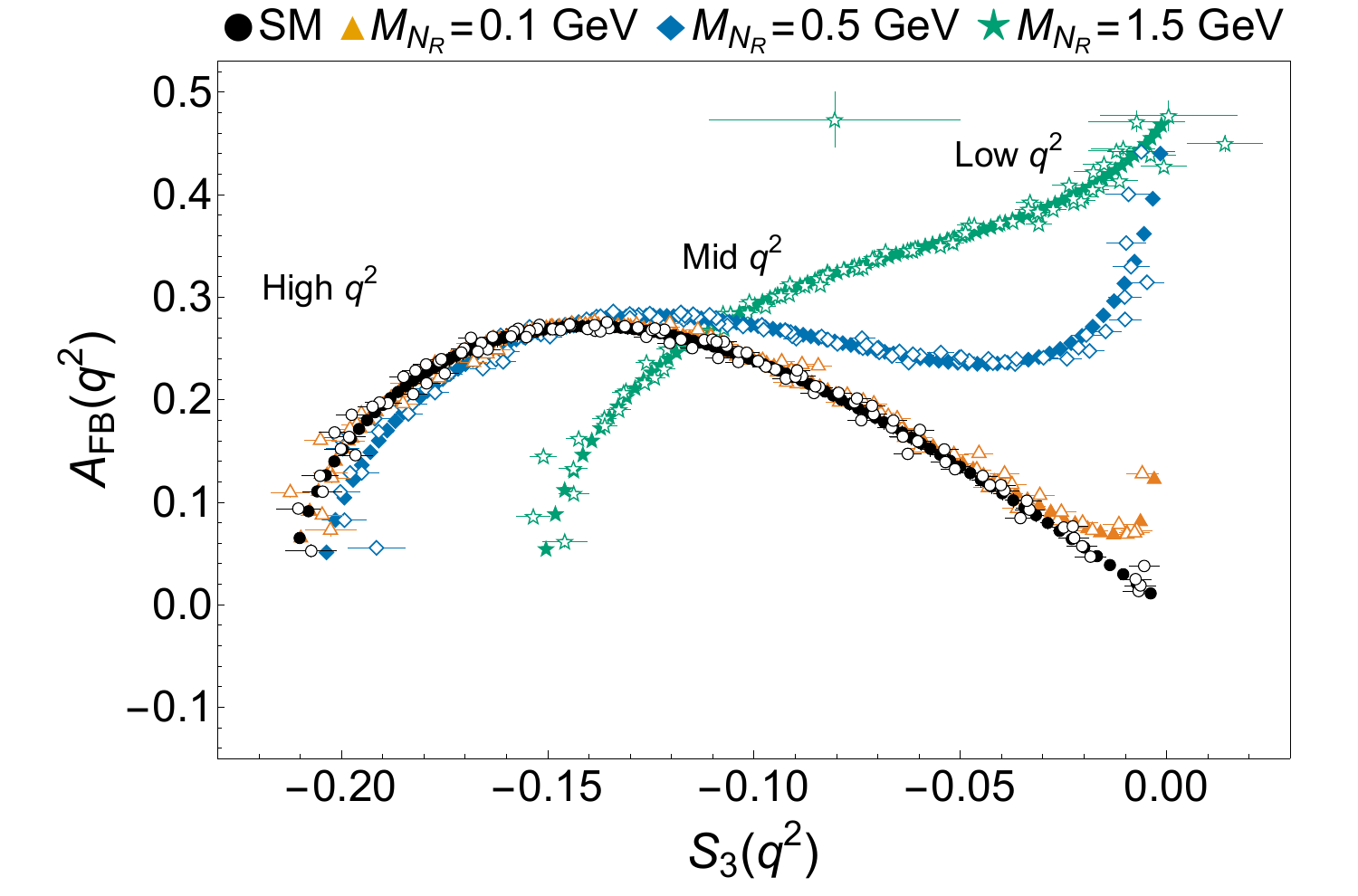}
\includegraphics[width=0.45\textwidth]{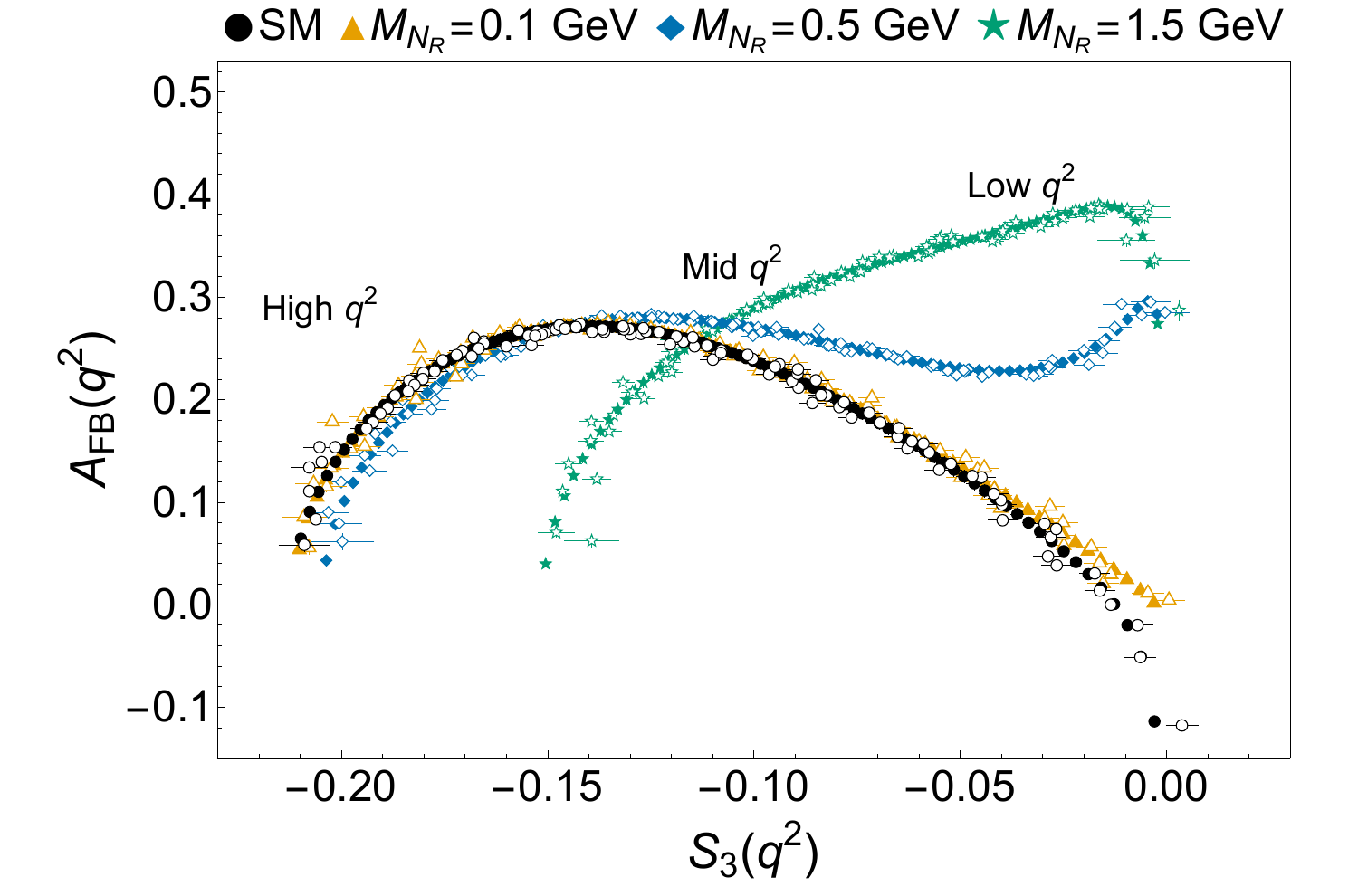}
\includegraphics[width=0.45\textwidth]{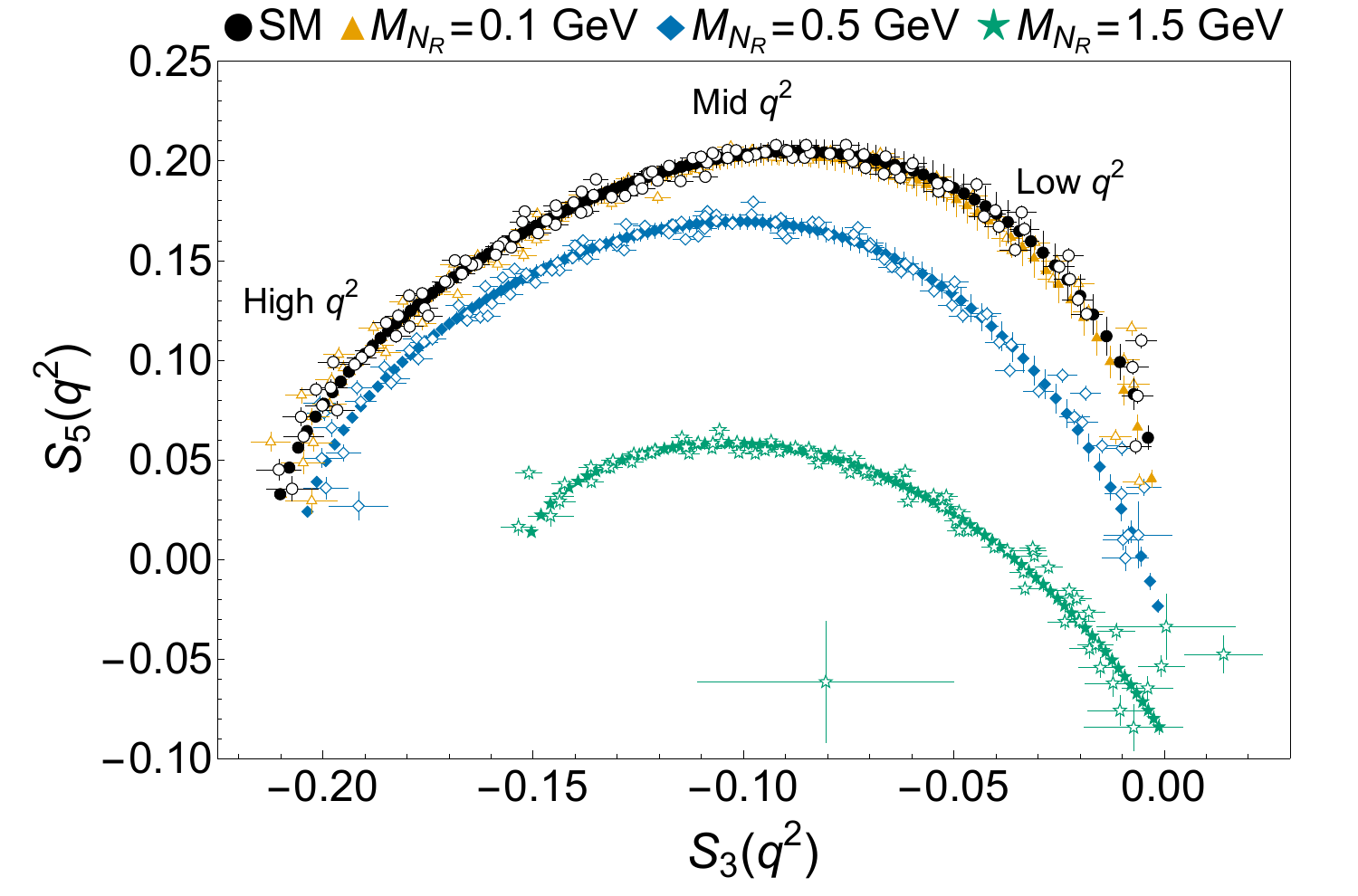}
\includegraphics[width=0.45\textwidth]{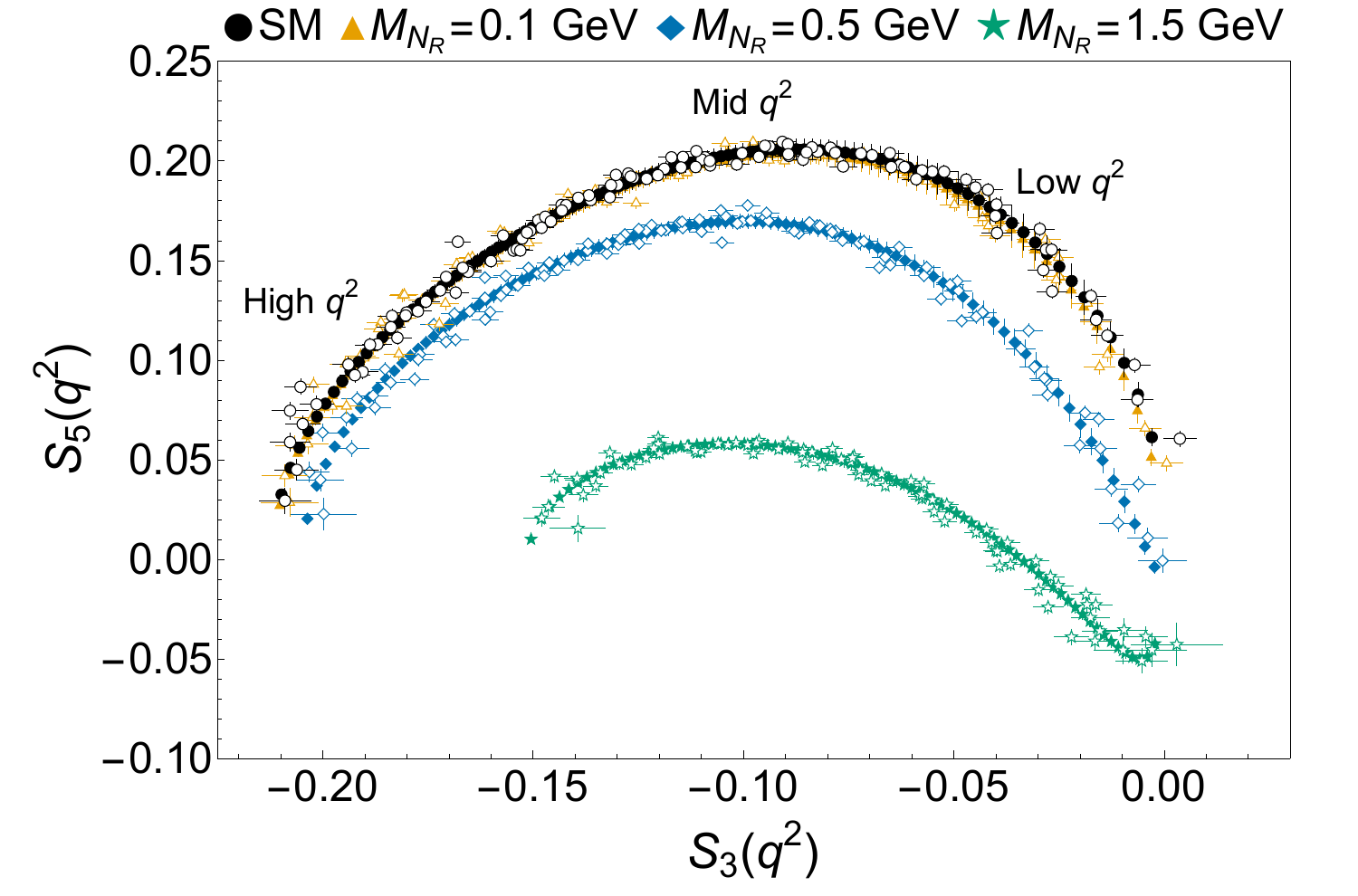}
\caption{Correlation between angular asymmetries $A_{FB}$ and $S_3$ (top panels) and $S_5$ and $S_3$ (bottom panels). The two left panels show the electron channel, while the two right panels show the muon channel. Results for the SM (black) and RHN with mass $M_{N_R} = $ 0.1 GeV (orange), 0.5 GeV (blue), and 1.5 GeV (teal) are shown. Open (closed) points represent results obtained from EvtGen simulations (theoretical expressions). We use 0.1 GeV$^2$-wide $q^2$ bins for both the theory and simulated results. \label{fig:afb_s3_corrEvt}}
\end{figure}
The plots in the top panel of Fig.~\ref{fig:afb_s3_corrEvt} show the correlation between $A_{FB}$ and $S_3$ for both the electron (left panel) and muon (right panel) channels. Similarly, the plots in the bottom panel of Fig.~\ref{fig:afb_s3_corrEvt} show the correlation between the two angular-asymmetry observables, $S_5$ and $S_3$. These plots demonstrate how the forward-backward asymmetry, $A_{FB}$, and the angular asymmetry, $S_5$, vary with the transverse angular asymmetry, $S_3$. These plots show the behavior in the SM and in the presence of an RHN. The largest deviations from the SM are observed for $M_{N_R} = 1.5$ GeV.

\bibliography{BDstarRHN.bib}

\end{document}